\documentclass[twocolumn]{aastex631}

\usepackage{amsmath}
\usepackage[toc,page]{appendix}
\shorttitle{Type Ia Supernova Nucleosynthesis}
\shortauthors{Keegans et al.}
\graphicspath{{./}{figures/}}

\begin{document}

\title{Type Ia Supernova Nucleosynthesis: 
Metallicity-Dependent Yields}

\author[0000-0002-7365-9813]{James D. Keegans}
\affiliation{Astrophysics Group, Lennard-Jones Laboratories, Keele University, Keele ST5 5BG, UK}
\affiliation{E.A. Milne Centre for Astrophysics, University of Hull, Hull, HU6~7RX, UK}
\affiliation{NuGrid Collaboration, \url{http://nugridstars.org}}

\author{Marco Pignatari}
\affiliation{Konkoly Observatory, Research Centre for Astronomy and Earth Sciences, Hungarian Academy of Sciences, Konkoly Thege Miklos ut 15-17, H-1121 Budapest, Hungary}
\affiliation{CSFK, MTA Centre of Excellence, Budapest, Konkoly Thege Miklós út 15-17, H-1121, Hungary}
\affiliation{E.A. Milne Centre for Astrophysics, University of Hull, Hull, HU6~7RX, UK}
\affiliation{NuGrid Collaboration, \url{http://nugridstars.org}}
\affiliation{Joint Institute for Nuclear Astrophysics - Center for the Evolution of the Elements, USA}

\author[0000-0002-6972-9655]{Richard J. Stancliffe}
\affiliation{H. H. Wills Physics Laboratory, University of Bristol, Tyndall Avenue, Bristol BS8 1TL, UK}
\affiliation{E.A. Milne Centre for Astrophysics, University of Hull, Hull, HU6~7RX, UK}
\affiliation{NuGrid Collaboration, \url{http://nugridstars.org}}

\author{Claudia Travaglio}
\affiliation{INAF-Astrophysical Observatory, Turin, Italy}

\author[0000-0003-3970-1843]{Samuel Jones}
\affiliation{Theoretical Division, Los Alamos National Laboratory, Los Alamos, NM 87545, USA}

\author{Brad K. Gibson}
\affiliation{E.A. Milne Centre for Astrophysics, University of Hull, Hull, HU6~7RX, UK}

\author{Dean M. Townsley}

\author{Broxton J. Miles}

\author[0000-0002-9632-6106]{Ken J.\ Shen}
\affiliation{Department of Astronomy and Theoretical Astrophysics Center, University of California, Berkeley, CA 94720, USA}

\author{Gareth Few}
\affiliation{E.A. Milne Centre for Astrophysics, University of Hull, Hull, HU6~7RX, UK}



\begin{abstract}

Type Ia supernova explosions (SNIa) are fundamental sources of elements for the chemical evolution of galaxies. They efficiently produce intermediate-mass (with Z between 11 and 20) and iron group elements - for example, about 70$\%$ of the solar iron is expected to be made by SNIa. In this work, we calculate complete abundance yields for 39 models of SNIa explosions, based on three progenitors - a 1.4M$_{\odot}$ deflagration detonation model, a 1.0$_{\odot}$ double detonation model and a 0.8 M$_{\odot}$ double detonation model - and 13 metallicities, with $^{22}$Ne mass fractions of 0, 1$\times$10$^{-7}$, 1$\times$10$^{-6}$, 1$\times$10$^{-5}$, 1$\times$10$^{-4}$, 1$\times$10$^{-3}$, 2$\times$10$^{-3}$, 5$\times$10$^{-3}$, 1$\times$10$^{-2}$, 1.4$\times$10$^{-2}$, 5$\times$10$^{-2}$, and 0.1 respectively. Nucleosynthesis calculations are done using the NuGrid suite of codes, using a consistent nuclear reaction network between the models. Complete tables with yields and production factors are provided online at \dataset[Zenodo: Yields]{https://doi.org/10.5281/zenodo.8060323
}. 
We discuss the main properties of our yields in the light of the present understanding of SNIa nucleosynthesis, depending on different progenitor mass and composition. Finally, we compare our results with
a number of relevant models from the literature.

\end{abstract}



\section{Introduction} \label{sec:intro}

Type Ia supernovae (SNIa) are explosions ignited from electron degenerate carbon-oxygen (CO) white dwarf (WD) progenitors. 
Thermonuclear detonation of degenerate CO cores can generally reproduce the main features of SNIa explosions. Among other features, they reproduce the SN light-curve luminosities observed, the characteristics of the Philips relationship, and they efficiently produce iron group elements as required from galactic chemical evolution (GCE) simulations \citep[see, for example:][]{sim2010detonations,nomoto2017thermonuclear,nomoto2018single,gronow2020sne,matteucci:21}. 
However, the dominant mechanisms driving the SNIa explosions remain uncertain. The explosion may be the outcome of mass transfer in either a single degenerate (one compact object) or double degenerate (two compact objects) system, and for each of these cases the compact object accreting material may explode either near the Chandrasekhar mass in a deflagration to detonation transition, or as a sub-Chandrasekhar double detonation.
\citet{livio2018progenitors} presents a comprehensive review of the possible pathways to explosion for SNIa describing these and other pathways to explosion. 

As WDs are unconditionally stable as isolated objects, a companion star is necessary to initiate the explosion. The nature of the companion influences the nucleosynthesis and ejected yields from the SNIa. As SNIa have been used for decades as standardisable candles for extra-galactic distance measurements and for GCE modeling, unambiguous identification of the progenitors of SNIa is crucial to reduce systematic errors of GCE models and improve their predictive abilities \citep{hillebrandt2013towards, livio2018progenitors}.

At present, it is a matter of debate as to the relative contributions of Chandrasekhar and sub-Chandrasekhar mass progenitors. Based on GCE calculations of [Mn/Fe] in the Milky Way disk and in the solar neighborhood, \citet{seitenzahl2013solar} find that 50\% of SNIa should originate from sub-Chandrasekhar mass WDs, \citet{scalzo2014ejected} obtain a range of between 25-50\%, and \citet{eitner2020observational} about 75\%. Late time spectroscopy also favors sub-Chandrasekhar progenitors, with up to 85\% of WDs predicted to be of this type \citep{flors2020sub}. In addition to this, there are difficulties in the observation of Mn. Most studies assume Local Thermal Equilibrium (LTE) when calculating abundances, however, \citet{bergemann2008nlte} find that the assumption of LTE for heavier elements can increase the inferred abundance of Mn significantly. \citet{mishenina2015mn} confirm that most Mn is produced by thermonuclear supernovae, but do not identify a statistical difference between the LTE or non-LTE assumptions.

There also exists tension between the single and double degenerate models of progenitors - \citet{woods2014uv} find that the contribution of single degenerate progenitors can be restricted significantly by considering the upper limits of detection of He II and C II emission lines in UV emission spectra of early type galaxies. They find from these observations that contributions from the single degenerate scenario should account for less than 10\% of observed SNIa.

There are a number of potential progenitors of SNIa which are discussed extensively in the literature and are reviewed in, for example, \citet{maoz2012type}. They are most commonly split into two categories - the single and the double degenerate scenarios. In the single degenerate scenario, a WD primary accretes material from a main sequence or helium burning companion. 
Depending on the evolutionary pathway of the SNIa event, the ignition of the WD 
may occur through several different mechanisms. For example, the primary C/O white dwarf may reach the Chandrasekhar mass or the carbon may be ignited in the core of the WD through a detonation in an accreted helium layer. In addition to these commonly considered scenarios, there are other less favored mechanisms - for a full discussion see \citet{livio2018progenitors}. In the double degenerate model, both the primary and secondary stars are C/O white dwarfs. In this case, the secondary is either disrupted to form an accretion disk or may collide with the primary \citep{raskin201056ni}.

The energy generation during the SNIa explosion is largely insensitive to the choice of nuclear reaction rates. \citet{bravo2012sensitivity} find there is a variation of less than around 4\% in energy generation when varying the $^{12}$C $^{12}$C or $^{16}$O $^{16}$O reaction rates by a factor of 10 or 0.1. They also observe a small impact on the abundance of iron group isotopes when a large range of charged particle reactions are varied, although a larger difference is observed in the intermediate mass region for the same variations in reaction rates. \citet{parikh2013effects} performed a sensitivity study on over 2300 reaction rates, finding that only 53 reaction affect the yield of any isotope with an ejected mass of more than 10$^{-8}$M$_{\odot}$ by more than a factor of 2 when varied by a factor of ten. This further indicates that the nucleosynthesis during the SNIa explosion is robust in the face of reaction rate changes, however they do identify some key reaction rates (for example $\alpha$ capture rates on $^{12}$C and $^{20}$Ne along with some electron capture rates) which do impact the isotopic yields of the system. The role of radionuclide production and the impact of reaction rates in the ratios of observable radionuclides is also discussed in \citet{parikh2013effects} who find that variations in the ratio of certain radionuclides can be up to a factor of 10 different depending on the choice of reaction rates. This suggests that though the isotopic and elemental yields of SNIa are largely insensitive to the reaction rates, simulations of the SNIa light curve may depend more sensitively on the reaction rates chosen. In the more recent \citet{bravo2019sensitivity}, a focus on varying the weak reaction rates during the explosion confirms that weak reaction rate increases cannot lead to production of neutron rich isotopes in sub-Chandrasekhar mass progenitors. \citet{fitzpatrick2021impacts} also find that the iron group abundances are relatively insensitive to reaction rate uncertainties, however there are a number of rates identified which impact the abundances of intermediate mass elements. Whilst variations in reaction rates are clearly important for investigating the production of some isotopes during the explosion, we do not include such a study in this work.

Each of the various formation channels presents challenges, as none reproduces all of the observational data of a typical SNIa - double degenerate merger light curves with a total mass larger than the Chandrasekhar mass
are too broad \citep{fryer2010spectra} but their calculated rate better matches the observed frequency of SNIa \citep{nomoto1997type} and explain the absence of hydrogen in the spectra \citep{hillebrandt2000type}. For the case where the double degenerate double detonation occurs before the formation of a remnant - with an explosion mass less than the Chandrasekhar mass - simulated light curves are similar to the sub-Chandrasekhar case. 
For the single degenerate case growing the white dwarf mass to the Chandrasekhar mass through accretion is difficult except for a narrow range of accretion rates \citep{cassisi1998hydrogen,hillebrandt2000type} although forming a helium shell through accretion is easier as no stable burning is required.
Observed delay time distributions of SNIa explosions cannot be explained solely by the single degenerate scenario \citep{mennekens2010delay}, and require at least a component from the double degenerate model.

In this work, we investigate the Chandrasekhar mass model of \citet{townsley2016tracer} (T1.4) which undergoes a deflagration to detonation transition, and the two sub-Chandrasekhar models of \citet{shen2018sub} and \citet{miles2019quantifying} (S1.0 and M0.8 respectively) which are both pure detonations. We focus in this work on the effect of the progenitor mass, rather than the pathway to explosion as, 
while there are a number of important observables governed by the path to explosion such as radio emission from circumstellar material \citep{kool2023radio}, gamma ray observations of radionucleids \citep{summa2013gamma}, high velocity features of ejecta \citep{silverman2015high}, early lightcurve observations \citep{hosseinzadeh2017early}, and other features \citep{maoz2014observational,hillebrandt2013towards,livio2018progenitors}, the bulk nucleosynthesis in the models investigated here depends on the peak temperature and central density only. 
Initial mass is not the only parameter which effects the prediction of GCE models, however. \citet{matteucci2009effect} show that the fraction of prompt SNIa explosion (those occurring in the first 100 Myr) has a larger impact in GCE simulations than the nature of the SNIa progenitor and whether it is a single or double degenerate model.

Observed isotopic abundances from SNIa are not readily available - they are widely regarded to not produce grains in their ejecta and so there is no single source isotopic abundance data available for them, other than for $^{55}$Mn, as stable Mn is monoisotopic. We can deduce the contribution to the solar budget of various isotopes by eliminating contributions from other stellar sources; however, this limits our predictive power to those isotopes which are dominated by the contribution from SNIa or which have well constrained contributions from all other sources. $\gamma$-ray spectroscopy presents an alternative probe of the nucleosynthesis in the SNIa explosions and therefore radionuclides with large production differences between models and with appropriate half-lives provide a potential tracer of the explosion conditions of the SN and the progenitors, although this is not discussed in this work.

The main products of SNIa explosions are in the intermediate-mass elements and in the iron group. \citet{matteucci1986relative} compare the relative contributions of type I and II supernovae and find that SNIa are the bulk producers of iron, producing around 70\% of solar iron, and argue for a significant contribution to the elemental abundance of Si of around 42\%. SNIa explosions contribute significantly to the production of the isotopes $^{50,52,54}$Cr, $^{55}$Mn, $^{54,56,57,58}$Fe, $^{59}$Co, $^{58,60,61}$Ni and $^{64}$Zn, as well as isotopes of Ne, Na, Mg and Si as discussed in, e.g.  \citet{woosley2002evolution,nomoto1984accreting,travaglio2011type}. 
Many sources of SNIa yields are available in the literature, including an extensive suite of post processed explosions in the Heidelberg Supernova Model Archive (HESMA) - e.g. \citet{fink2010double}, \citet{kromer2010double}, \citet{pakmor2010sub}; \citet{travaglio2004nucleosynthesis} and \citet{travaglio2011type}, where detailed modeling of the accreted He envelope shows the production of p-nuclei; and the yields of \citet{nomoto1982accreting} (W7),
and \citet{leung2018explosive}; the yields of \citet{shen2018sub}; and the two-dimensional yields of \citet{boos2021multidimensional}

In this work, we present new nucleosynthesis yields for 39 SNIa models. We consider three progenitor masses 
- the 1.4 M$_{\odot}$ mass deflagration detonation model of \citet[][T1.4]{townsley2016tracer}, the 1.0 M$_{\odot}$ double detonation model of \citet[][S1.0]{shen2018sub} and the 0.8 M$_{\odot}$ double detonation model of \citet[][M0.8]{miles2019quantifying}. 
By varying the $^{22}$Ne initial abundance, we provide the yields for 13 metallicities for each progenitor, between Z$_{met}$ = 0 and Z$_{met}$ = 0.1. 
A large number of different SNIa yields prescriptions have been used for GCE simulations in the literature, depending on the type of analysis and on the main research area. For example, \cite{mishenina:17} compare GCE calculations with the Milky Way disk stars using four different codes. They adopted different SNIa yields and different implementations for them with respect to metallicity, but they were all based on Chandrasekhar-mass progenitors only. More recently, the same approach is used by \cite{prantzos:18}. \citet{kobayashi2020origin} perform galactic chemical evolution studies using two SNIa progenitors and metallicities spanning from Z$_{met}$ = 0 to 0.1. The models presented in this work are therefore aligned with current GCE modeling, although yields from more progenitor systems may be necessary in specific works which use a larger number of progenitor models for their investigations (see, for example, \citet{lach2020nucleosynthesis,gronow2021metallicity}). Note also that the 0.8 M$_{sun}$ presented in this work is not expected to resemble a typical SNIa and may be too dim to be observed, it may therefore not be suitable for GCE studies but has been included here as a limiting case.

\section{Nucleosynthesis models and stellar simulations}
\label{sec: methodology}

\subsection{Tppnp - Parallelised Post-Processing}

The post-processing nucleosynthesis calculations presented in this work are completed using the NuGrid code tppnp: tracer particle post-processing network - parallel. A full description of the tppnp code can be found in \citet{jones201960fe}, along with details about the nuclear reaction network in \citet{ritter2018nugrid} and \citet{pignatari2016nugrid}. 
Key improvements implemented in \citet{jones201960fe} include the introduction of a variable-order semi-implicit integrator \citep[the Bader-Deulflhard method:][]{bader1983semi,deuflhard1983order,timmes1999integration}, nuclear statistical equilibrium (NSE) state solutions are coupled with weak reaction rates using an integration method outlined in \citet{cash1990variable} (Cash-Carp Runge-Kutta). The NSE solver is consistent with the full reaction network, and a treatment of electron screening taken from \citet{chugunov2007coulomb} is implemented. Our reaction network consists of over 5200 isotopes and over 75,000 reactions.
After the post-processing of tracer particles, all radioactive products are allowed to fully decay to stability before the final yields are presented. 

The sets of trajectories T1.4, S1.0 and M0.8 are based on hydrodynamics simulations using more limited networks. We find in this work that those networks in the published works \citet{townsley2016tracer,shen2018sub,miles2019quantifying} are sufficient to capture the broad nucleosynthesis in these models, however for some isotopes our results differ by more than a factor of 2-3 (see section \ref{sec:lit_yield_comparison} for a full discussion thanks to the change in reaction network. The network used in the post-processing of T1.4 uses 225 nuclides in the reconstruction of the thin flame front \citep{townsley2016tracer}. These include weak reactions discussed in \citet{calder2007capturing}, which are necessary for the computation of the neutronisation in the flame front and are taken from \citet{langanke2000shell} and \citet{langanke2001rate}. Where newer rates are not available, they use those found in \citet{fuller1985stellar} and \citet{oda1994rate}. The standard TORCH network is extended in their work, including 25 additional reactions pertaining to neutron rich Fe-group isotopes. 
We also see production of trace amounts of higher-Z material not found in \citet{townsley2016tracer} because of our more extended network, with total ejected mass of order 10$^{-9}$ M$_{\odot}$. The network in \citet{shen2018sub} used for the hydrodynamical modeling consists of 41 isotopes and 190 reactions, all of which were taken from the JINA reaclib library \citep{cyburt2010jina}. With errors of only a few percent in energy generation, they find that this network is sufficient to follow the explosion dynamics. Post-processing in this work was carried out with a 205 isotope network using MESA \citep{paxton2010modules}, again with all reactions being taken from JINA reaclib. Finally, the network in \citet{miles2019quantifying} is again 205 isotopes, also with JINA reaclib rates. In each case, the hydrodynamics have been calculated separately, with the final yields being post-processed.

Variations arise from a difference in the choice of reaction rates used between the NuGrid post-processing codes and those used in the original publications. Post-processing these models again with a consistent nuclear reaction network between them allows us to disentangle the nuclear physics uncertainties from hydrodynamical modeling uncertainties.  

Reaction rates for the tppnp code are taken from a variety of different sources: the \citet{rauscher2000astrophysical} JINA reaclib, \citet{cyburt2010jina} Basel reaclib, \citet{dillmann2006kadonis} KADoNIS, \citet{angulo1999compilation}, \citet{caughlan1988thermonuclear} and \citet{iliadis2001proton} compilations between them constitute the majority of reactions other than the weak reaction rates, which are compiled from \citet{fuller1985stellar}, \citet{oda1994rate}, \citet{goriely1999uncertainties} and \citet{langanke2000shell}, along with rates from \citet{takahashi1987beta} as discussed in \citet{jones201960fe} and \citet{pignatari2016nugrid}. The NSE solver is based on the theory described in \citet{seitenzahl2008proton}, and care was taken to include coulomb screening and ensure consistency between the reaction network and NSE solution as has been described by \citet{calder2007capturing}.

Following a common approach in the literature, we select an initial abundance of metals determined by the mass fraction of $^{22}$Ne. $^{22}$Ne acts as a proxy for the metallicity of the progenitor in this work, as the weak interaction $^{18}$F($\beta^{+}$)$^{18}$O in the CNO cycle converts protons to neutrons and changes the electron fraction of the system. The electron fraction can also change through carbon simmering or freezout from neutron rich NSE \citep{martinez2017observational} in near Chandrasekhar mass progenitors.

The initial composition, beyond the electron fraction, is not important in those particles which reach NSE and the composition of the outer, cooler layers of the white dwarf are unlikely to resemble scaled solar abundances. We have therefore chosen to treat our white dwarf as having neutron excess only in the form of $^{22}$Ne.

We note that the parent model for the S1.0 case does not include an accreted helium layer from a companion star, and the detonation is ignited centrally. \citet{shen2018sub} present the limiting case for the explosion where the accreted helium shell is minimally thin, and as such the conditions of the S1.0 case are representative of a single detonation scenario. Without the helium ashes of this accreted shell, the ejected abundances fail to take into account possible p-nuclei production, as discussed in \citet{travaglio2005metallicity}, as well as an increase in the ejected abundances of some key iron group isotopes as discussed in, for example, \citet{magee2021exploring}.

T1.4 choose a scaled solar abundance for their models, except for the CNO material, which is treated as $^{22}$Ne as in \citet{timmes2003variations}, with a constant abundance throughout the WD. The rest of the material is composed of C and O, and convective burning ashes: $^{20}$Ne, $^{16}$O, $^{13}$C and $^{22}$Ne. S1.0 includes $^{22}$Ne and $^{56}$Fe as proxies for the abundances of non-C/O isotopes and as a tracer of metallicity, again with a uniform distribution of abundances in the WD. M0.8 use solar abundances taken from \citet{asplund2009chemical}.

\begin{table}[]
\centering
    \begin{tabular}{llll}
    \hline
        Parameter                          & T1.4     & S1.0     & M0.8  \\
    \hline
        No. of Particles Post Processed    & 7856     & 107      & 9996  \\
        T$_{peak}$ Max (GK)                & 10.28    & 6.32     & 5.21  \\
        No. Timesteps                      & 6872     & 58738    & 1347  \\

    \end{tabular}
\caption{Number of particles, maximum peak temperature and number of timesteps for each of the three classes of model investigated in this work.}
\label{tab:model_param}
\end{table}

Table \ref{tab:model_param} shows some of the key parameters of each of the three models investigated in this work; namely the number of post-processed tracer particles, the maximum peak temperature of any particle in that set of tracer particles and the number of timesteps in each trajectory. M0.8 has the greatest number of particles, with approximately 10$^{4}$ being extracted for the post-processing step. We have extracted a similar number of particles from the T1.4 model, after performing a convergence study to ensure that our results were consistent with the full 10$^5$ particles of the T1.4 model. This was achieved by running the full particle set and comparing with yields from reduced numbers of particles selected from every 10th, 15th, 20th and 100th particle in the data set. Convergence of our results was seen for the case where one in ten particles was selected.; S1.0 has only 107 tracers in the parent model however \citet{shen2018sub} have shown that this is sufficient to model the SNIa explosion. Due to the sparse nature of the S1.0 tracer particles, many of the diagnostic plots presented later in this work appear to have missing data, but this is an artifact of the smaller number of tracers in the S1.0 models. 

Time resolution in the three models varies significantly. The S1.0 model is highly resolved, with close to 6$\times$10$^{5}$ timesteps. 
In this case, the trajectories remain as they were when extracted from the hydrodynamical modeling. This is in order to not introduce inconsistencies in the yields with respect to those published for S1.0. The number of timesteps for the M0.8 and T1.4 models is comparable at around 10$^{3}$ per trajectory.

For the T1.4 model, resolution in mass 
in the hotter parts of the explosion is  high; however, the resolution in the outer layers of the WD is not as high as that found in \citet{travaglio2004nucleosynthesis}. 
For instance, we can make no predictions about the abundances of p-nuclei in any of our models as the CO WDs of the S1.0 and M0.8 models are treated as having a minimally thin accreted He shell (i.e. no shell). The production of p-nuclei as described in \citet{travaglio2015testing} depends sensitively on the nucleosynthesis during the He shell flashes and on the heavy element abundances which are built during the accretion phase. They subsequently form the seeds of the p-process activated during the supernova explosion \citep[][]{battino2020heavy}. Without this component, our calculations do not include p-process yields.

In the models of \citet{townsley2016tracer}, based on the earlier work of \citet{calder2007capturing} and \citet{townsley2007flame}, the explosion in the WD progenitor is initiated through the introduction of an artificial hotspot at a predetermined radius. 
In \citet{townsley2016tracer}, there are a number of important improvements to the original models, including the ability to change the Y$_{e}$ of the fuel. The most notable feature of the \citet{townsley2016tracer} models, however, is the introduction of reconstruction of the trajectories.  
The flame front during the SNIa explosion is artificially thick in the raw data from the hydrodynamic modeling. In order to account for this, the reconstruction process narrows the flame by reducing the effective time that the material spends at the peak temperatures in the simulation. 

The S1.0 models utilize the FLASH code \citep{fryxell2000flash} in order to model 1D  explosion, using density profiles from MESA for the initial conditions of their WD  progenitor. The explosion in this model is also initialized by a hotspot, here with a central temperature of 2 GK. The hotspot is relatively large compared to the minimum detonatable region; however, the mass of the spot is small. \citet{shen2018sub} argue that this also therefore minimizes the impact on the ejected yields. The M0.8 model uses a uniform composition throughout the WD progenitor, with 1.4\% of metals. The ignition of this model is again achieved with the introduction of a hotspot, this time at 1.98 GK. FLASH is used to evolve the system, through the initial detonation, leading to the triggering of the C/O detonation.

\section{SNIa Yields} \label{sec:yields}

\begin{table}[]
    \centering
    \begin{tabular}{lll}
    \hline
    T1.4Z0         & S1.0Z0         & M0.8Z0        \\   
    T1.4Z0.0000001 & S1.0Z0.0000001 & M0.8Z0.0000001\\
    T1.4Z0.000001  & S1.0Z0.000001  & M0.8Z0.000001 \\
    T1.4Z0.00001   & S1.0Z0.00001   & M0.8Z0.00001  \\
    T1.4Z0.0001    & S1.0Z0.0001    & M0.8Z0.0001   \\   
    T1.4Z0.001     & S1.0Z0.001     & M0.8Z0.001    \\   
    T1.4Z0.002     & S1.0Z0.002     & M0.8Z0.002    \\   
    T1.4Z0.005     & S1.0Z0.005     & M0.8Z0.005    \\   
    T1.4Z0.014     & S1.0Z0.014     & M0.8Z0.014    \\   
    T1.4Z0.01      & S1.0Z0.01      & M0.8Z0.01     \\   
    T1.4Z0.02      & S1.0Z0.02      & M0.8Z0.02     \\   
    T1.4Z0.05      & S1.0Z0.05      & M0.8Z0.05     \\   
    T1.4Z0.1       & S1.0Z0.1       & M0.8Z0.1      \\ \hline

    \end{tabular}
    \caption{List of SNIa model sets presented in this work. Models are labeled according to the initial mass of the progenitor and the metallicity. T1.4 corresponds to the \citet{townsley2016tracer} 1.4 M$_{\odot}$ deflagration detonation model, S1.0 the \citet{shen2018sub} 1.0 M$_{\odot}$ double detonation model, and M0.8 the \citet{miles2019quantifying} 0.8 M$_{\odot}$ model. Metallicities range from Z$_{met}$ = 0 to Z$_{met}$ = 0.1, with the initial fraction of metals being represented by the mass fraction of $^{22}$Ne with a uniform distribution through all tracer particles.
    }
    \label{tab:models_explained}
\end{table}

The full list of models is shown in table \ref{tab:models_explained}. For each of the three model configurations, a set of 13 metallicities is considered, providing a total of 39 distributions of SNIa yields.
In general, we can identify three broad burning regimes in SNIa ejecta, similar to explosive burning conditions in massive stars, as discussed in \citet{iliadis2015nuclear}. The high temperature component of these models, with SN temperature peaks in the order of or above 7 GK, are present only in the Chandrasekhar mass explosions. It is characterized by an efficient production of iron-group material, 
formed by the freezeout from nuclear statistical equilibrium (NSE). 
The intermediate component, with peak temperatures between 4.5 and 7 GK, is present in all models. However, the M0.8 suite of yields has a smaller contribution from this region. Production in this peak temperature range gives rise to burning products which resemble incomplete explosive silicon burning \citep[][]{iliadis2015nuclear}. 

Due to the lower peak temperature range in the M0.8 models and the subsequent small contribution to iron group material, the mass of $^{56}$Ni ejected by the M0.8 models is lower when compared to the S1.0 and T1.4 models by orders of magnitude, and a typical SNIa explosion is not expected from these progenitors. 

\begin{figure*}
    \gridline{\fig{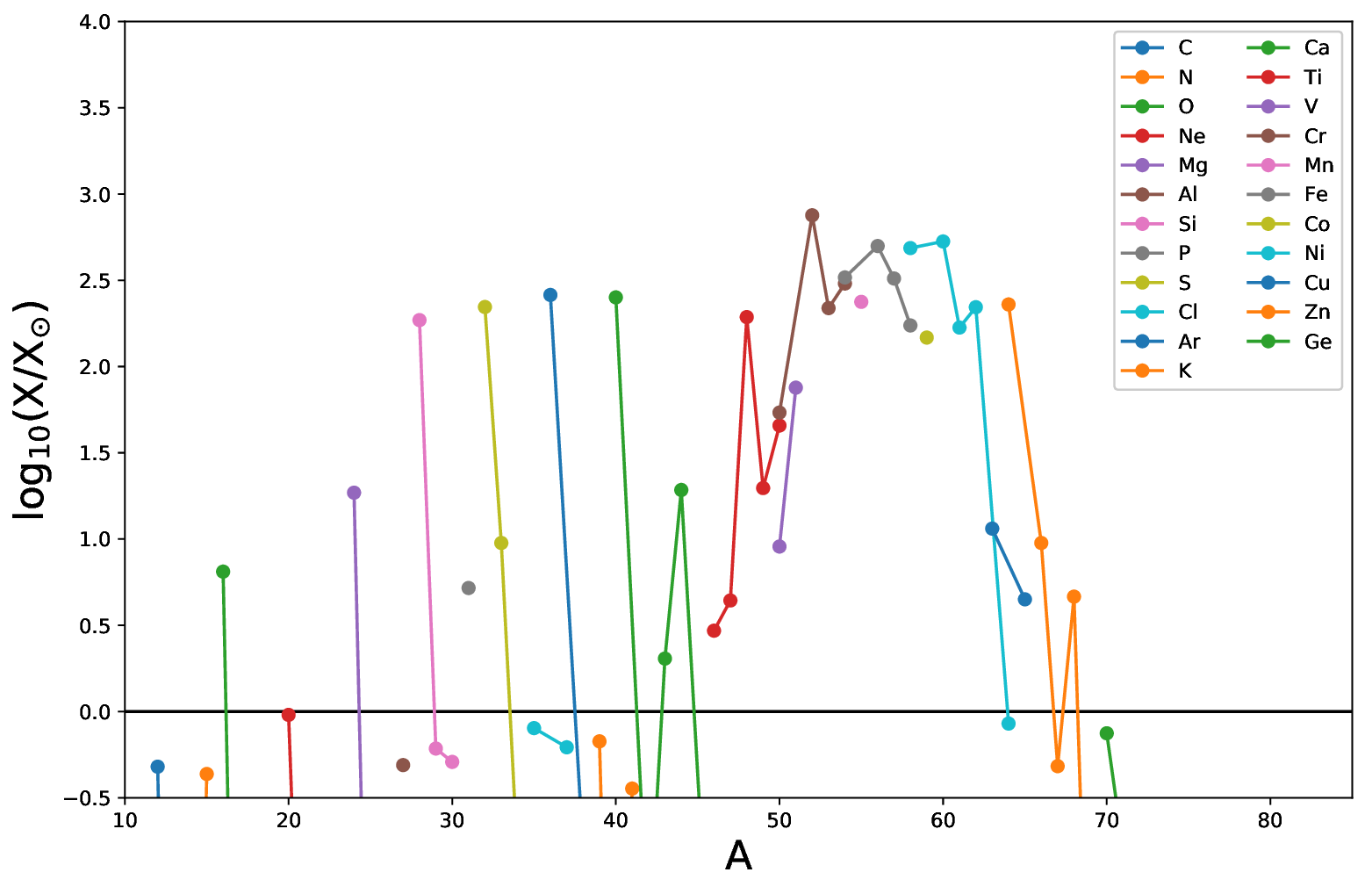}{0.54\textwidth}{T1.4Z0}}
    \gridline{\fig{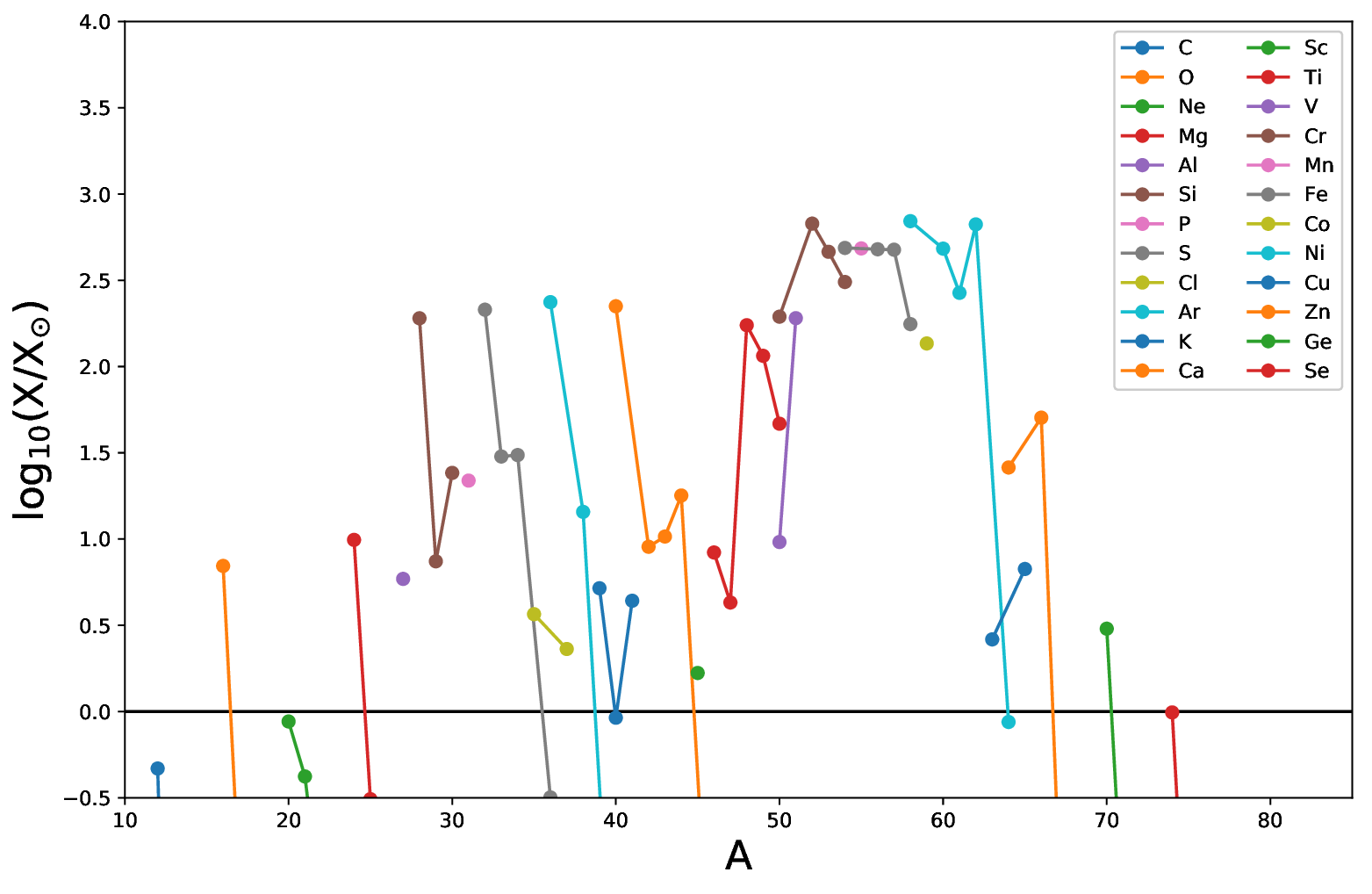}{0.54\textwidth}{T1.4Z0.014}}
    \gridline{\fig{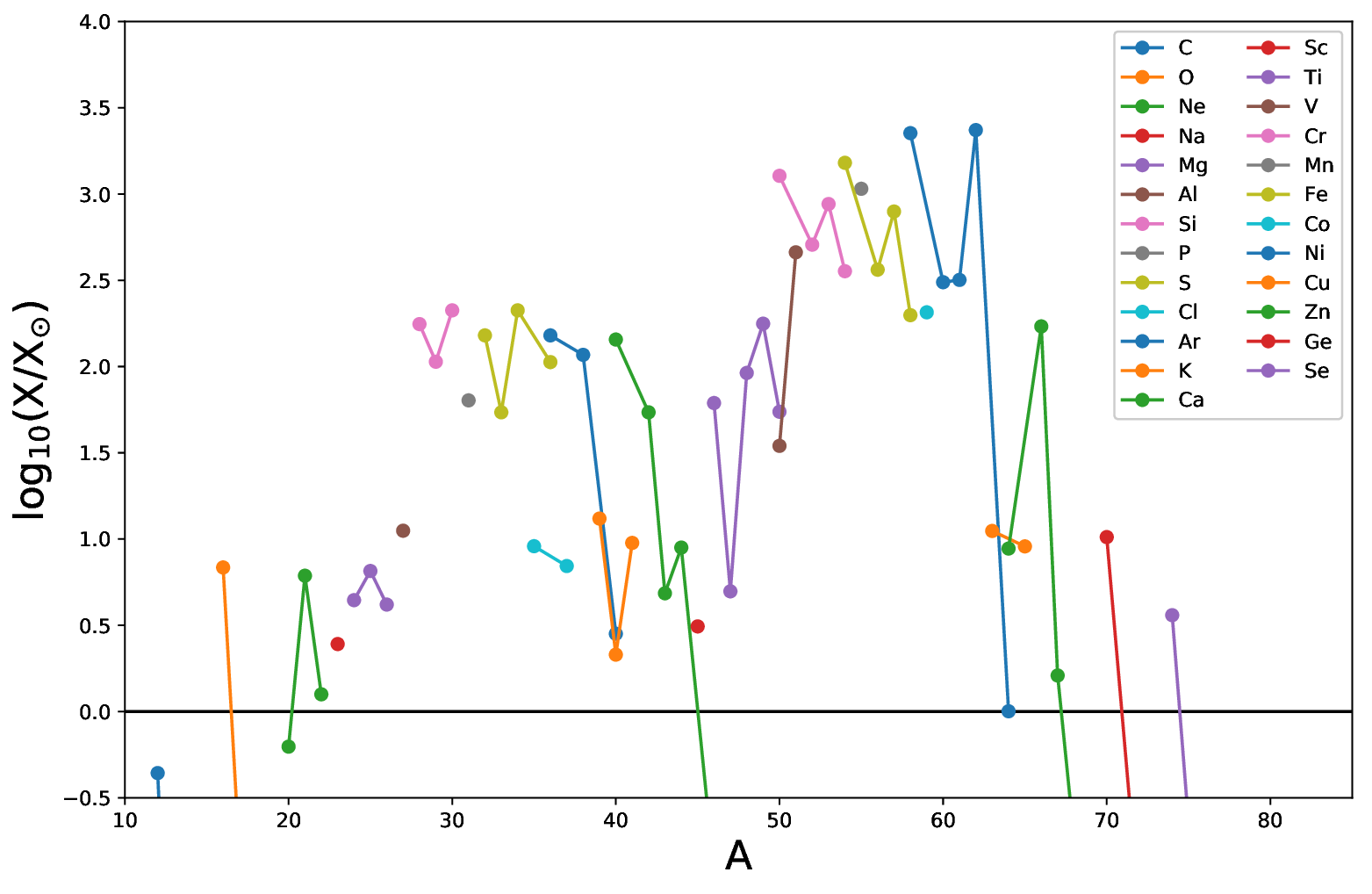}{0.54\textwidth}{T1.4Z0.1}}
    \caption{Production factors for models T1.4Z0 (top), T1.4Z0.014 (middle) and T1.4Z0.1 (bottom). 
    }
    
\label{fig:dean_production_factors}
\end{figure*}

Absolute yields and production factors for the three model sets T1.4, S1.0 and M0.8 at all metallicities, are provided for isotopes between C (Z = 6) and As (Z = 33) as supplementary material. The isotopic production factors for models T1.4Z0, T1.4Z0.014 and T1.4Z0.1 are shown in figure \ref{fig:dean_production_factors}.  
As expected, large over-productions in the iron group mass region and for the most abundant isotopes of the $\alpha$-elements Si, S, Ca and Ti (i.e., $^{28}$Si, $^{32}$S, $^{40}$Ca and $^{48}$Ti) are present at all metallicities. Isotopic production of odd-Z elements and neutron-rich isotopes for light and intermediate-mass even-Z elements increase with increasing metallicity in the SNIa progenitor.

The large peak in production in the iron group arises from the complete destruction of the $^{12}$C and $^{16}$O nuclei to $\alpha$ particles, and their subsequent recombination as the temperature drops. For a given Z in the low to intermediate mass isotopes (A $\lesssim$ 45), the lightest stable nucleus of that element is the most overproduced at low metallicities. As the metallicity of the system increases, we see a boost in production of the heavier isotopes of a given element due to the decreasing Y$_{e}$. The bulk of the iron group material is formed in the freeze out from NSE, at high central densities. These high density regions modify the Y$_{e}$ of the material undergoing nucleosynthesis through electron capture reactions, resulting in a less pronounced dependence on the initial metallicity of the progenitor \citep[e.g.,][]{brachwitz2000role}; however, contributions in the less dense outer regions of the SNIa explosion still depend on the initial metallicity of the progenitor.

In particular, as described in \citet{brachwitz2000role}, odd-odd nuclei (e.g. $^{40}$K) and odd-A (e.g. $^{55}$Co, $^{55}$Fe) nuclei have the largest effect on the Y$_{e}$ of the SNIa explosion. The choice of electron capture reactions for these nuclei in particular are therefore relevant for nucleosynthesis calculations.
Our weak rates are taken from \citet{langanke2000shell} and \citet{langanke2001rate}. Gamow-Teller back resonances in astrophysical conditions significantly boost electron capture rates \citep{brachwitz2000role}. Due to the dense conditions in the SNIa explosion, forbidden transitions are negligible and these resonances dominate \citep{fuller1982stellar}. In addition to this \citet{fuller1982stellar} also highlight the importance of the neutron closed shells in SNIa nucleosynthesis, which block further electron captures.

The middle panel of Figure \ref{fig:dean_production_factors} illustrates the effect of the decreased Y$_{e}$ in this model compared to the analogous model at Z$_{met}$ = 0. In particular, through the span of our models we range from an initial Y$_{e}$ of 0.5 at Z$_{met}$ = 0, to a value of Y$_{e}$ of 0.495 at Z$_{met}$ = 0.1. We see a significant boost in the production of more neutron rich isotopes in the A $\lesssim$ 65 region, with production of, for example, $^{50}$Cr, $^{58}$Fe and $^{64}$Ni increasing by several orders of magnitude. 
The odd-even effect is more pronounced in the T1.4Z0.1 run, and the nucleosynthesis shifts in all regions to more neutron rich isotopes, due to the decreased Y$_{e}$. 

\begin{figure*}
    \gridline{\fig{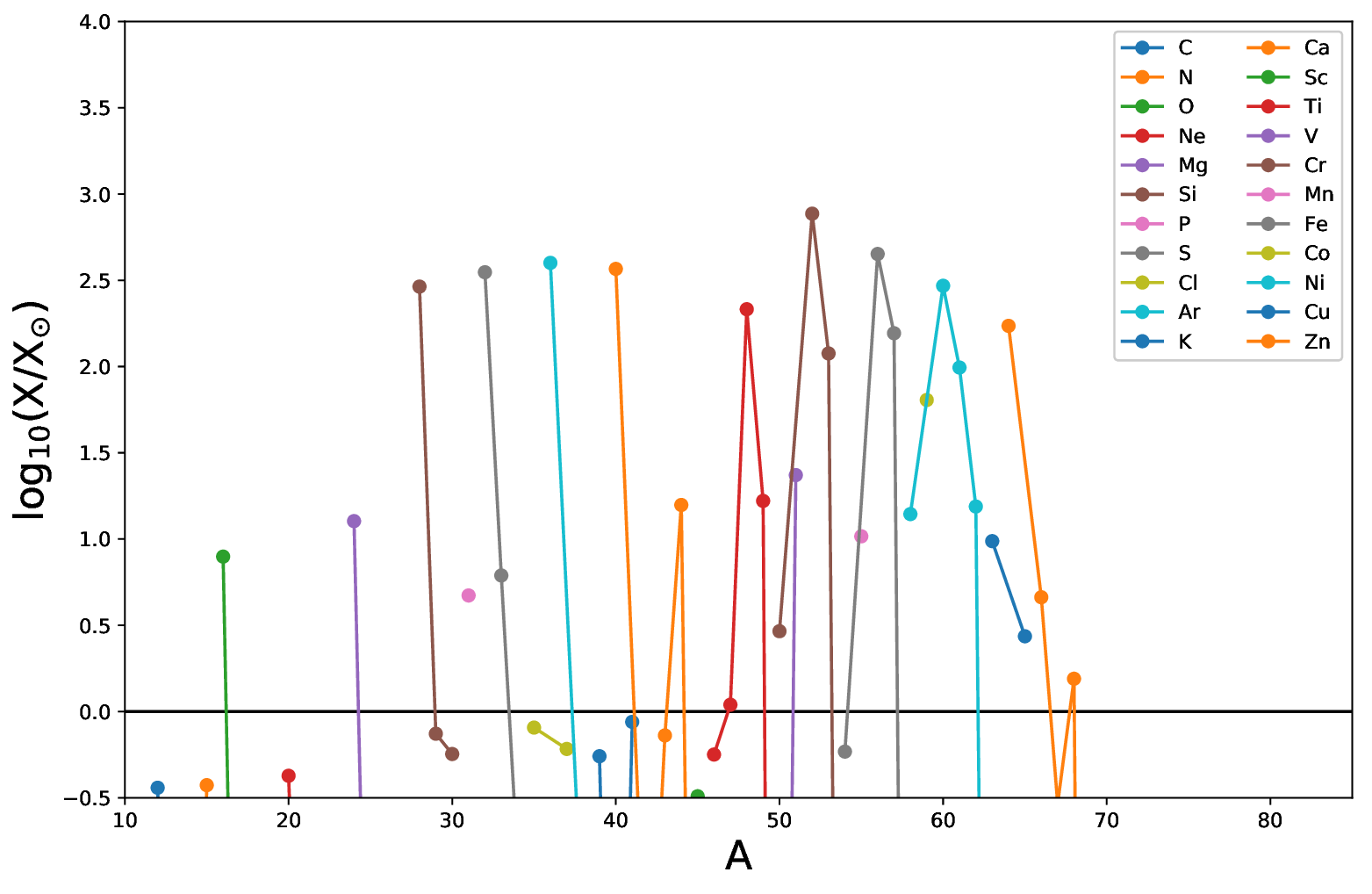}{0.54\textwidth}{S1.0Z0}}
    \gridline{\fig{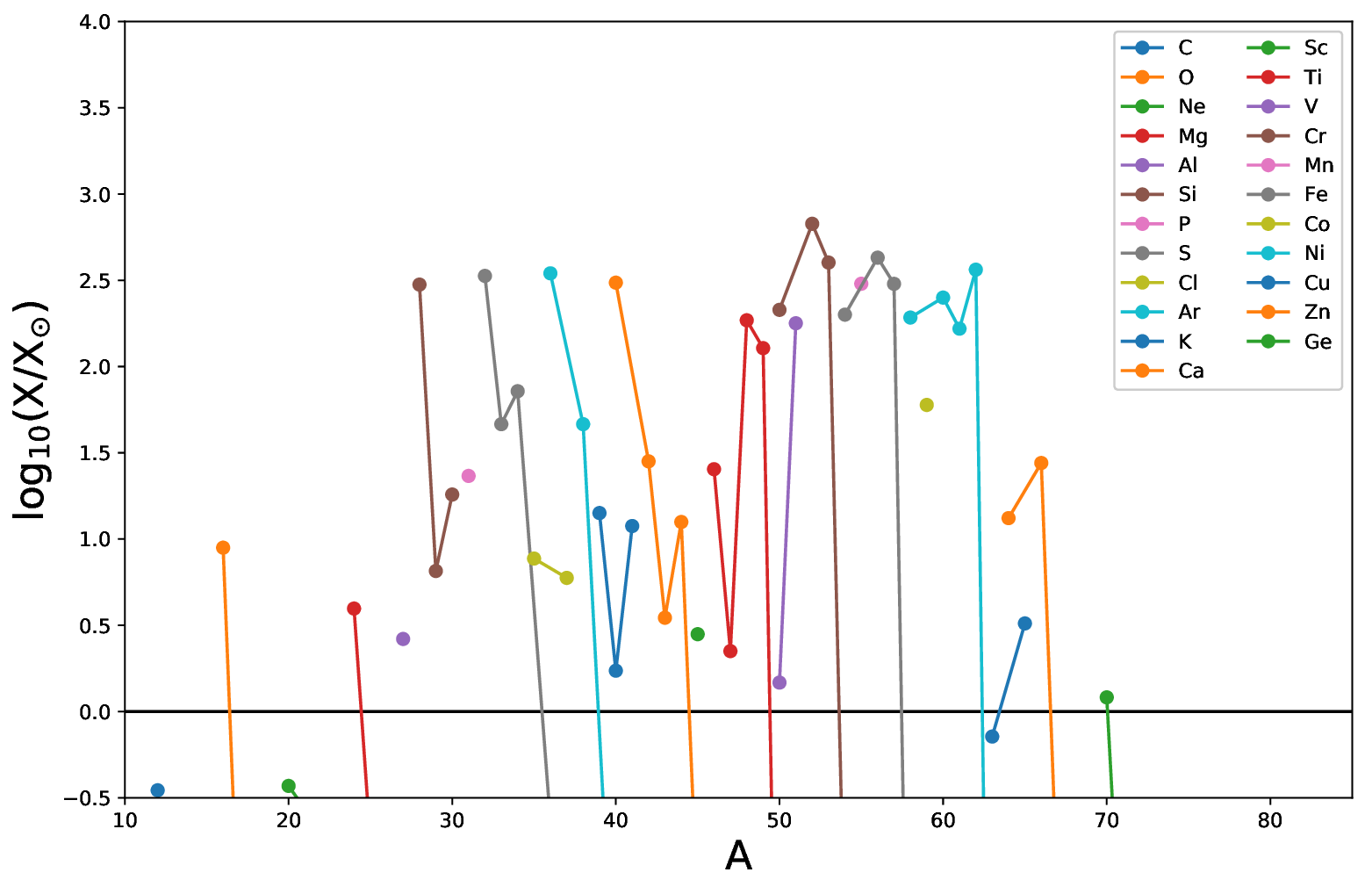}{0.54\textwidth}{S1.0Z0.014}}
    \gridline{\fig{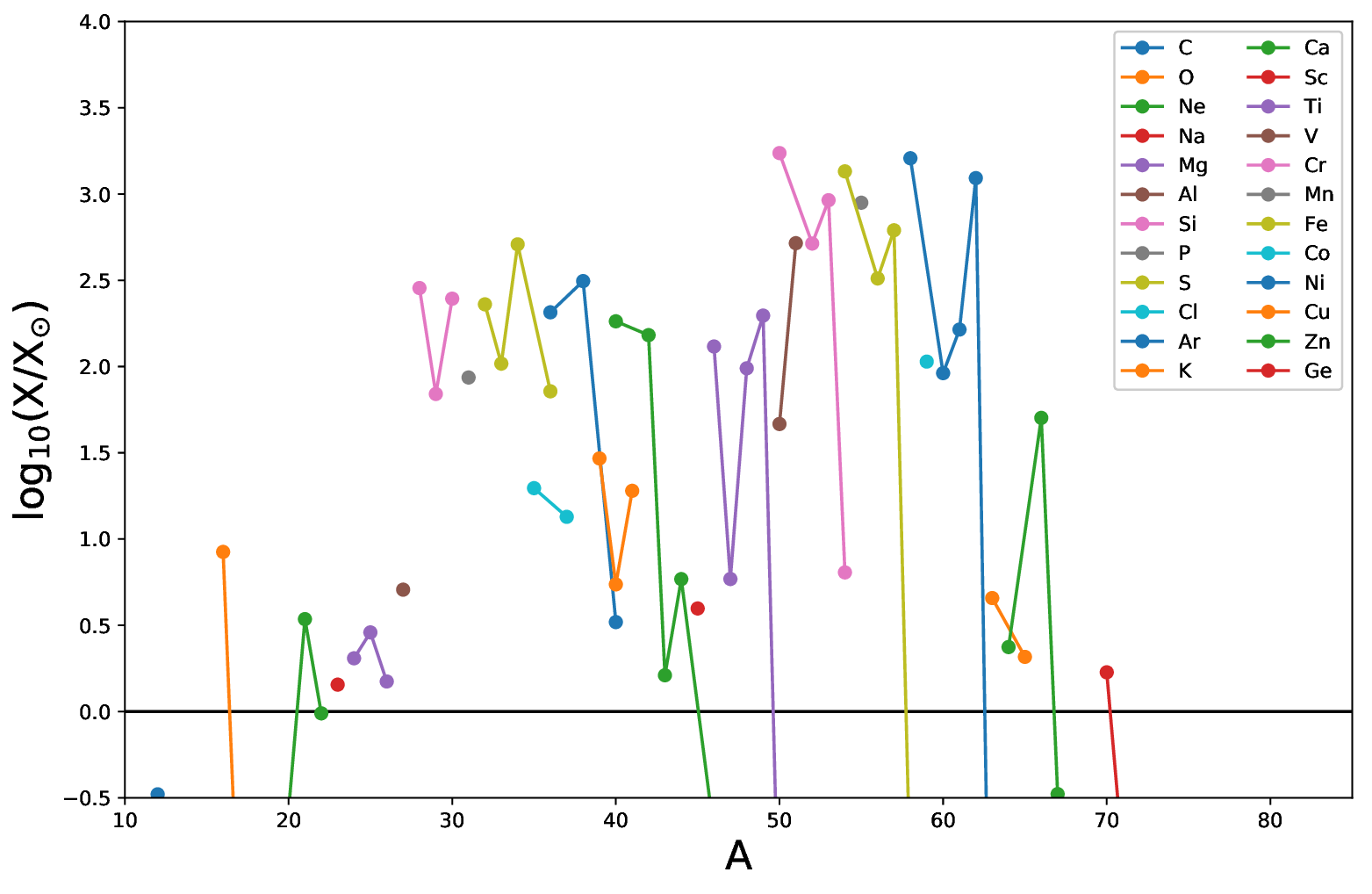}{0.54\textwidth}{S1.0Z0.1}}
    \caption{Isotopic production factors for models S1.0Z0, S1.0Z0.014 and S1.0Z0.1.}
\label{fig:shen_production_factors}
\end{figure*}

Comparing figures \ref{fig:dean_production_factors} and \ref{fig:shen_production_factors}, the abundance distribution of the T1.4 and the S1.0 sets of models are similar at all metallicities.   
For the iron group, the most abundant isotopes of a given element are produced in similar proportion in the two sets of models, but the neutron-rich isotopes are more produced in the T1.4 model. For instance, $^{64}$Ni has a production factor of close to the solar value in the T1.4Z0 model, 17 orders of magnitude larger than the S1.0Z0 model. There are 11 orders of magnitude difference in the $^{58}$Fe production factor.

\begin{figure*}
    \gridline{\fig{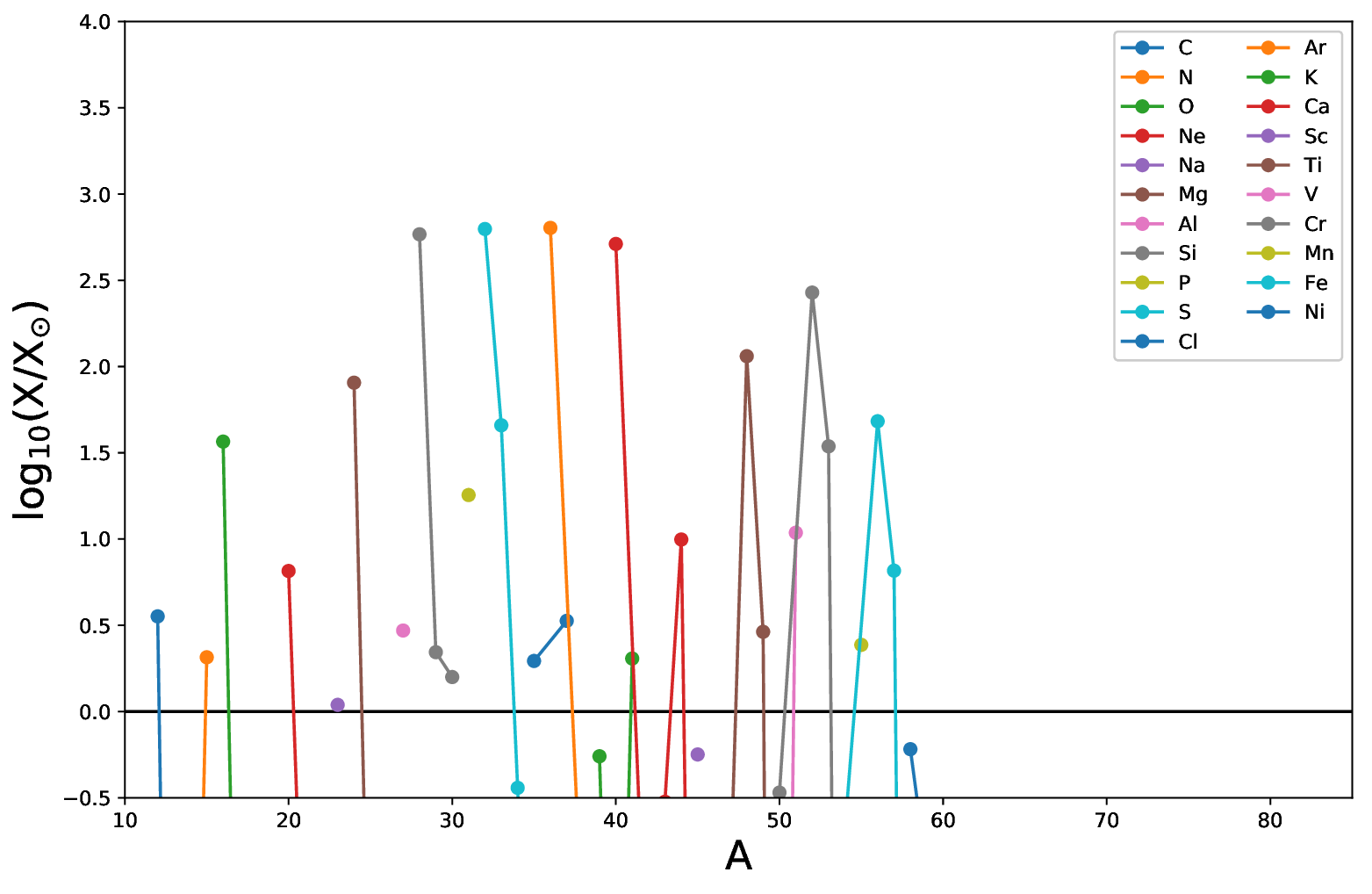}{0.54\textwidth}{M0.8Z0}}
    \gridline{\fig{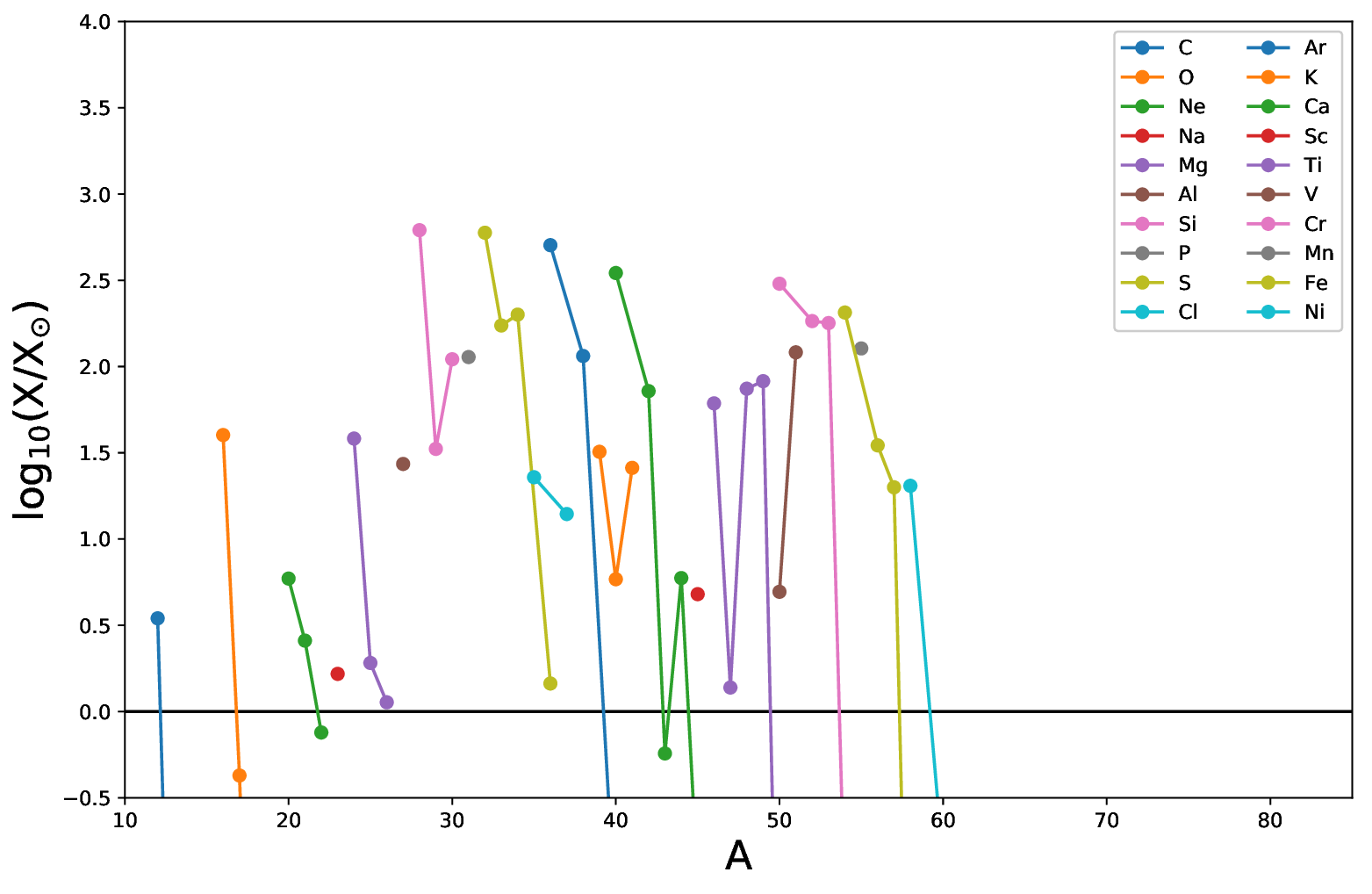}{0.54\textwidth}{M0.8Z0.014}}
    \gridline{\fig{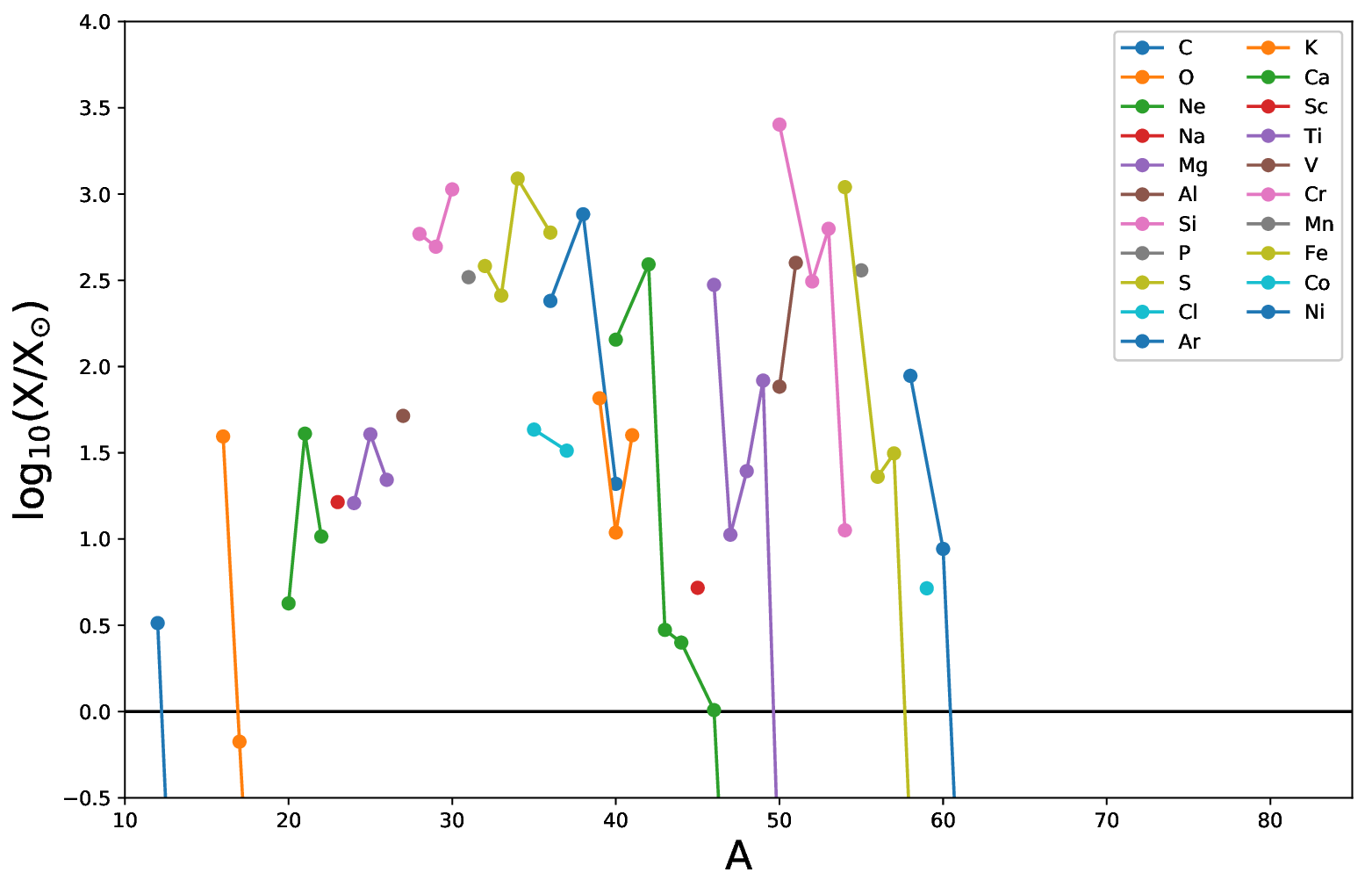}{0.54\textwidth}{M0.8Z0.1}}
    \caption{Production factors for models M0.8Z0, M0.8Z0.014 and M0.8Z0.1.}
\label{fig:brox_production_factors}
\end{figure*}

The M0.8 set shows a large reduction in the production of iron group isotopes compared with T1.4 and S1.0 sets (Figure \ref{fig:brox_production_factors}), with typical differences of between 0.2 times the abundances of the same isotope between the M0.8 and T1.4 cases however for some isotopes - all those more massive than $^{58}$Ni - the difference is much larger, with no production of these isotopes outside the higher temperature regions of the T1.4 and S1.0 models. Iron is also under-produced by more than a factor of 15 across all isotopes heavier than $^{54}$Fe. Instead, the production of intermediate mass elements like Si and S and their isotopes is enhanced compared to the heavier models.

\section{Yields Analysis By Element} \label{section:yields_analysis_by_element}

In this section, we describe the production of each stable isotope produced in the SNIa explosion. We select for elements which have been identified in the literature to have a significant contribution from SNIa. We primarily use the following publications to identify these elements - \citet{kobayashi2020origin}, \citet{pignatari2016nugrid}, \citet{woosley2002evolution} - however, other references are given on an element-by-element basis. All other isotopic abundances are available in the yield tables provided as supplementary material. It should be noted that some isotopes may have a larger contribution from SNIa where the elemental contribution is negligible, however these are not discussed in this current work. Where the percentage of an isotope as a fraction of the solar abundance of the element is presented, the data has been taken from the NNDC NuDat 3.0 website \citep[][https://www.nndc.bnl.gov/nudat3/]{kinsey1997nudat}.

For each set of models, we discuss isotopic yields at metallicities corresponding to a mass fraction of $^{22}$Ne of 0, 0.014 and 0.1. The figures with the detailed yield distributions in the SNIa ejecta at these three representative metallicities, including the detailed radiogenic contribution from their parent radioactive species, and for all other isotopes not discussed in the text, are available online at \dataset[Zenodo: Figures]{https://doi.org/10.5281/zenodo.8059974
}.


\subsection{Silicon}

\begin{figure*}
    \gridline{\fig{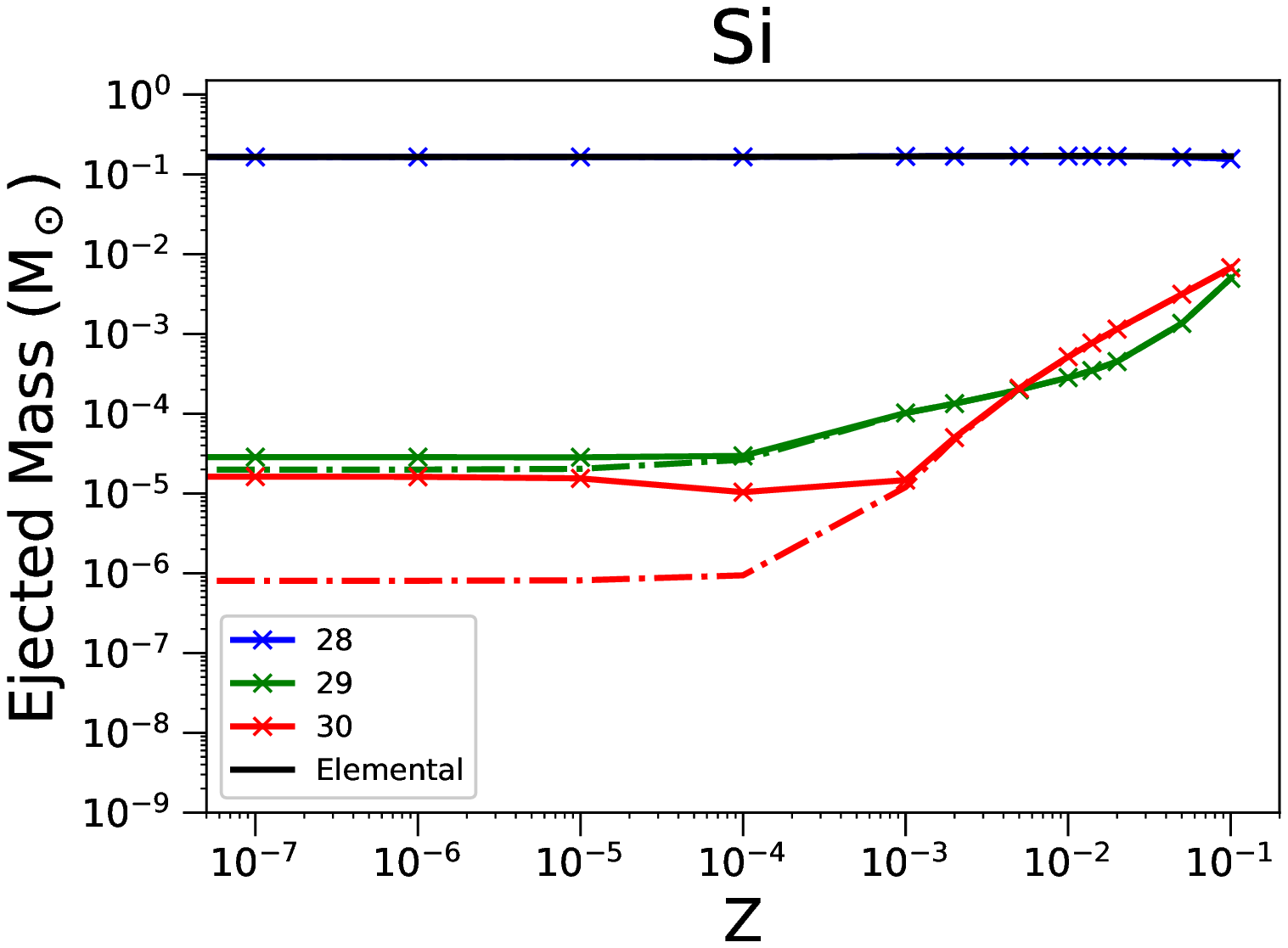}{0.45\textwidth}{T1.4}}
    \gridline{\fig{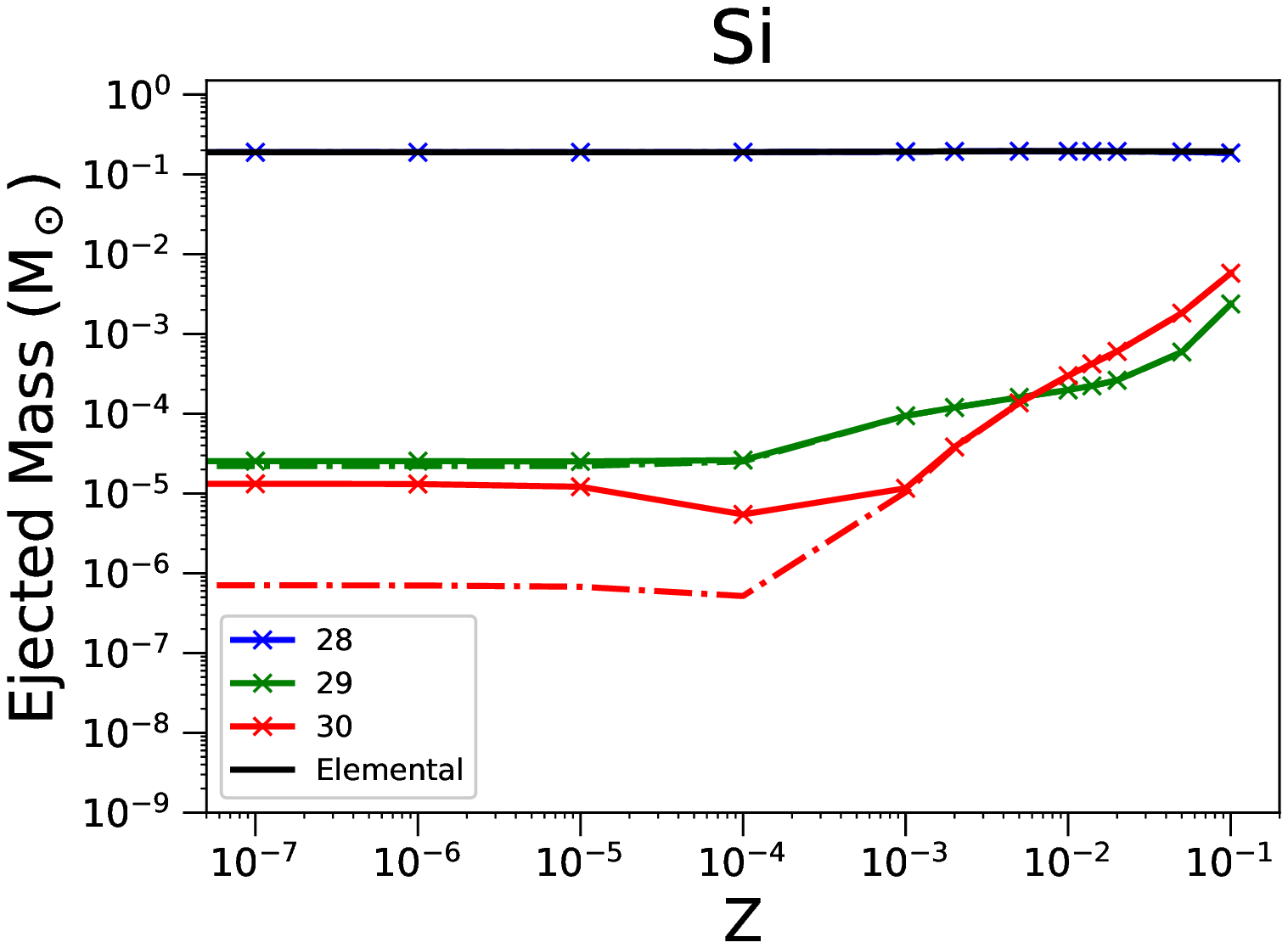}{0.45\textwidth}{S1.0}}
    \gridline{\fig{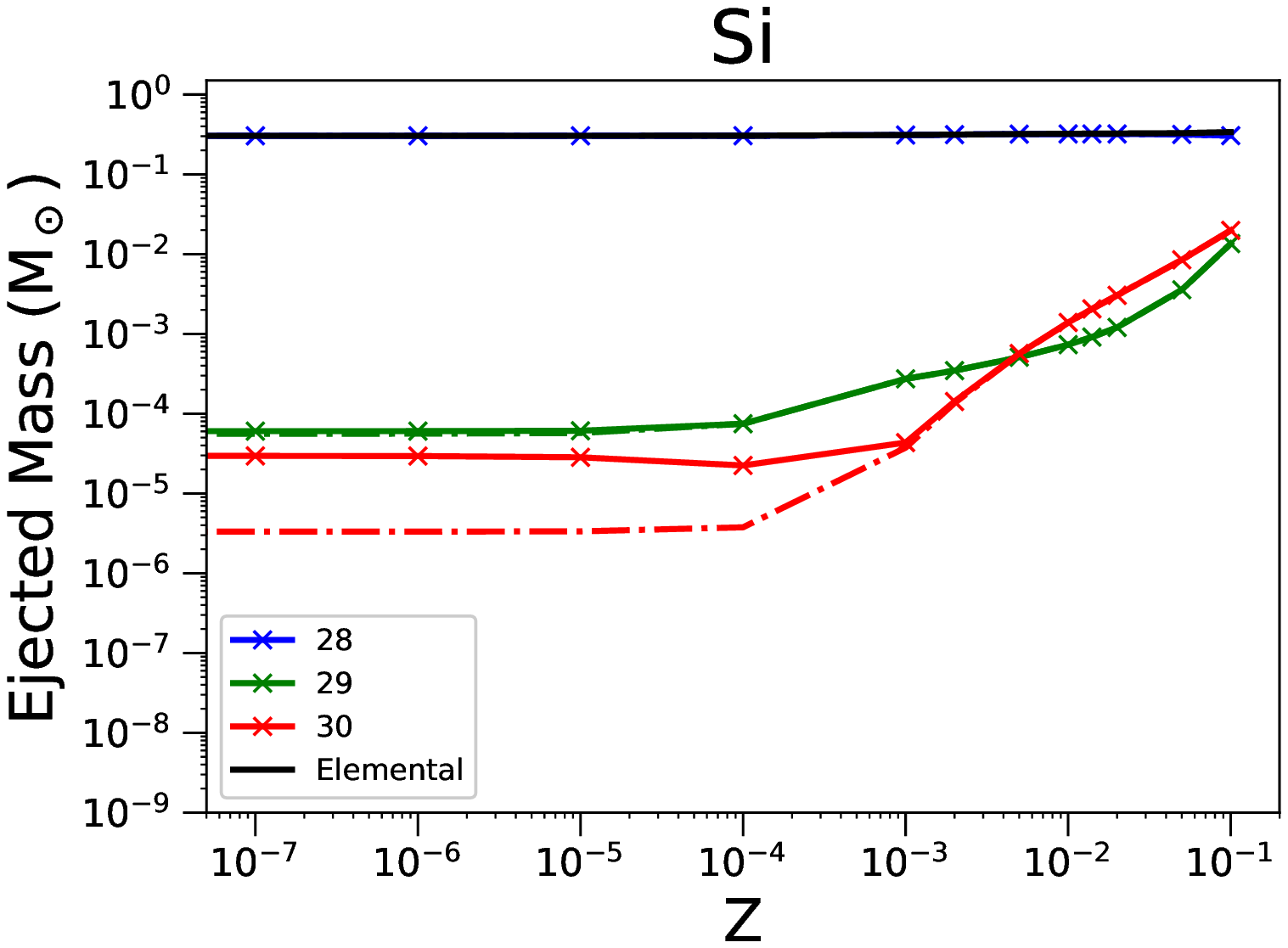}{0.45\textwidth}{M0.8}}
    \caption{The ejected mass for Si and its stable isotopes $^{28}$Si, $^{29}$Si and $^{30}$Si are shown at different metallicities. The radiogenic contributions are included in the elemental yields. Decayed and undecayed abundances are shown for isotopes (continuous lines with crosses and dot-dashed lines with empty circles, respectively). Data is presented for sets T1.4 (upper panel), S1.0 (middle panel) and M0.8 (lower panel).
    }
\label{fig:Silicon_metalicity_multi}
\end{figure*}

Silicon consists of three stable isotopes: $^{28}$Si, $^{29}$Si and $^{30}$Si. These are mostly made in CCSN, however contributions from SNIa events are needed to fit the observed solar system abundances \citep[see, for example][]{seitenzahl2013three,kobayashi2020origin}. $^{28}$Si is the most abundant naturally occurring stable Si isotope (92\% of the solar abundance), and is a product of oxygen fusion through the reaction $ ^{16}\mathrm{O}({}^{16}\mathrm{O},\alpha){}^{28}\mathrm{Si}.$ $^{29}$Si is the next most abundant at 4.7\%, and $^{30}$Si at 3.1\%.

In Figure \ref{fig:Silicon_metalicity_multi}, for all models, Si is produced with ejected masses above 0.2-0.3 M$_{\odot}$. The largest producers of Si are the M0.8 models, with a factor of around 2 larger ejected mass than compared with the other two sets of SNIa yields across all metallicities. $^{28}$Si has no trend with metallicity below the super-solar metallicity models. For those, we obtain a small dip in production of all classes of model, arising from reduced production of $^{28}$Si at the lowest peak temperatures due to neutron captures. The $^{29}$Si and $^{30}$Si yields strongly depend on the initial metallicity of the SNIa progenitors however, since these isotopes contribute less than 10\% of the ejected silicon, they do not affect the overall elemental abundance trend.

\begin{figure*}
    \gridline{\fig{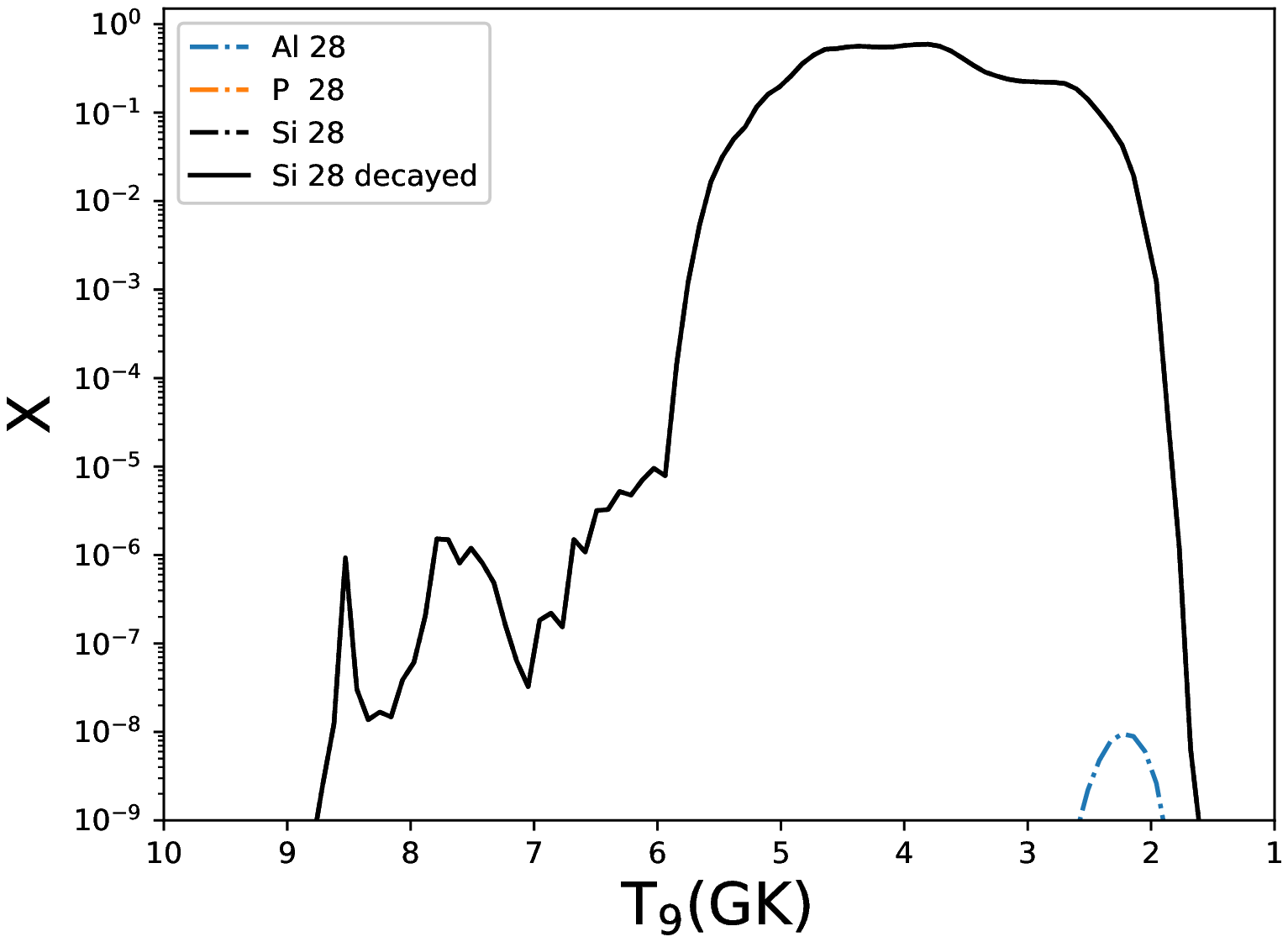}{0.3\textwidth}{T1.4Z0}
              \fig{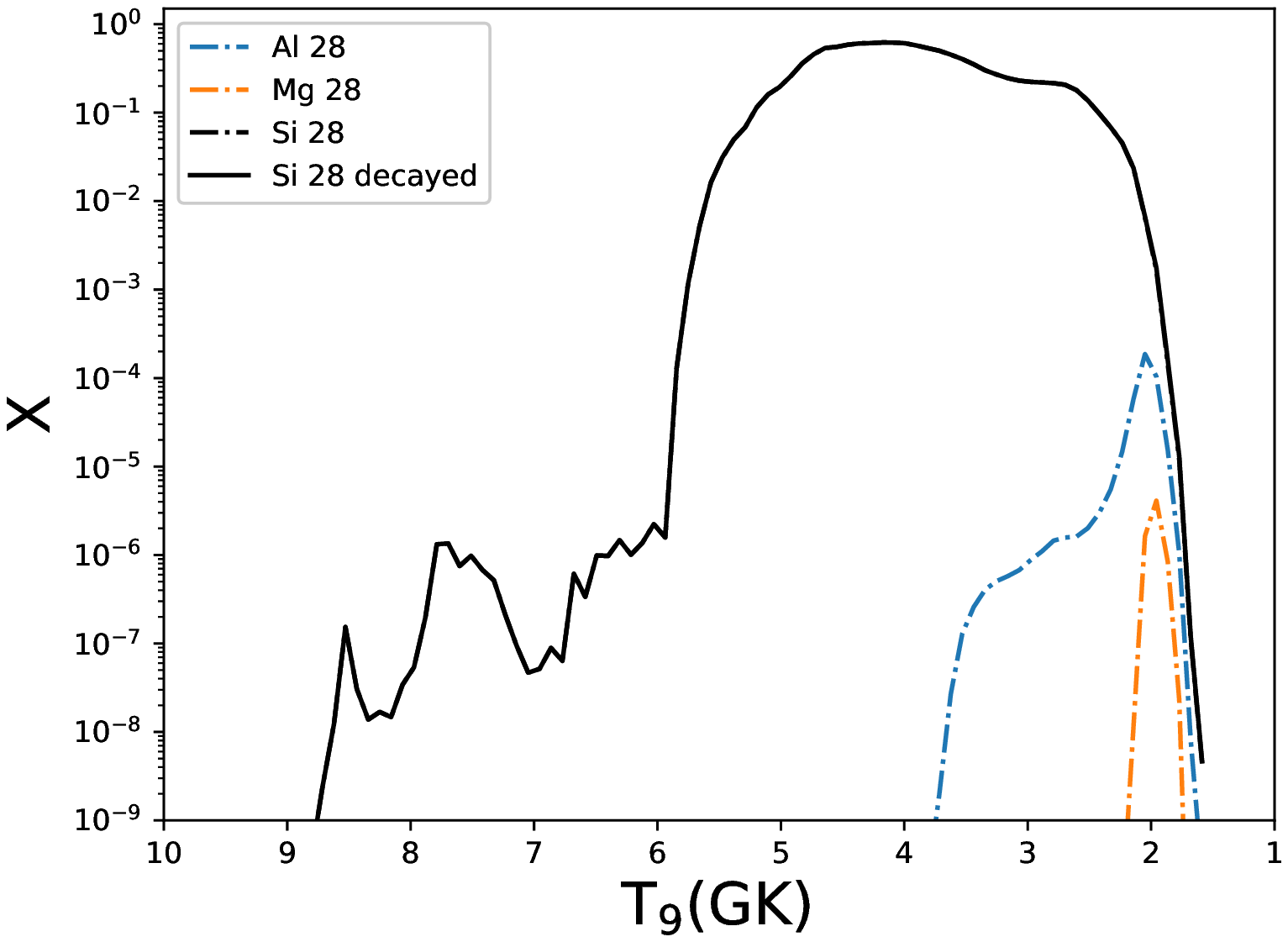}{0.3\textwidth}{T1.4Z0.014}
              \fig{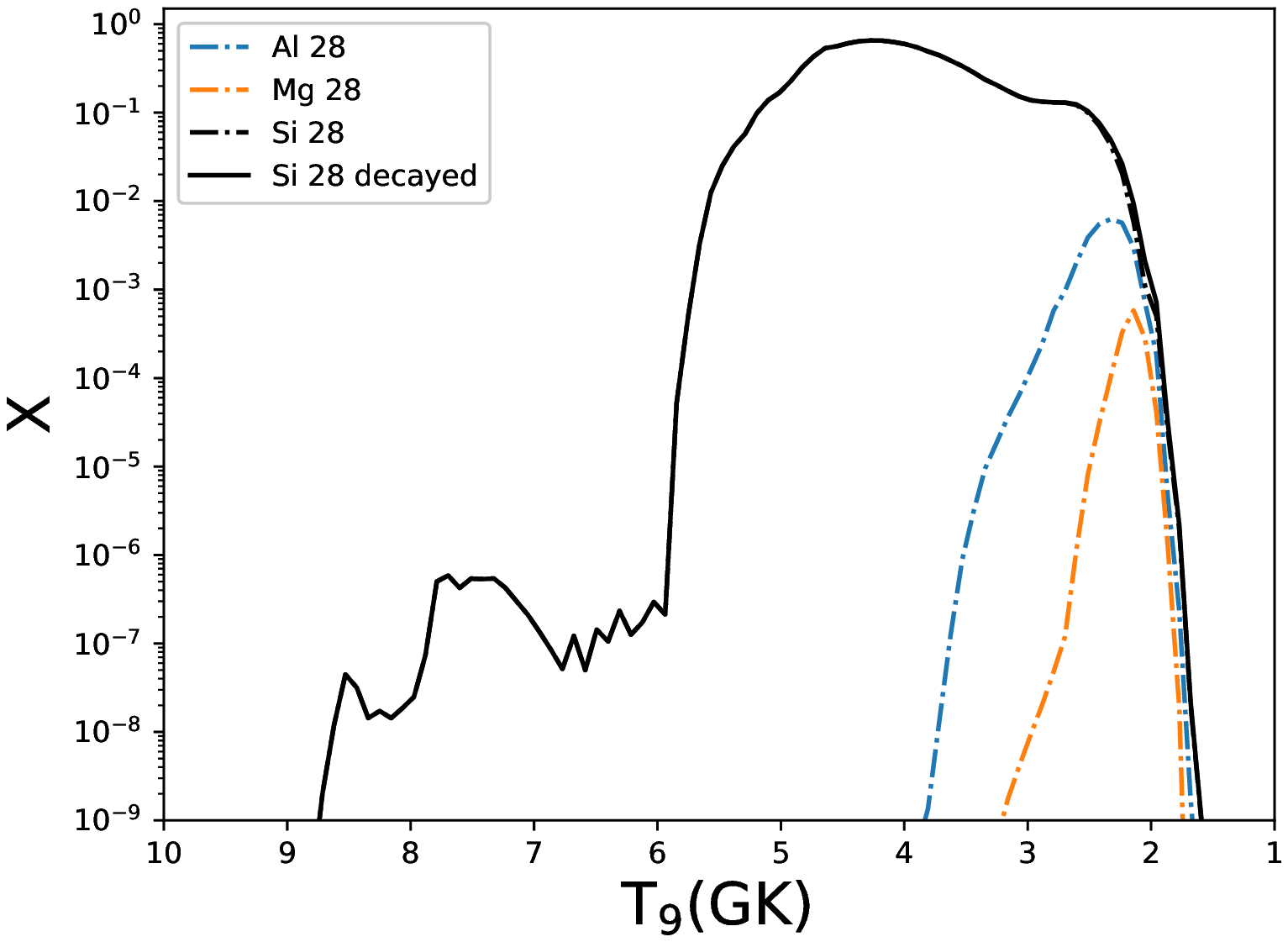}{0.3\textwidth}{T1.4Z0.1}
             }
    \gridline{\fig{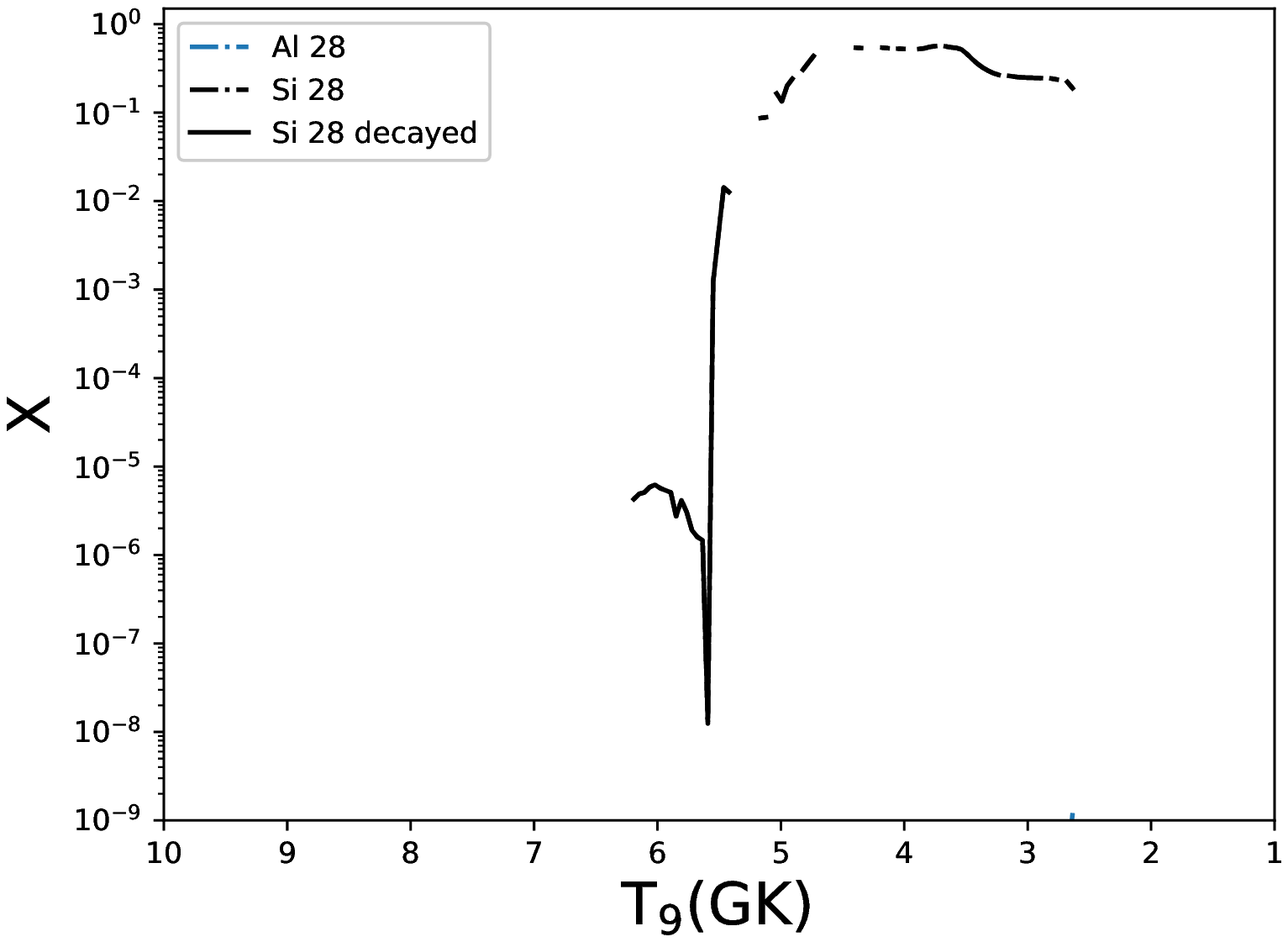}{0.3\textwidth}{S1.0Z0}
              \fig{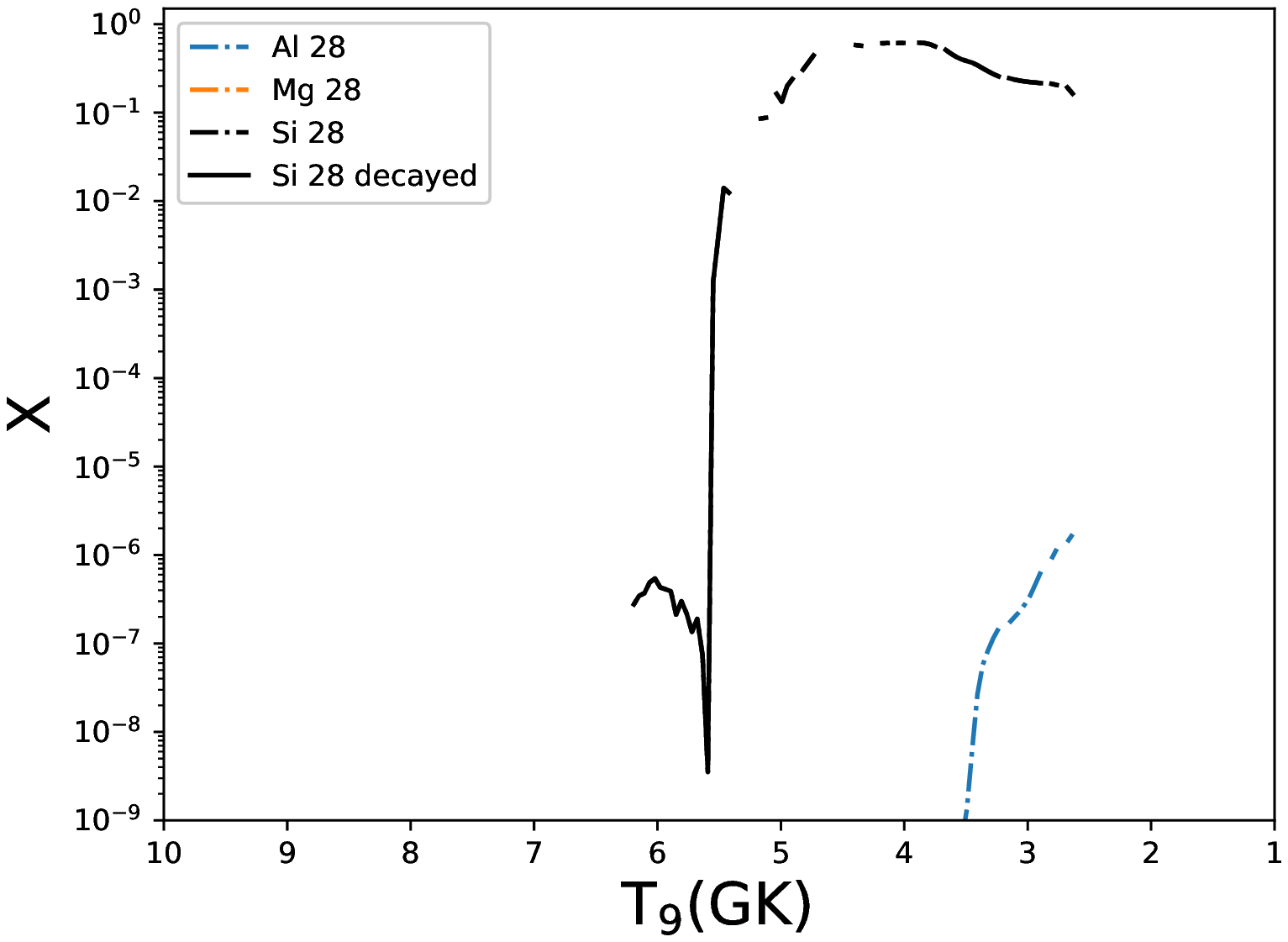}{0.3\textwidth}{S1.0Z0.014}
              \fig{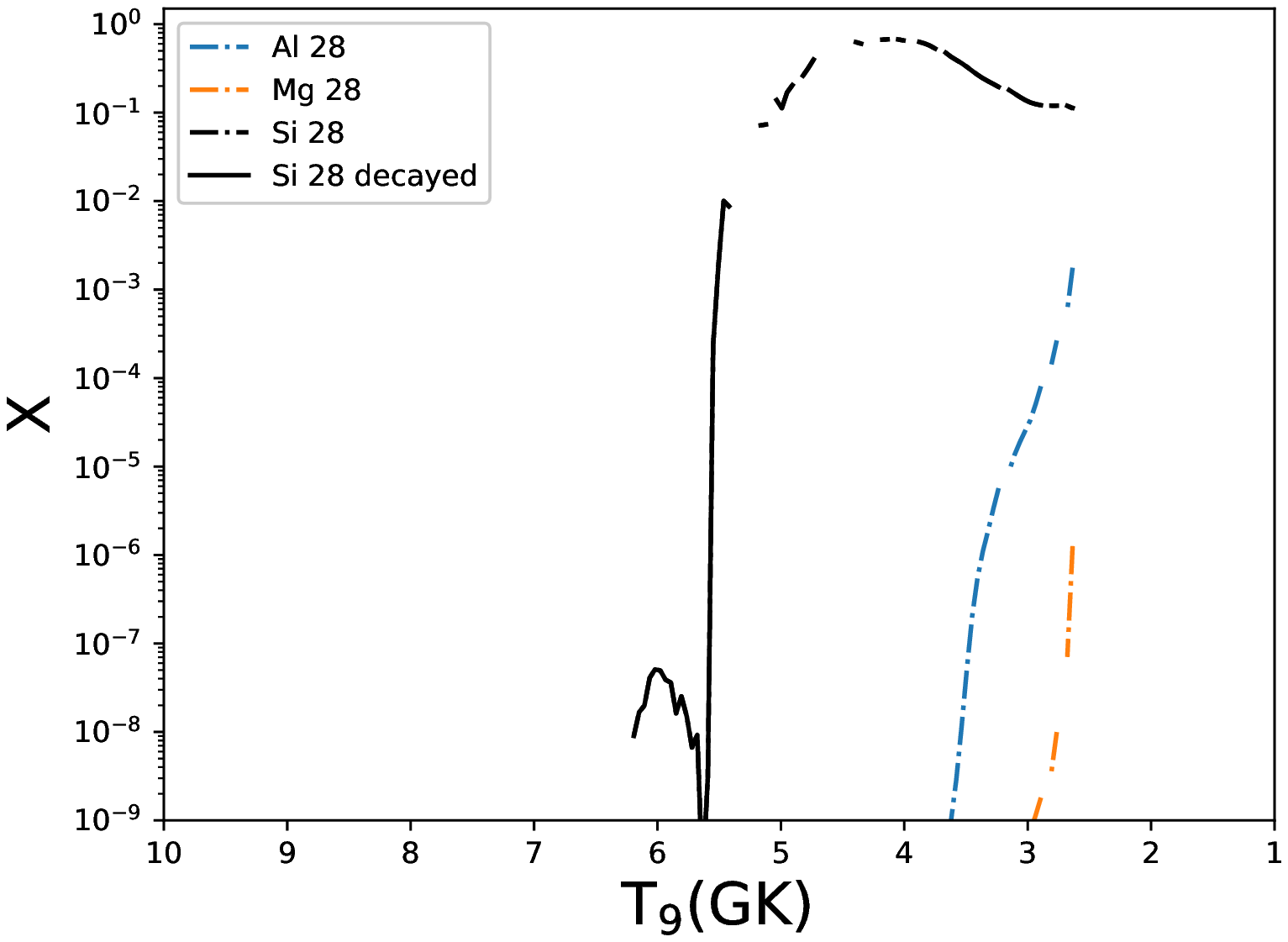}{0.3\textwidth}{S1.0Z0.1}
             }
    \gridline{\fig{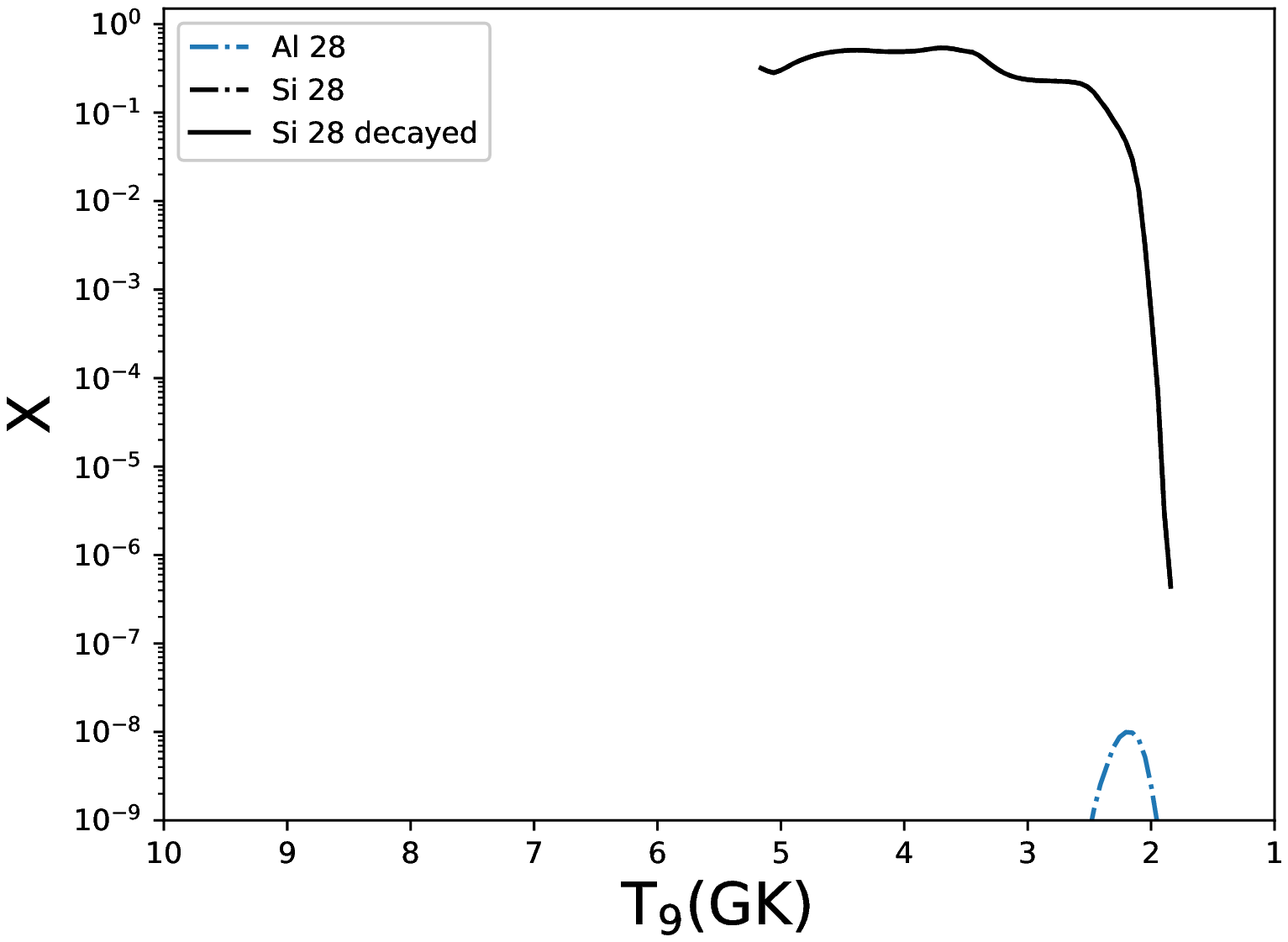}{0.3\textwidth}{M0.8Z0}
              \fig{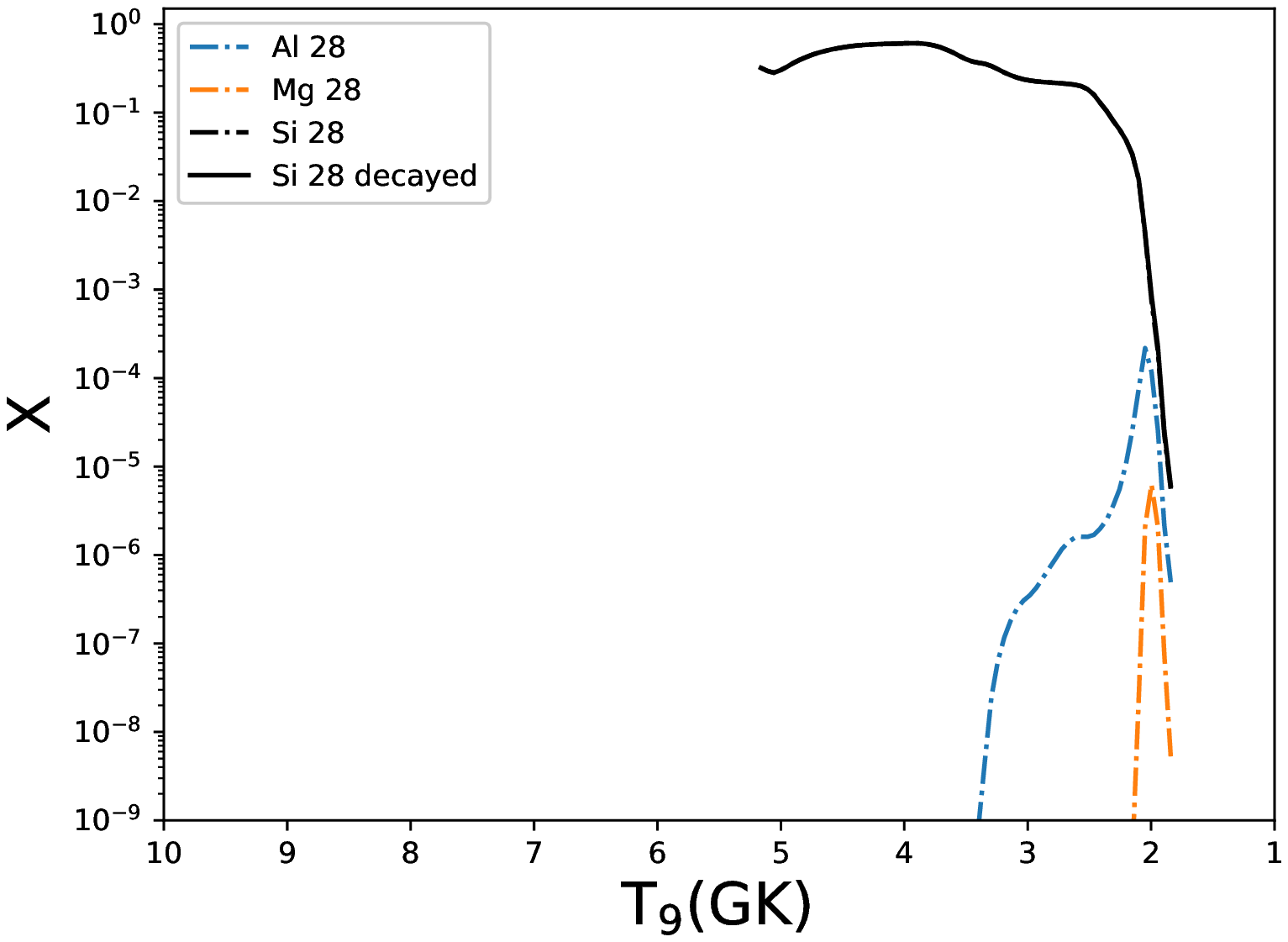}{0.3\textwidth}{M0.8Z0.014}
              \fig{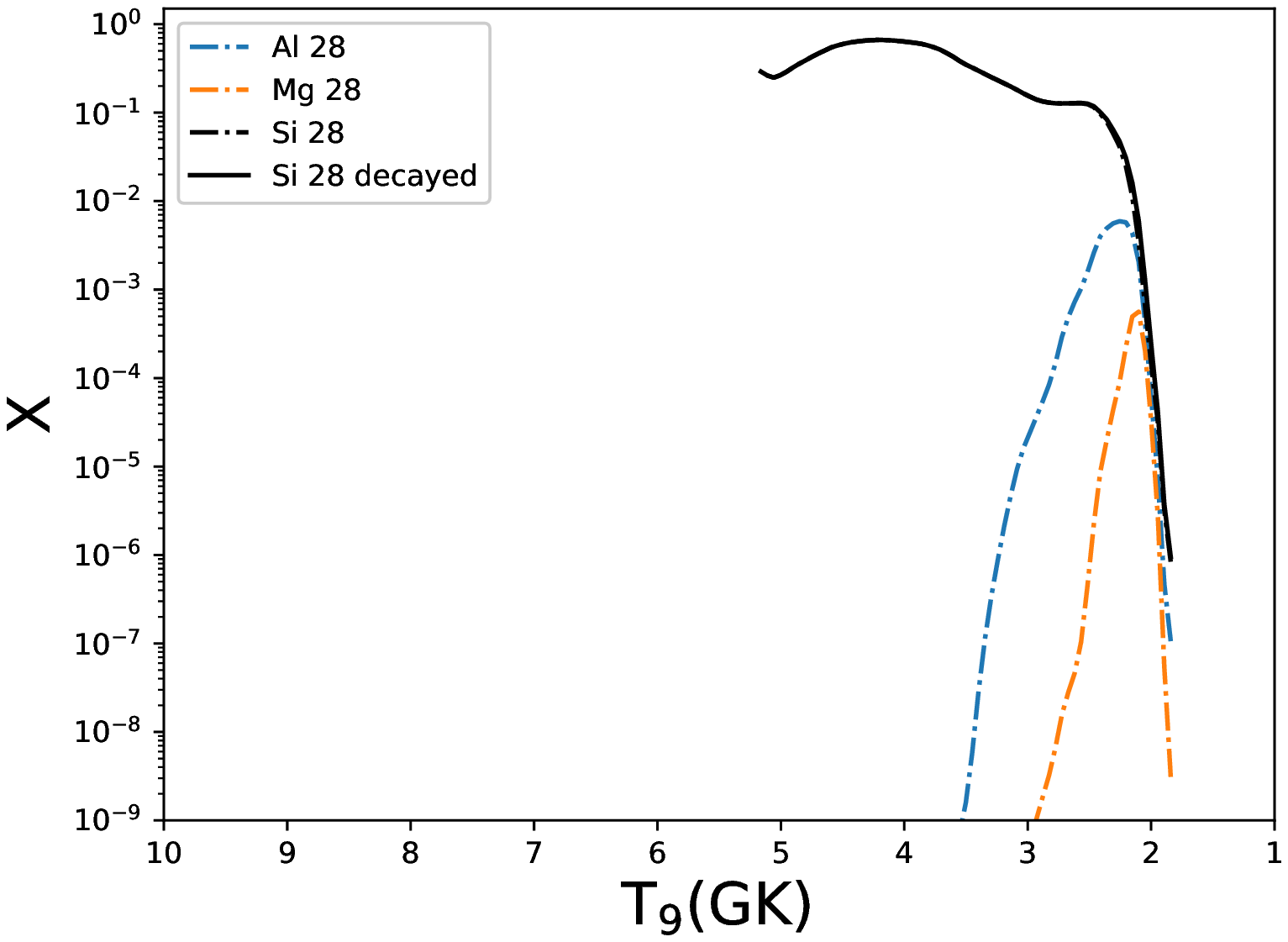}{0.3\textwidth}{M0.8Z0.1}
             }
    \caption{Top panels: The $^{28}$Si abundances in the ejecta are shown with respect to the explosion temperature peak (continuous black line) for the models T1.4Z0 (Z$_{met}$ = 0, left), T1.4Z0.014 ( Z$_{met}$ = 0.014, middle) and T1.4Z0.1 (Z$_{met}$ = 0.1, right). $^{28}$Si abundances are shown with contributions from radioactive species, which are negligible in this case.
    Middle Panels: As for top panels, but for models S1.0Z0, S1.0Z0.014 and S1.0Z0.1, respectively.
    Bottom Panels: As for top panels, but for models M0.8Z0, M0.8Z0.014 and M0.8Z0.1, respectively.}
\label{fig:si_28_radiogenic}
\end{figure*}

The ejected abundance distribution for $^{28}$Si is shown in detail for three representative metallicities in Figure ~\ref{fig:si_28_radiogenic}. All other such isotopic production plots are available on \dataset[Zenodo: Figures]{https://doi.org/10.5281/zenodo.8059974
}, and their features discussed below. For each model, all tracer particles are plotted without binning of the data.

$^{28}$Si shows a broad peak of production in all models, stretching from 1.8 to 6 \,GK (slightly lower in the S1.0 and M0.8 models, due to the lower density of these tracer particles). Production of $^{28}$Si is dominated by direct production - radiogenic contributions are negligible.

Starting in the deepest ejecta (Figure~\ref{fig:si_28_radiogenic}), Models T1.4Z0, T1.4Z0.014 and T1.4Z0.1 have a smaller production peak at higher temperatures, which is relatively insensitive to the initial metallicity. This is due to the high density region in the SNIa causing electron capture reactions. For stellar conditions where nuclear reaction rates are approaching NSE, charged particle reaction rates are in equilibrium. Therefore, the weak interactions govern the shift in distribution from normal NSE to a more neutron rich distribution \citep{langanke2000shell}. The density of these hottest particles ensures that the Y$_{e}$ is similar between all T1.4 models in the high temperature region. On the other hand, for all models the bulk of the ejected $^{28}$Si comes from trajectories at lower peak temperatures and this high temperature region makes a negligible contribution.

$^{29}$Si is produced for a broad range of conditions in T1.4Z0, up to 6\,GK. Such a small component to the ejected $^{29}$Si is not present at higher metallicities as production moves to more neutron rich isotopes, where the secondary production at lower peak temperatures instead dominates the isotopic yields. This same trend is seen in the S1.0 and M0.8 models, with a reduction in the production at intermediate peak temperatures accompanied by a boost in production at lower temperatures. The radiogenic $^{29}$P abundance is a significant contributor to $^{29}$Si in the zero-metallicity models T1.4Z0, S1.0Z0 and M0.8Z0. On the other hand, there is no significant $^{29}$P radiogenic contribution for higher metallicities. Only at super solar metallicities do we see a small contribution to the ejected $^{29}$Si mass of the order of a few percent.

The production of $^{30}$Si is more complex than $^{29}$Si, especially for model T1.4Z0. In this case, broad radiogenic contributions are obtained from two isotopes - $^{30}$P is dominant at most peak temperatures, while there is equal production of $^{30}$P and $^{30}$S at temperatures above 5\,GK. At higher metallicities, for all models the radiogenic contribution becomes negligible, and most $^{30}$Si is ejected directly with a dominant contribution between 2 and 4 GK in all models.


\subsection{Sulfur}

\begin{figure*}
    \gridline{\fig{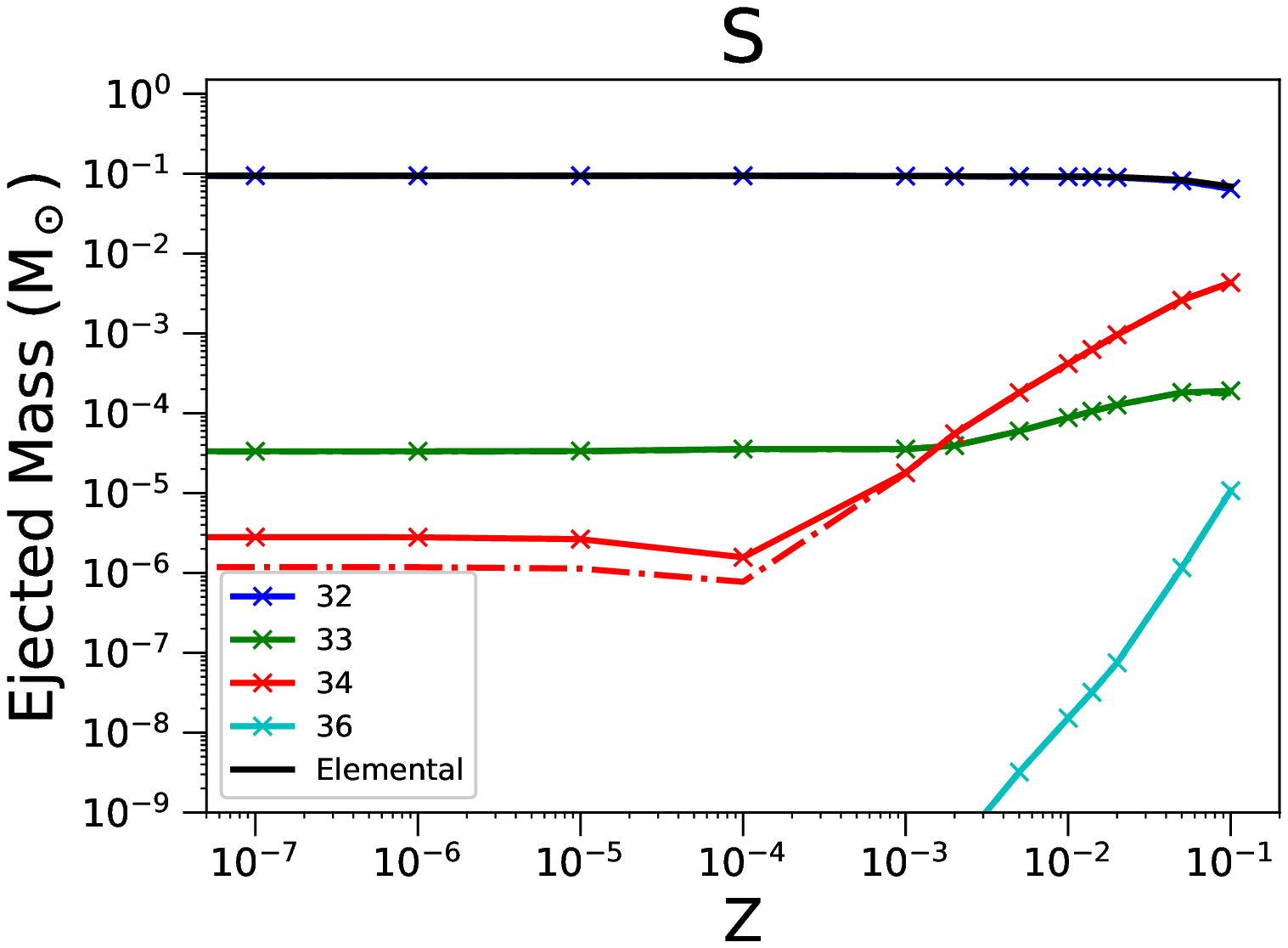}{0.45\textwidth}{T1.4}}
    \gridline{\fig{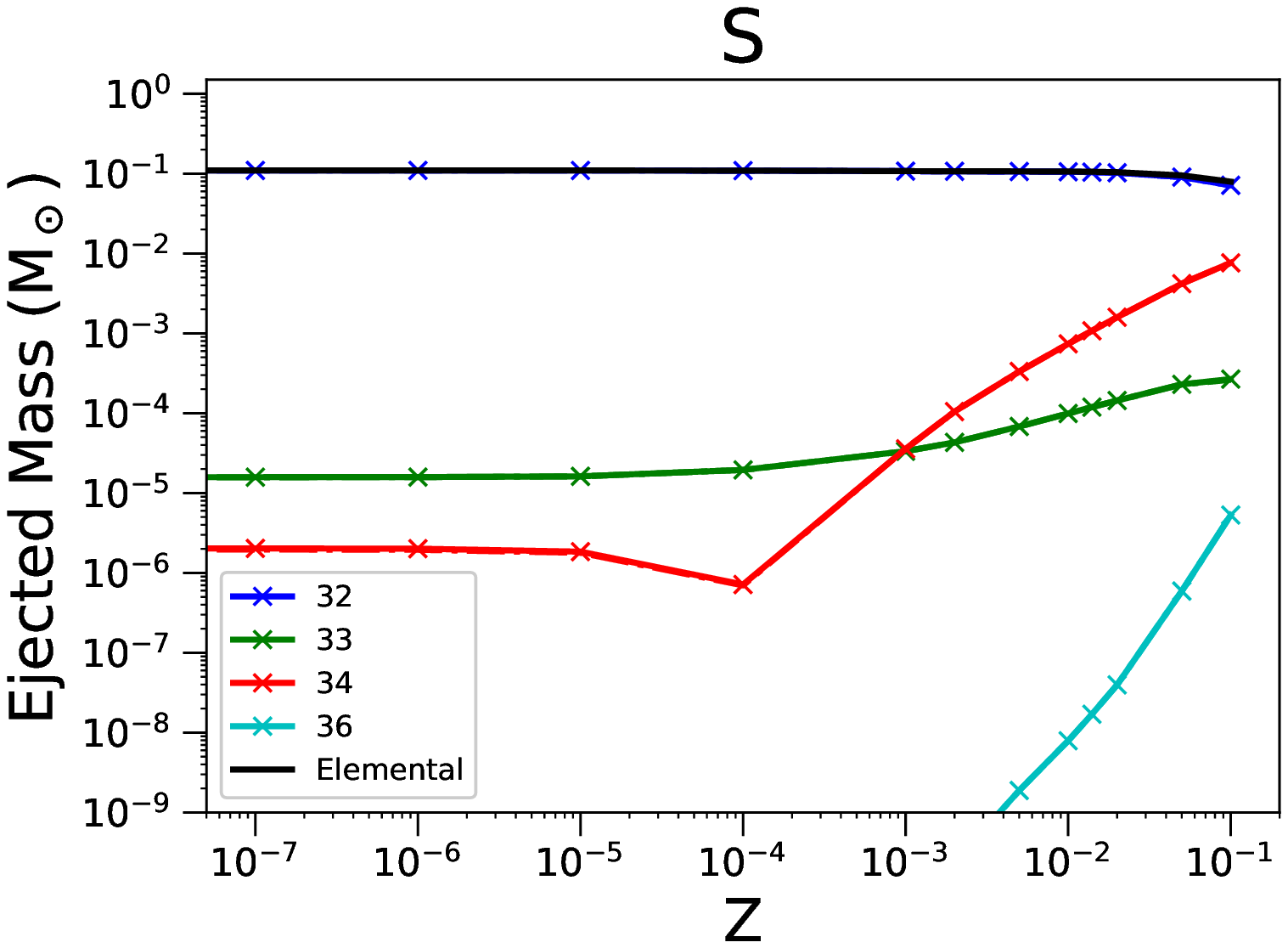}{0.45\textwidth}{S1.0}}
    \gridline{\fig{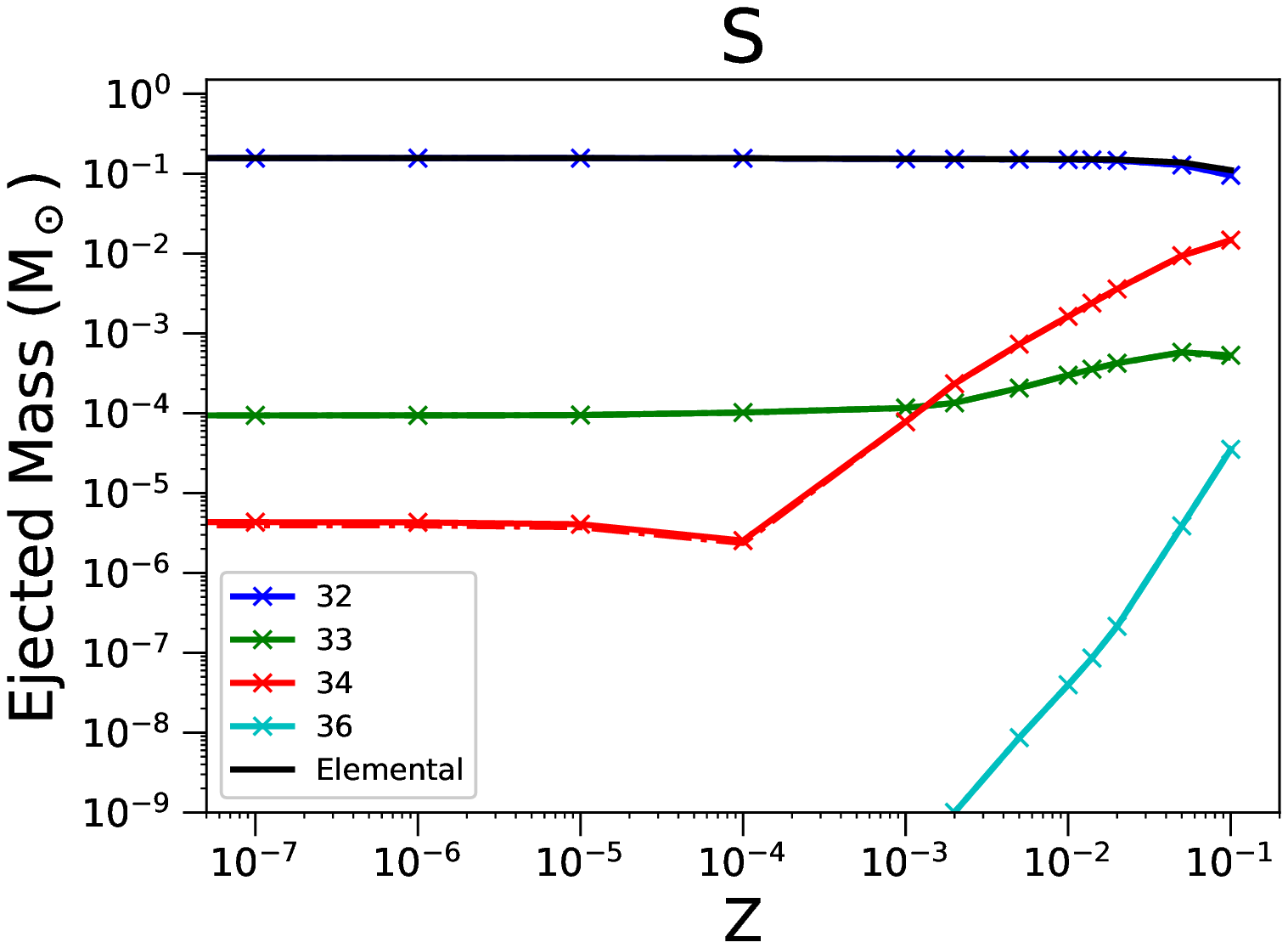}{0.45\textwidth}{M0.8}}
    \caption{As in Figure \ref{fig:Silicon_metalicity_multi} but for ejected mass of S, and its stable isotopes $^{32}$S, $^{33}$S, $^{34}$S and $^{36}$S.}
\label{fig:Sulphur_metalicity_multi}
\end{figure*}

Sulfur consists of 4 stable isotopes: $^{32}$S, $^{33}$S, $^{34}$S and $^{36}$S. $^{32}$S (the most abundant S isotope in the solar system comprising 94.99\% of the solar budget of sulfur) and $^{34}$S (4.25\% of the solar budget of sulfur) are mostly produced in explosive and hydrostatic oxygen burning in CCSNe. In addition to this component,
$^{33}$S (0.75\% of the solar budget) is also produced by explosive neon burning \citep[e.g.,][]{woosley2002evolution}. Finally, the isotope $^{36}$S (the least abundant S isotope in the solar abundances) is made by the s-process in AGB (asymptotic giant branch) stars \citep[][]{kahane:00} and by CCSNe, in either hydrostatic helium burning or carbon oxygen burning \citep{pignatari2016nugrid,woosley2002evolution}. \citet{kobayashi2020origin} identify a contribution to the elemental abundance of S from SNIa.

From figure \ref{fig:Sulphur_metalicity_multi}, the trends of each S isotope are very similar between different sets of model, and  show similar trends with respect to metallicity. On the absolute ejected mass of the various S isotopes, the models differ
by about a factor of 2 except for $^{33}$S, which is more strongly produced in the M0.8 models, at some metallicities by nearly an order of magnitude.

Concerning the detailed production of S species, $^{32}$S is directly ejected for all models at low metallicity, with a similar production profile below 5\,GK. 
In the T1.4Z0 model, there is a negligible $^{32}$S contribution to the final ejecta from a high peak temperature tail. At solar and super-solar metallicities, we also observe the rise of a secondary production of radiogenic $^{32}$S due to the decay of $^{32}$P and $^{32}$Si. The isotope $^{33}$S is mostly made directly in all models. Some contribution is seen from the decay of $^{33}$Cl in the models at zero metallicity, and from $^{33}$P in solar-metallicity models. The isotope $^{34}$S can be made directly in our SNIa models with a significant contribution from $^{34}$Cl at zero metallicity, and from $^{34}$P in solar-metallicity conditions. Finally, the $^{36}$S abundance ejecta is not shown here as it is negligible, but it is produced directly in all models.


\subsection{Argon}

\begin{figure*}
    \gridline{\fig{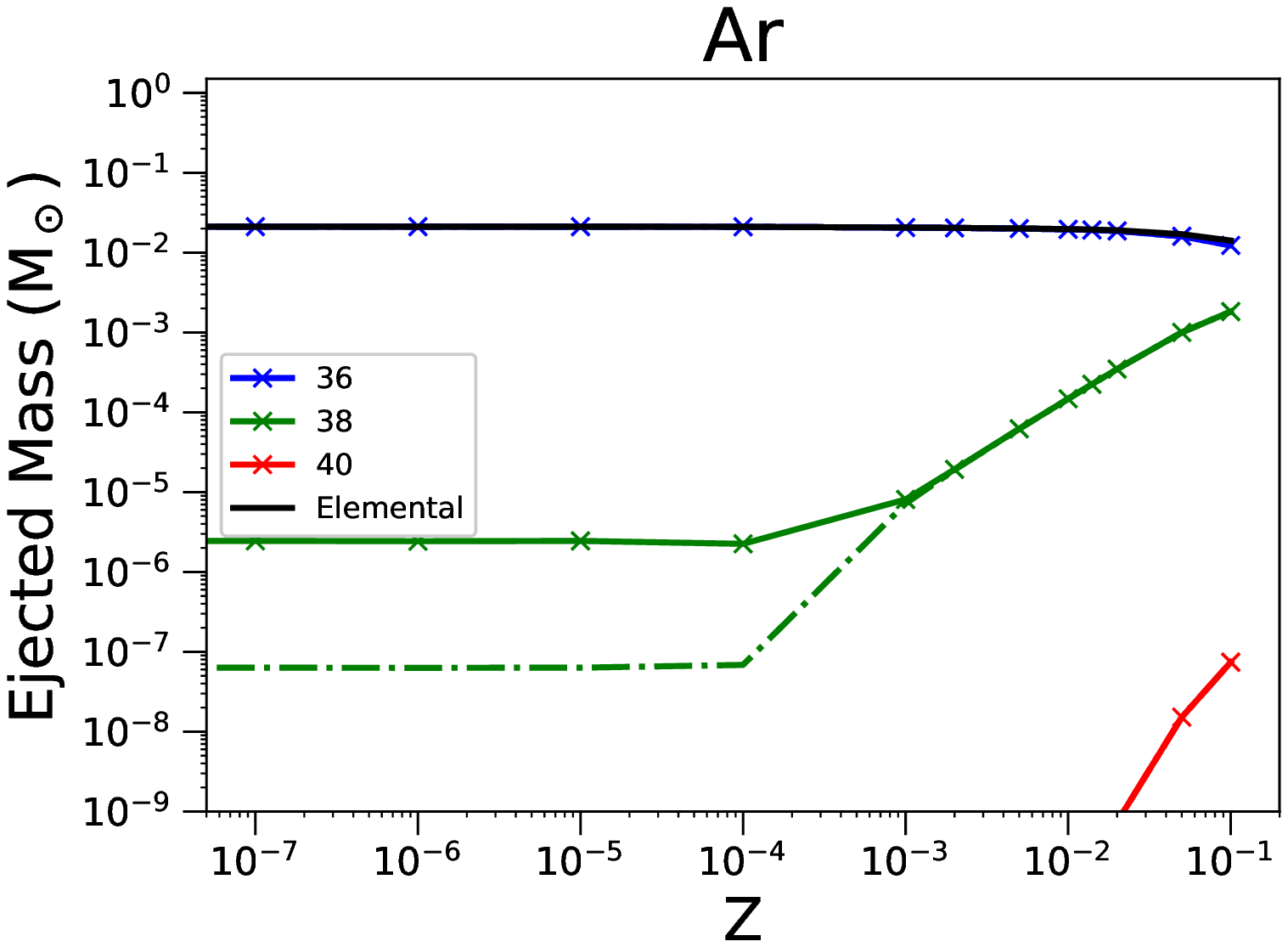}{0.45\textwidth}{T1.4}}
    \gridline{\fig{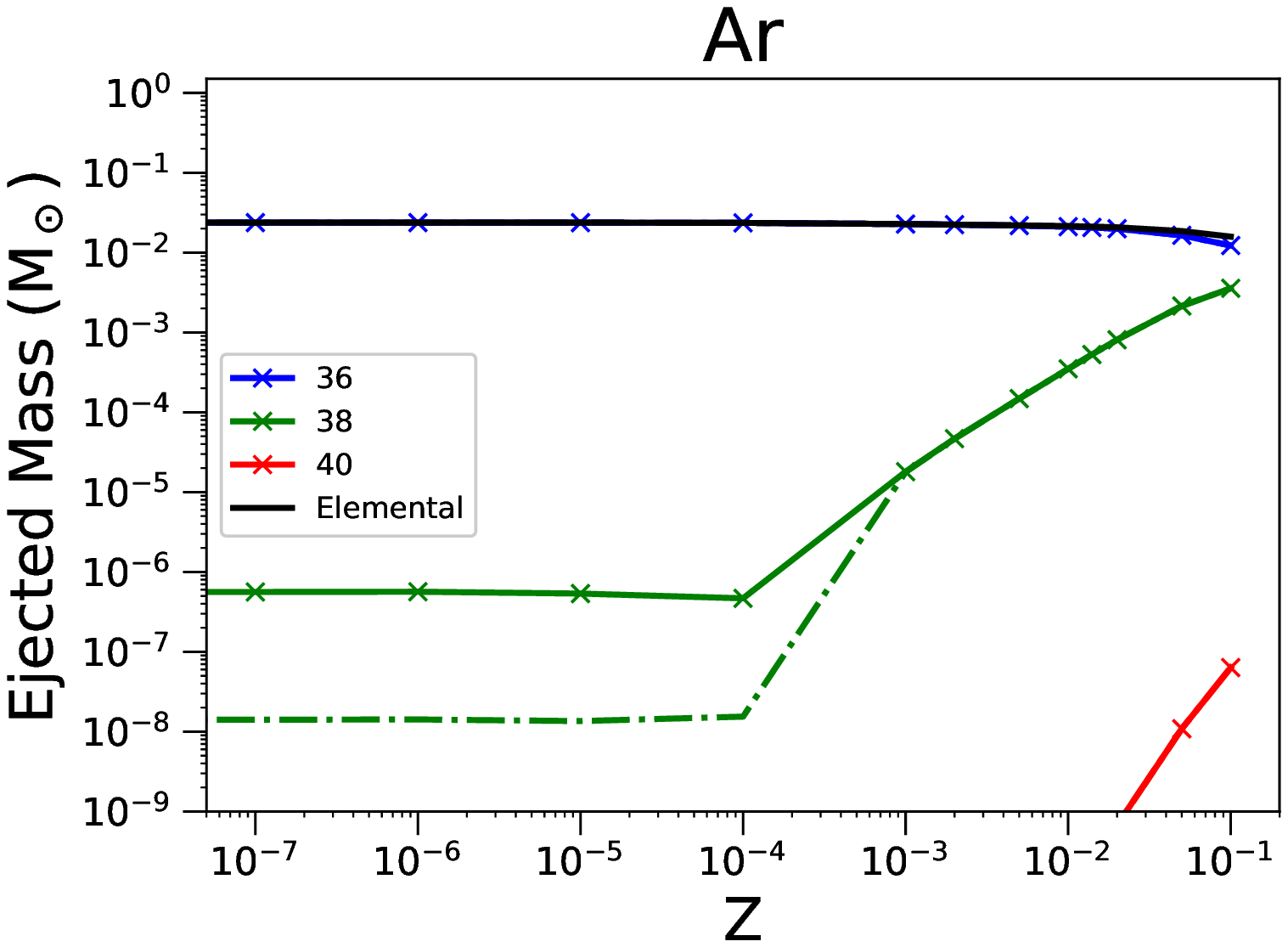}{0.45\textwidth}{S1.0}}
    \gridline{\fig{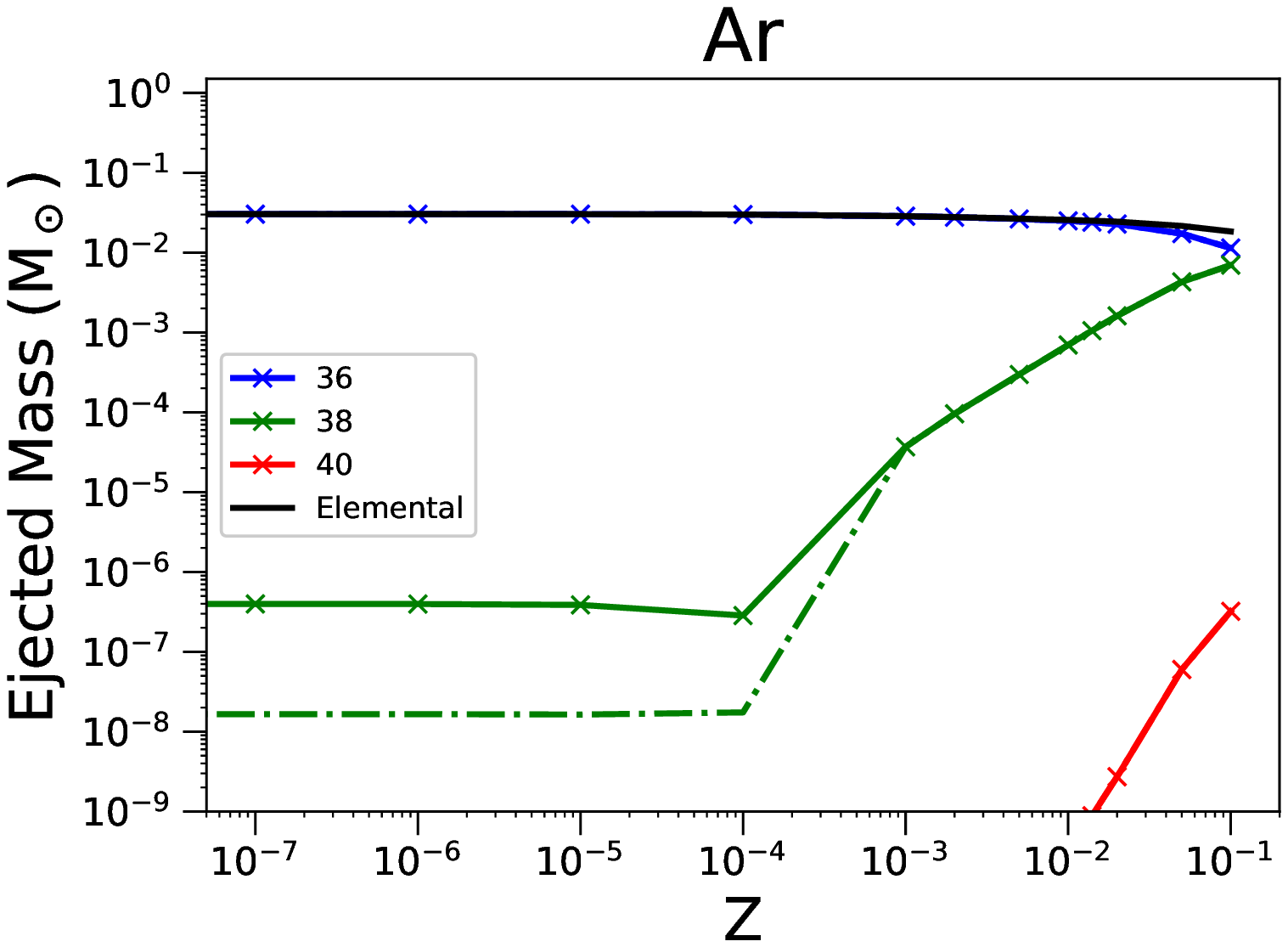}{0.45\textwidth}{M0.8}}
    \caption{As in Figure \ref{fig:Silicon_metalicity_multi} but for ejected mass of Ar, and its stable isotopes $^{36}$Ar, $^{38}$Ar and $^{40}$Ar.}
\label{fig:Argon_metalicity_multi}
\end{figure*}

Argon consists of 3 stable isotopes: $^{36}$Ar, $^{38}$Ar and $^{40}$Ar. $^{36}$Ar and $^{38}$Ar are produced in CCSN during oxygen burning and explosive oxygen burning \citep{woosley2002evolution}. $^{40}$Ar (99.6\% of the solar Ar budget) is produced either during explosive helium burning by the neutron burst triggered by the supernova shock passage at the bottom of the hydrostatic He shell \citep[][]{rauscher:02, pignatari:18,lawson:22} or by the s-process and then ejected in the convective C-burning shell \citep[e.g.,][]{the:07, pignatari:10, pignatari2016nugrid}. 
Based on GCE simulations, \citet{kobayashi2020origin} also identify a proportion of elemental Ar as being produced in SNIa explosions.

In all three sets of models argon production is mostly primary\footnote{Here, primary denotes that material which is synthesized, and its abundance yields do not depend on the initial composition of the stellar SNIa progenitor, whereas secondary is material which production depends on the seed abundances of other nuclei in the stellar progenitor (ultimately in this work originating with the initial abundance of $^{22}$Ne).}, with a slight reduction in the total ejected mass at super-solar metallicities (Figure \ref{fig:Argon_metalicity_multi}). This is due to the decrease in production of $^{36}$Ar with increasing metallicity, which is partially compensated for at the highest metallicities with the increase of the $^{38}$Ar yields. In the case of the M0.8 models, $^{38}$Ar becomes almost as abundant as $^{36}$Ar. Indeed, the ratio of $^{36}$Ar to $^{38}$Ar at super-solar metallicities approaches unity as the mass of the progenitor decreases, whilst there is a factor of 5 difference in the T1.4Z0.1 model. Radiogenic $^{38}$Ar has a primary production up to Z$_{met}$ = 10$^{-4}$, while its direct production drives the increase in yields above Z$_{met}$ = 10$^{-3}$.

For $^{36}$Ar, all three classes of models show similar conditions of production and similar trends with metallicity. At Z$_{met}$ = 0, $^{36}$Ar is primarily produced at temperatures between 4 and 5.5 \,GK. At temperatures between 2.5 and 3\,GK a small amount of $^{36}$Ar is ejected as secondary $^{36}$Cl, which remains a small component, even at super-solar metallicities. Overall, the production of $^{36}$Ar is not affected by the initial composition of the progenitor.

At zero metallicity, For all the SNIa model sets, $^{38}$Ar is mainly made as the radiogenic product of $^{38}$K, while for solar and super-solar metallicity $^{38}$Ar is directly produced. In the T1.4Z0 model, there is a large $^{38}$Ar contribution in the higher temperature range of 5.5 - 7GK, temperatures which are not reached in the M0.8 models, and only at the lower end (around 6\,GK) in the S1.0 models. $^{38}$Ar has an ejected mass approximately 5$\times$ greater in the low metallicity models of T1.4 as compared with S1.0 and M0.8 due to this higher temperature production range. 
At higher metallicities, the production in the most internal ejecta becomes negligible compared to the metallicity-dependent colder component. The nucleosynthesis in the lower temperature region is driven by C-fusion and O-fusion. 

Not shown in these figures, $^{40}$Ar does not show any high temperature component in SNIa ejecta. The $^{40}$Ar production is highly dependent on the initial metallicity, and no models show production of $^{40}$Ar at Z$_{met}$ = 0. Only trace amounts of $^{40}$Ar are synthesized between 2 and 3\,GK in the Z$_{met}$ = 0.014 models. We instead see a sharp rise in the production of $^{40}$Ar at super-solar metallicities (Figure~\ref{fig:Argon_metalicity_multi}).


\subsection{Calcium}

\begin{figure*}
    \gridline{\fig{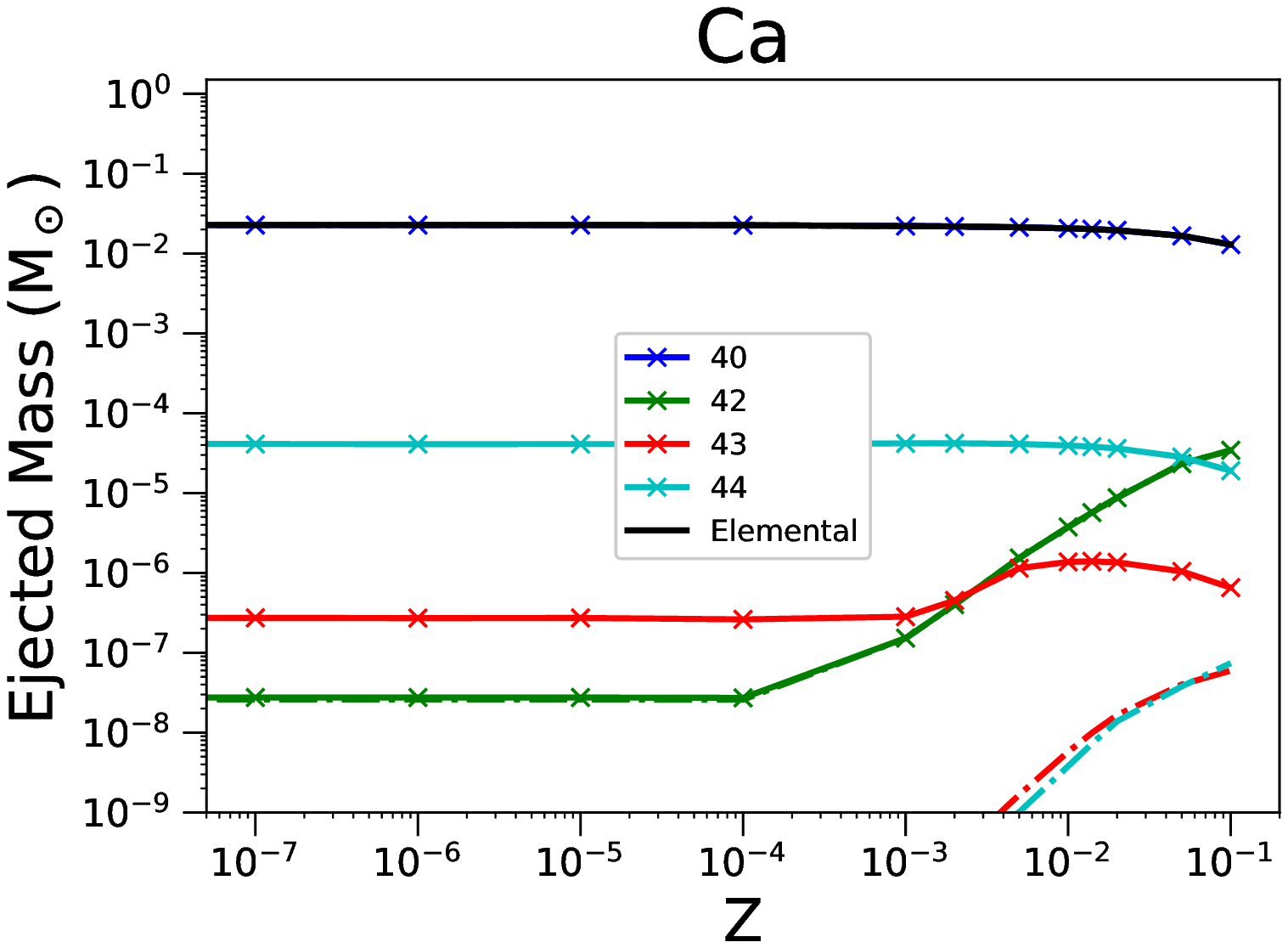}{0.45\textwidth}{T1.4}}
    \gridline{\fig{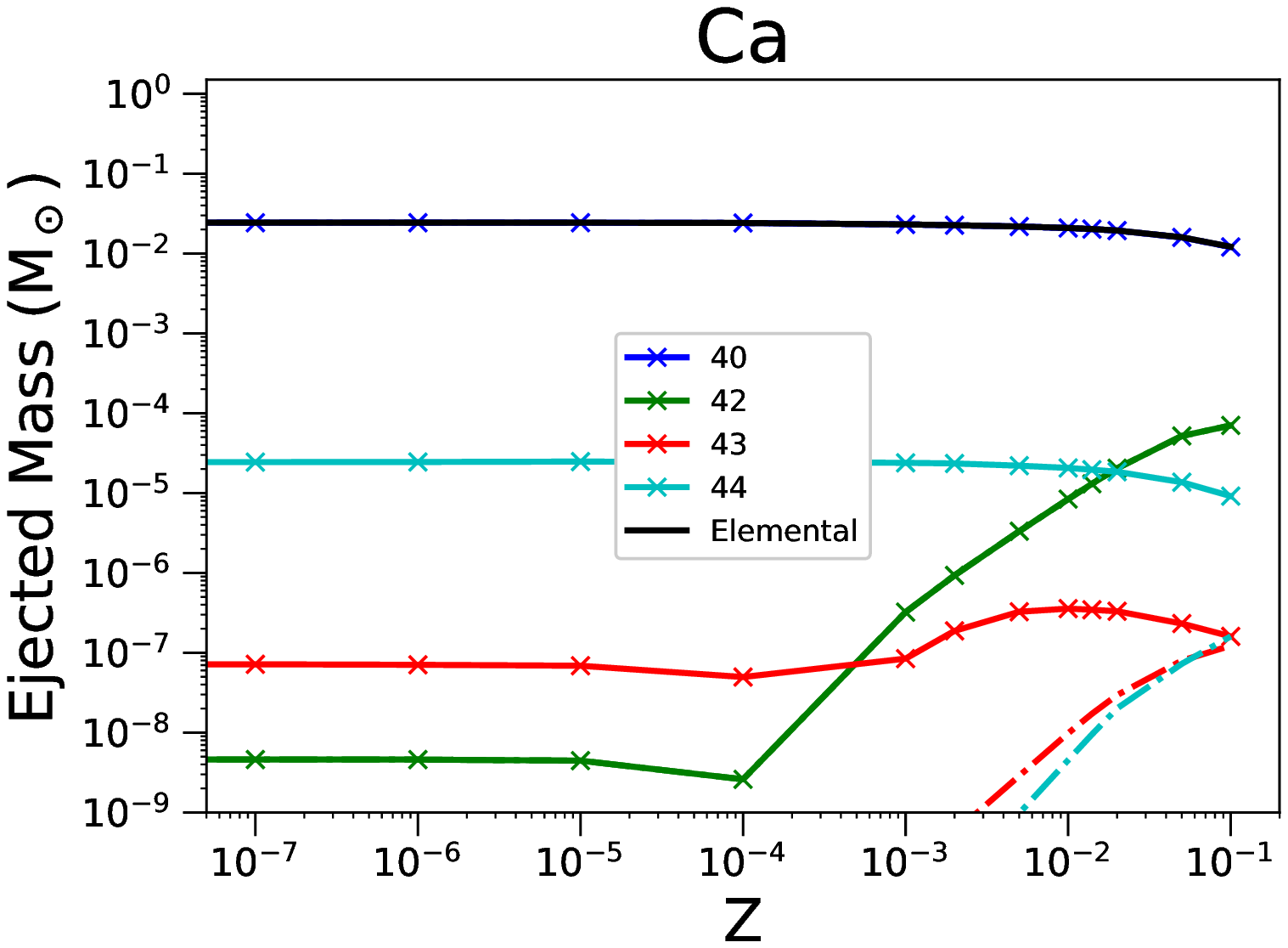}{0.45\textwidth}{S1.0}}
    \gridline{\fig{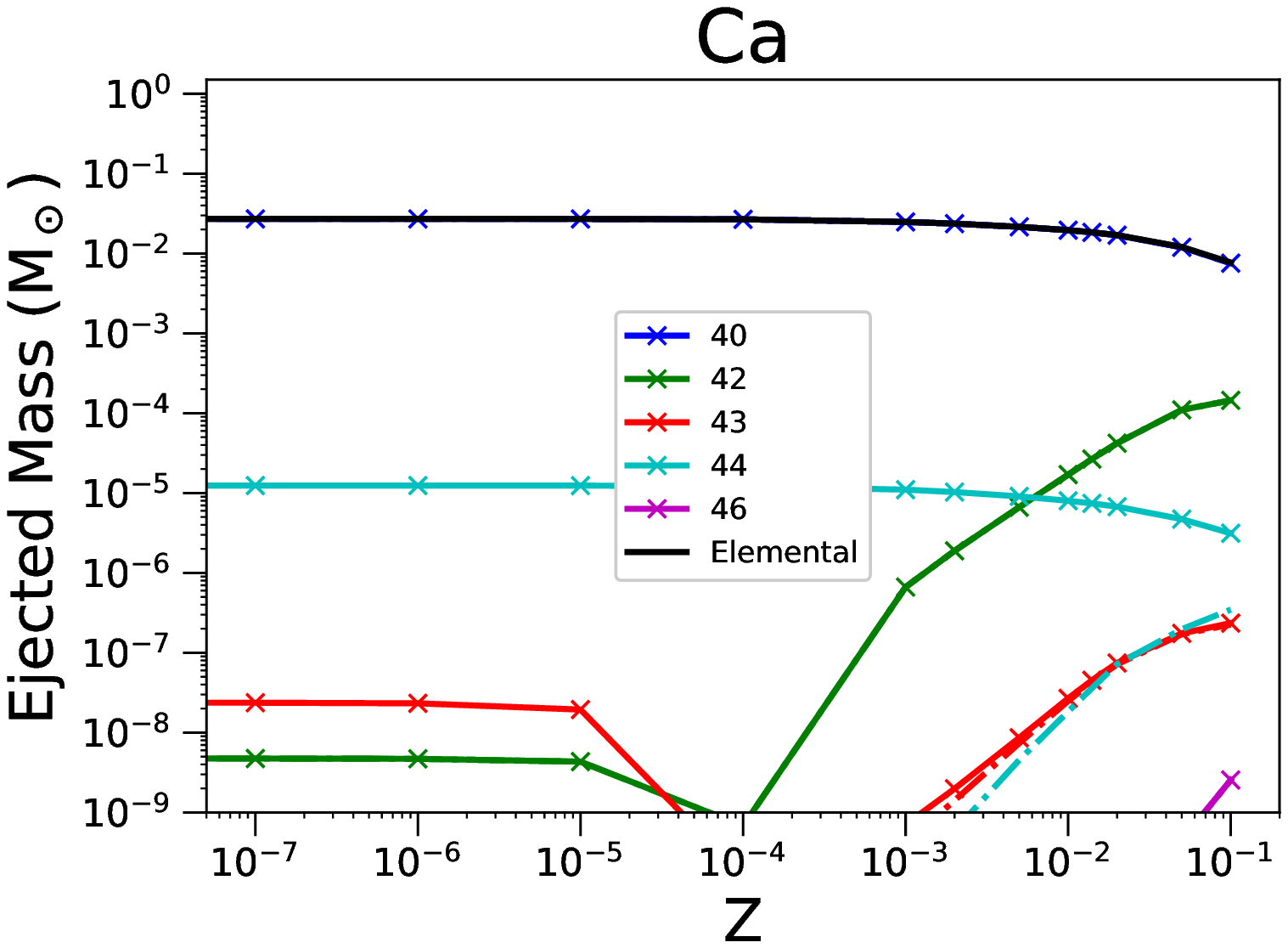}{0.45\textwidth}{M0.8}}
    \caption{As in Figure \ref{fig:Silicon_metalicity_multi} but for ejected mass of Ca, and its stable isotopes  $^{40}$Ca, $^{42}$Ca, $^{43}$Ca and $^{44}$Ca.
    }
\label{fig:Calcium_metalicity_multi}
\end{figure*}

Calcium consists of 6 stable isotopes: $^{40}$Ca, $^{42}$Ca, $^{43}$Ca and $^{44}$Ca, $^{46}$Ca and $^{48}$Ca. CCSNe provide a significant contribution to the Ca galactic inventory: in these stars both $^{40}$Ca (96.9\% of the solar budget) and $^{42}$Ca (0.6\%) are mostly produced by explosive oxygen burning, and $^{43}$Ca (0.1\%) is produced in carbon and neon burning, and in $\alpha$-rich freezeout following nuclear statistical equilibrium at high temperatures \citep{woosley2002evolution}. 
$^{44}$Ca (2\%) is predicted to be produced efficiently from SNIa
or from $\alpha$-rich freezeout in CCSNe, where it is formed as $^{44}$Ti \citep{pignatari2016nugrid,magkotsios2010trends}. $^{46}$Ca (0.004\%) has a contribution from AGB stars, and is also produced in carbon and neon burning in CCSNe. Finally, the origin of $^{48}$Ca (0.1\%) in nature is unclear, since in most neutron-rich stellar conditions it is difficult to produce without overproducing the rare $^{46}$Ca. \cite{jones:19} showed that with a small occurrence of thermonuclear electron-capture supernovae the full solar abundance of $^{48}$Ca could be explained. The effective relevance of such a stellar source is however still very uncertain.

For the three sets of models considered here, shown in Figure \ref{fig:Calcium_metalicity_multi}, we obtain similar production trends with metallicity for $^{40}$Ca, $^{42}$Ca and $^{44}$Ca. 
The $^{43}$Ca trend is similar for T1.4 and S1.0 models. On the other hand, between Z$_{met}$ = 10$^{-4}$ and Z$_{met}$ = 10$^{-2}$, its production drops significantly in the M0.8 models, and it increases again for high metallicities due to its direct production. Only a trace amount of $^{46}$Ca is produced in any model, and then only in the M0.8Z0.1 model. $^{48}$Ca is not produced in any of our post-processed models.

The production of $^{40}$Ca is fairly simple and consistent throughout the models. A broad production peak is observed in the Z$_{met}$ = 0 models in the peak temperature range between 4 and 5.5 GK. As the initial metallicity increases, the yields of $^{40}$Ca in this region decrease, resulting in the metallicity trend of $^{40}$Ca observed in Figure \ref{fig:Calcium_metalicity_multi}. Between Z$_{met}$ = 0 and Z$_{met}$ = 0.1, the ejected $^{40}$Ca decreases from 2.27x10$^{-2}$ to 1.29x10$^{-2}$ M$_{\odot}$ for the T1.4 models, from 2.43x10$^{-2}$ to 1.21x10$^{-2}$ M$_{\odot}$ in the S1.0 models, and from 2.71x10$^{-2}$ to 7.55x10$^{-3}$ M$_{\odot}$ in the M0.8 models. The larger effect in the M0.8 model is due to the absence of ejecta at high temperatures.

Not shown here in figures, the production of $^{42}$Ca in T1.4Z0 is dominated by a high temperature peak in the 6-7\,GK range, with smaller peaks between 4 and 4.3\,GK and 8 and 9\,GK. The lowest temperature peak is reproduced in the S1.0Z0 and M0.8Z0 models in the same temperature range. Direct synthesis of $^{42}$Ca is the main path for all three classes of model. Production of $^{42}$Ca is significantly boosted in the solar and super-solar metallicity models, where the low temperature abundance peak observed in the Z$_{met}$ = 0 models is increased by several orders of magnitude and covers a peak temperature range from 2 to 4\,GK. This temperature range covers both typical carbon fusion and oxygen fusion conditions, with the largest contribution being in the 3.5 - 5 GK range. At these temperatures, there is small radiogenic contribution from $^{42}$K and $^{42}$Ar.

Not shown in the figures presented here, the production of $^{43}$Ca in the T1.4Z0 is dominated by the 5.5 - 7 GK temperature range, where $^{43}$Ca is produced as $^{43}$Sc with a contribution on the order of 1\% from $^{43}$Ti. The majority of nucleosynthesis in this temperature range is primary as $^{43}$Sc production, and it does not change across metallicities; however, such a radiogenic contribution shows some secondary effects. For the T1.4Z0.014 model, the production below 5 GK shifts to lower peak temperatures and two components are ejected, peaking at about 2 and 4 GK. In these two regions, $^{43}$Ca is produced directly with a small contribution from $^{43}$Sc, and with a strong radiogenic contribution from $^{43}$K at the coldest peak temperatures.

In the S1.0Z0 model, the maximum peak temperature is approximately 6.2 GK. From around 5.5 to 6.2 GK we see the production of the radioisotope $^{43}$Sc - unlike in the T1.4Z0 case there is no contribution from $^{43}$Ti. A region of production in the  4 - 5 GK range is also present. In the model S1.4Z0.014, this lower temperature component resolves into a double peaked production region between 2 and 4 GK. Similar to the T1.4 case at analogous metallicity, there is a direct production of $^{43}$Ca with a small contribution from $^{43}$Sc, however the carbon region production is boosted in the S1.0Z0.014 model compared with T1.4Z0.014. S1.0Z0.1 also shows a double peak of production between 2 and 4 GK broadening with the radiogenic $^{43}$K contribution at lower temperatures. Production is mostly directly as $^{43}$Ca, with some contribution from $^{43}$K at the lowest peak of production in both the T1.4 and S1.0 cases, and a negligible contribution from $^{43}$Sc. Finally, the M0.8 models show the same trends in the 2 to 5 GK range as are observed in the T1.4 and S1.0 models.

$^{44}$Ca is produced primarily as $^{44}$Ti in all of the Z$_{met}$ = 0 models. In T1.4Z0, it is produced over a wide range of peak temperatures from 4 to 6.5\,GK, with a negligible contribution from a peak at 2.8\,GK. We see a similar distribution in the S1.0Z0 and M0.8Z0 models within the temperature range covered in their ejecta as discussed in the previous sections.
T1.4Z0.014 shows a small decrease in the production of $^{44}$Ti at the lower end of this broad range of production compared to the model at Z$_{met}$ = 0. This is partially compensated for by the increased secondary production in the small peak centered at 2.8\,GK. Across the rest of the temperature range, we see contributions only from $^{44}$Ca and $^{44}$Ti, with a small contribution from $^{44}$Sc at low peak temperatures. S1.0Z0.014 and M0.8Z0.014 show the same trend.
In the Z$_{met}$ = 0.1 models, a further suppression of the $^{44}$Ti production at the tail end of the broad production range is obtained, with the increase in direct synthesis of $^{44}$Ca compensating for this and leading to a continuation of the $^{44}$Ca production out to approximately 2.4 GK.

\subsection{Titanium}

\begin{figure*}
    \gridline{\fig{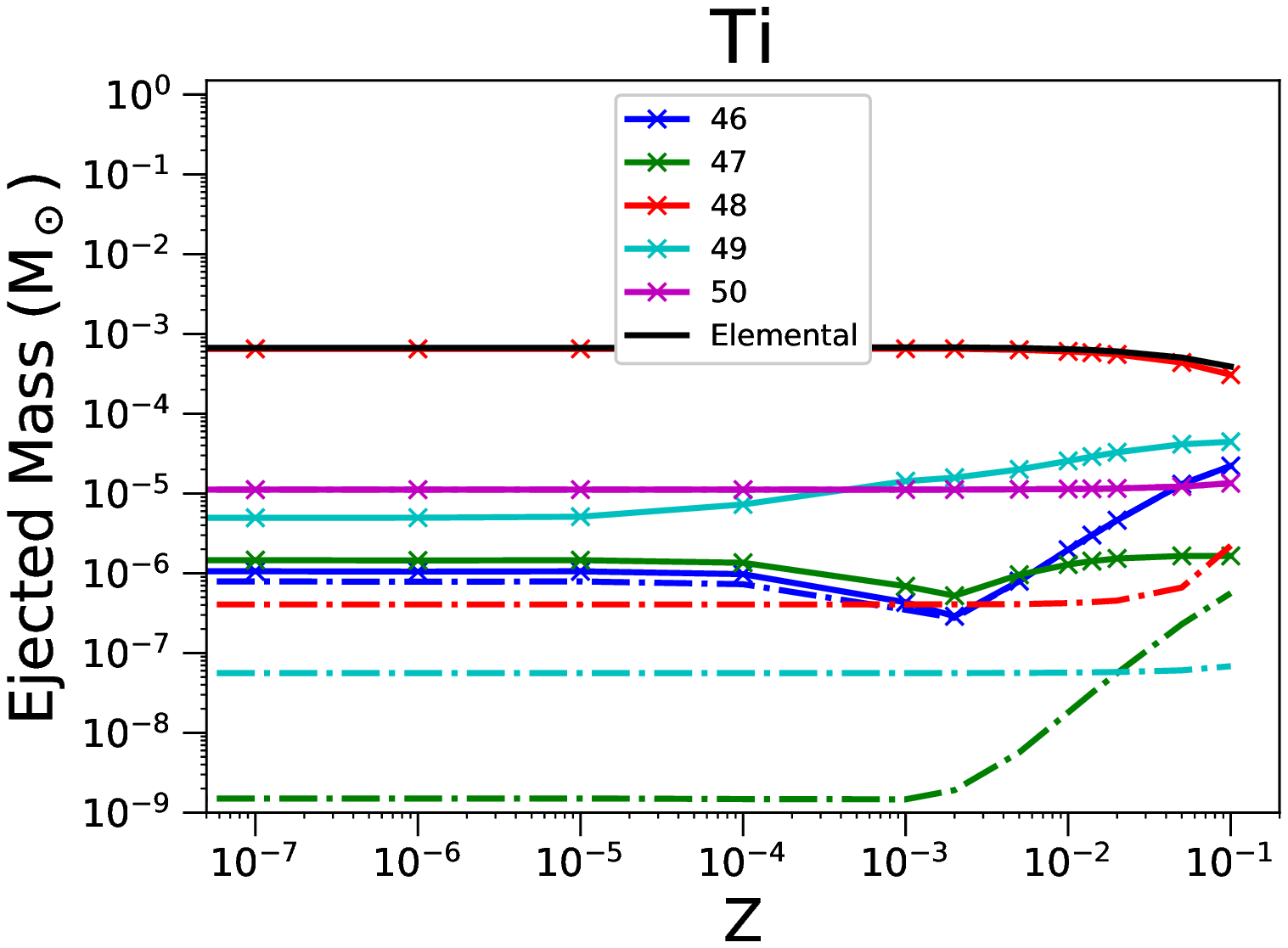}{0.45\textwidth}{T1.4}}
    \gridline{\fig{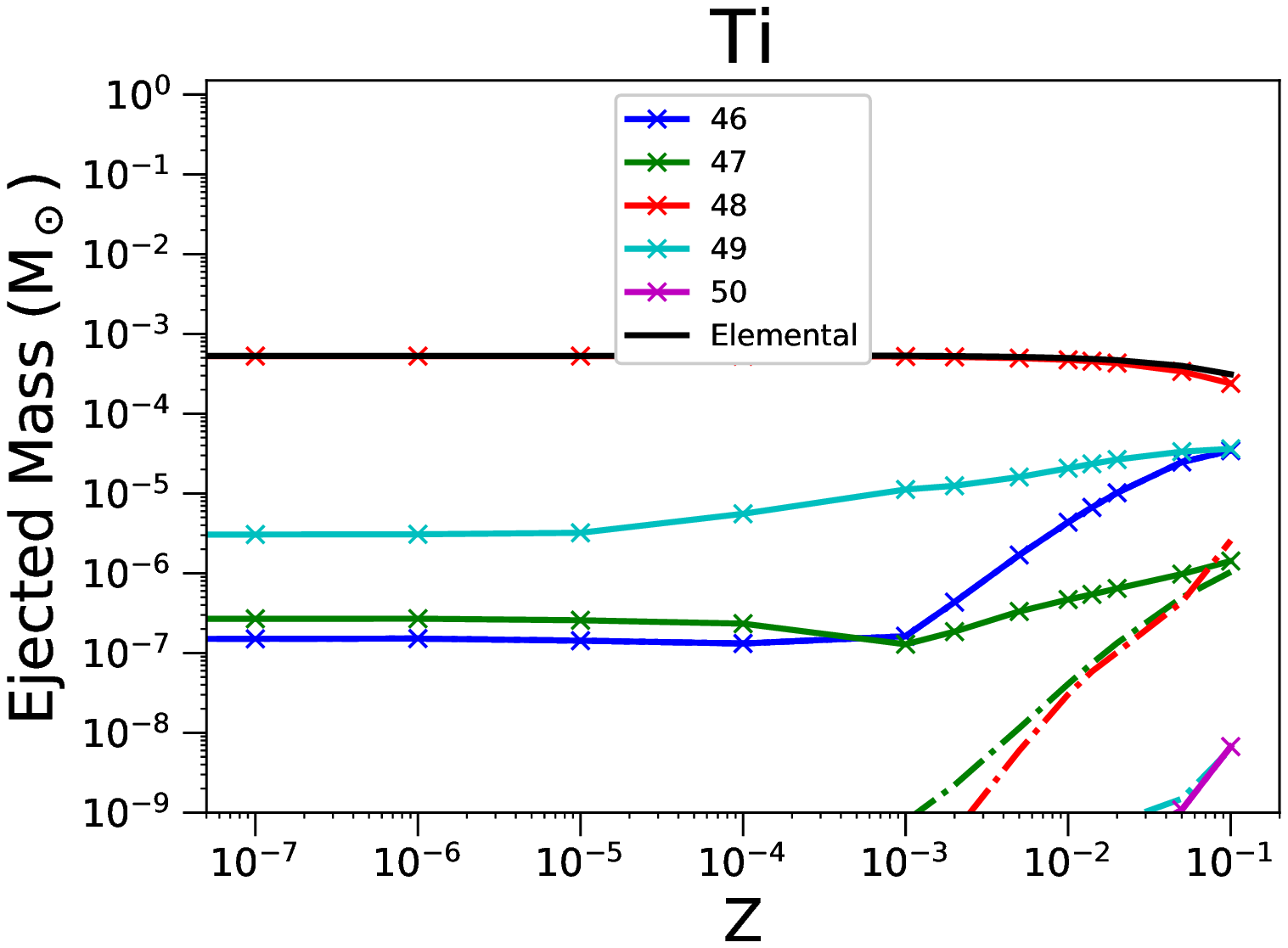}{0.45\textwidth}{S1.0}}
    \gridline{\fig{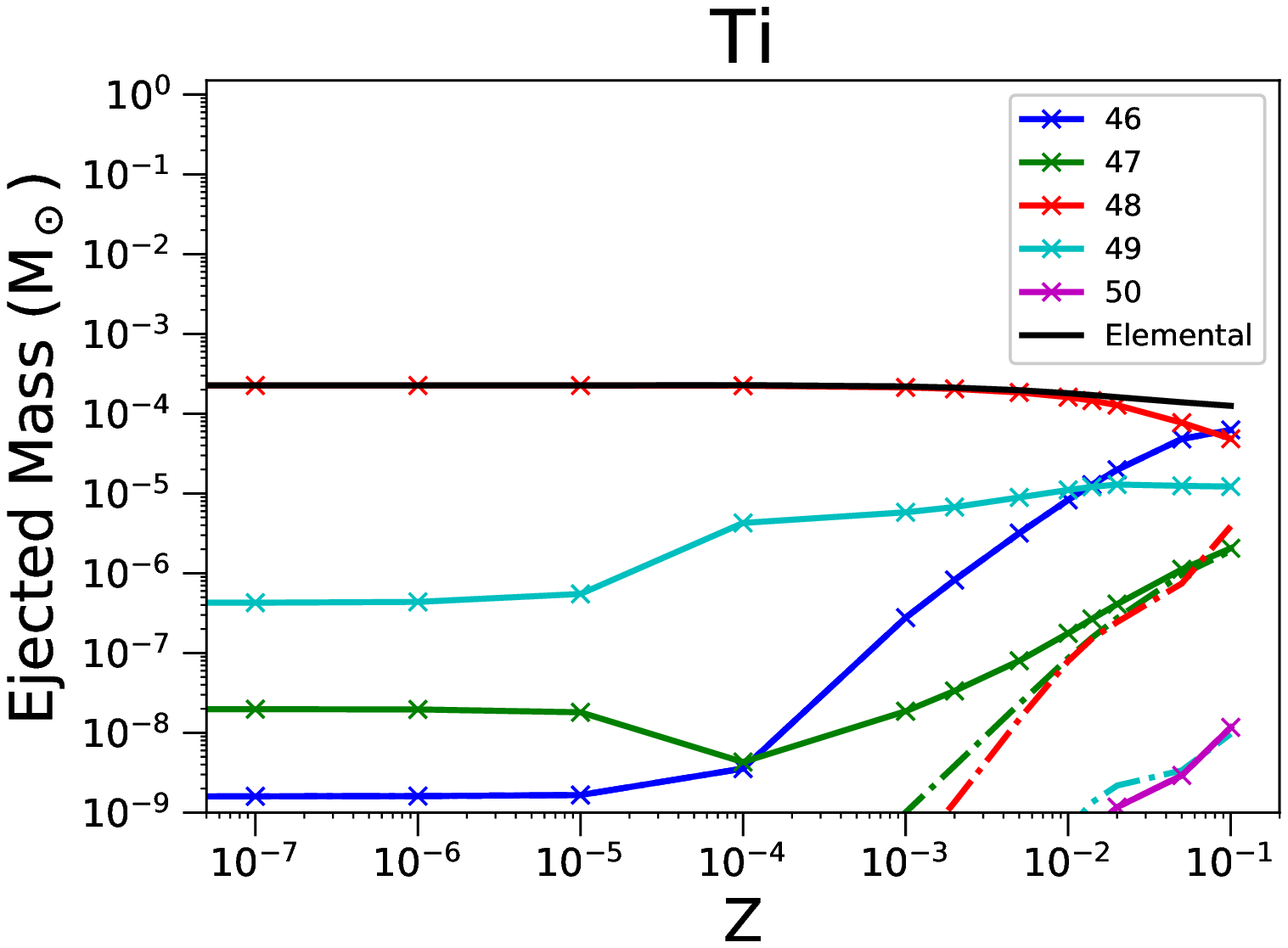}{0.45\textwidth}{M0.8}}
    \caption{As in Figure \ref{fig:Silicon_metalicity_multi} but for ejected mass of Ti, and its stable isotopes $^{46}$Ti, $^{47}$Ti, $^{48}$Ti, $^{49}$Ti and $^{50}$Ti. 
    }
\label{fig:Titanium_metalicity_multi}
\end{figure*}

Titanium has 5 stable isotopes: $^{46}$Ti, $^{47}$Ti, $^{48}$Ti, $^{49}$Ti and $^{50}$Ti. Ti isotopes are expected to be produced by both CCSNe and SNIa. In CCSNe, $^{46}$Ti (8.25\% of the solar budget) is made by explosive oxygen-burning, $^{47}$Ti (7.4\%) by explosive oxygen-burning and explosive silicon burning, $^{48}$Ti (73.7\%) and $^{49}$Ti (5.4\%) by explosive silicon burning. $^{50}$Ti (5.2\%) is mostly produced by the s-process in massive star progenitors \citep{woosley2002evolution}. While Ti has a significant contribution from SNIa, its full chemical inventory in the Galaxy is typically not reproduced by GCE simulations when compared to Fe for instance, making the production of Ti in stars still an open problem for nuclear astrophysics \citep[e.g.,][and references therein]{mishenina:17, prantzos:18, kobayashi2020origin}.

Figure \ref{fig:Titanium_metalicity_multi} presents our yields for these isotopes at different metallicities. $^{48}$Ti is the most abundant isotope of titanium in all three models for almost all metallicities. Overall, The ejected mass of elemental titanium decreases with increasing progenitor metallicity, driven by the $^{48}$Ti nucleosynthesis trend. Only in the M0.8Z0.1 model does $^{46}$Ti become more abundant than $^{48}$Ti. $^{46}$Ti production is primary at Z $<$ 
0.00001 and secondary above this, with the ejected mass of $^{46}$Ti increasing with metallicity. $^{47}$Ti has a strong secondary component in the three classes of model above Z$_{met}$ = 0.001 through direct production of $^{47}$Ti. $^{48}$Ti is flat with metallicity in the T1.4 models until Z $>$ 0.05, where a strong secondary contribution is seen. For the S1.0 and M0.8 models, this contribution starts at Z$_{met}$ = 0.001. The $^{50}$Ti yields are high in the T1.4 models and do not depend on the metallicity of the progenitor, with only trace amounts produced in the S1.0 and M0.8 models at Z $>$ 0.01. 
Therefore, the solar $^{50}$Ti/$^{48}$Ti isotopic ratio could be potentially used as a diagnostic to study the relative contribution of different SNIa progenitors to the solar composition by using GCE simulations.

The $^{46}$Ti production at Z$_{met}$ = 0 is much more effective in T1.4Z0 than in either of the other two models. This is due to $^{46}$Ti being synthesized in intermediate to high explosion temperatures (at 6\,GK or above), which do not exist in the S1.0 and M0.8 models. There is a small peak of production at approximately 4\,GK which is negligible.
In the Z$_{met}$ = 0.014 panels the production of $^{46}$Ti through direct nucleosynthesis is highly dependent on the initial metallicity, with the component at 4\,GK now dominating the production. The high temperature component of the T1.4 models is primary, and does not change with the progenitor composition.

$^{47}$Ti is produced over a broad range of peak temperatures in the T1.4 models. At Z$_{met}$ = 0, production spans the range from explosive oxygen fusion through to NSE, from around 4 GK to 9 GK, with peaks between 5.5 and 7 GK contributing to the majority of the ejecta. Across the whole temperature range 
$^{47}$Ti is mostly ejected as $^{47}$V, with a minor direct production between 8 GK to 9 GK and a negligible $^{47}$Sc production at around 9 GK. For the T1.4Z0.014 model, a direct $^{47}$Ti production peak is formed between 2.3 and 4.2 GK.  
In the ejecta of T1.4Z0.1, the $^{47}$Ti yields have increased further in the lower peak temperature regions, with $^{47}$Ti now accounting a significant fraction of the ejected Ti mass. 
Also in the S1.0 and M0.8 models, the radiogenic $^{47}$V contribution is important for the $^{47}$Ti yields, which are strongly metallicity dependent. In model S1.0Z0 the bulk of production is between 3.8 and 6.1 GK. In models S1.0Z0.014 and M0.8Z0.014 the direct $^{47}$Ti production becomes relevant at low peak temperatures, and it then dominates the isotope yields at Z$_{met}$ = 0.1.

In the T1.4 set, $^{48}$Ti is made over a broad rage of peak temperatures, ranging from about 4 GK to the high density NSE region at temperatures larger than 9 GK. Above 8.2 GK, $^{48}$Ti is made directly. Below 8.2 GK $^{48}$Ti is mostly produced as $^{48}$Cr, with a negligible contribution from $^{48}$V. 
In T1.4Z0.014 and T1.4Z0.1, the isotope yields extend into the temperature range between 2.3 and 3.8 GK. Here $^{48}$Ti is produced directly and it is metallicity dependent.
In the S1.0Z0 model, the $^{48}$Ti in the 3.5 - 6.1 GK temperature region is solely produced as $^{48}$Cr. Similar to the T1.4 and the M0.8 model sets, we obtain a direct production of $^{48}$Ti in the 2.8 - 3.8 GK range in the high metallicity models.

$^{49}$Ti production in T1.4Z0 ranges from 3.8 GK to $>$ 9GK, with three peaks of production between 3.8 and 6 GK, 7 and 8 GK and above 9 GK respectively. The majority of $^{49}$Ti is produced as $^{49}$Cr, particularly below 8 GK, where almost 100\% of the ejecta is radiogenic $^{49}$Cr. Only trace amounts of radiogenic $^{49}$Mn and $^{49}$V are present. Between 8 and 9 GK, there is a small region where production is dominated by contributions from radiogenic $^{49}$V, and above 9 GK we obtain a direct production of $^{49}$Ti. These two components, however, are small compared to the $^{49}$Cr yields. In the T1.4Z0.014 model, we see a significant increase in the 4 - 6 GK temperature region. Peak production in these particles increases by an order of magnitude between these two models. We also see an increase in the contribution from $^{49}$V, between 2.3 and 3.5 GK. 
In the T1.4Z0.1 model, this low temperature contribution is depleted, due to increased neutron captures shifting material to more neutron rich isotopes. We also see a further increase in the 4.5 - 6 GK region, which drives the increasing ejected mass of $^{49}$Ti with initial metallicity. The models of the S1.0 and M0.8 sets show a similar behavior for the two lower temperature components shared with the T1.4 models.

Regarding $^{50}$Ti, in T1.4Z0 the only relevant contribution is from the particles with peak temperature T $>$ 8 GK. As such, neither the S1.0 nor M0.8 model produce $^{50}$Ti at Z$_{met}$ = 0. As the metallicity increases, we see a small production of $^{50}$Ti between 3 and 3.5 GK for all models. This is reproduced in S1.0Z0.014 and M0.8Z0.014, where production is slightly larger than in the T1.4 model. In all cases, however, the $^{50}$Ti yields at lower peak temperatures remains small.


\subsection{Vanadium}

\begin{figure*}
    \gridline{\fig{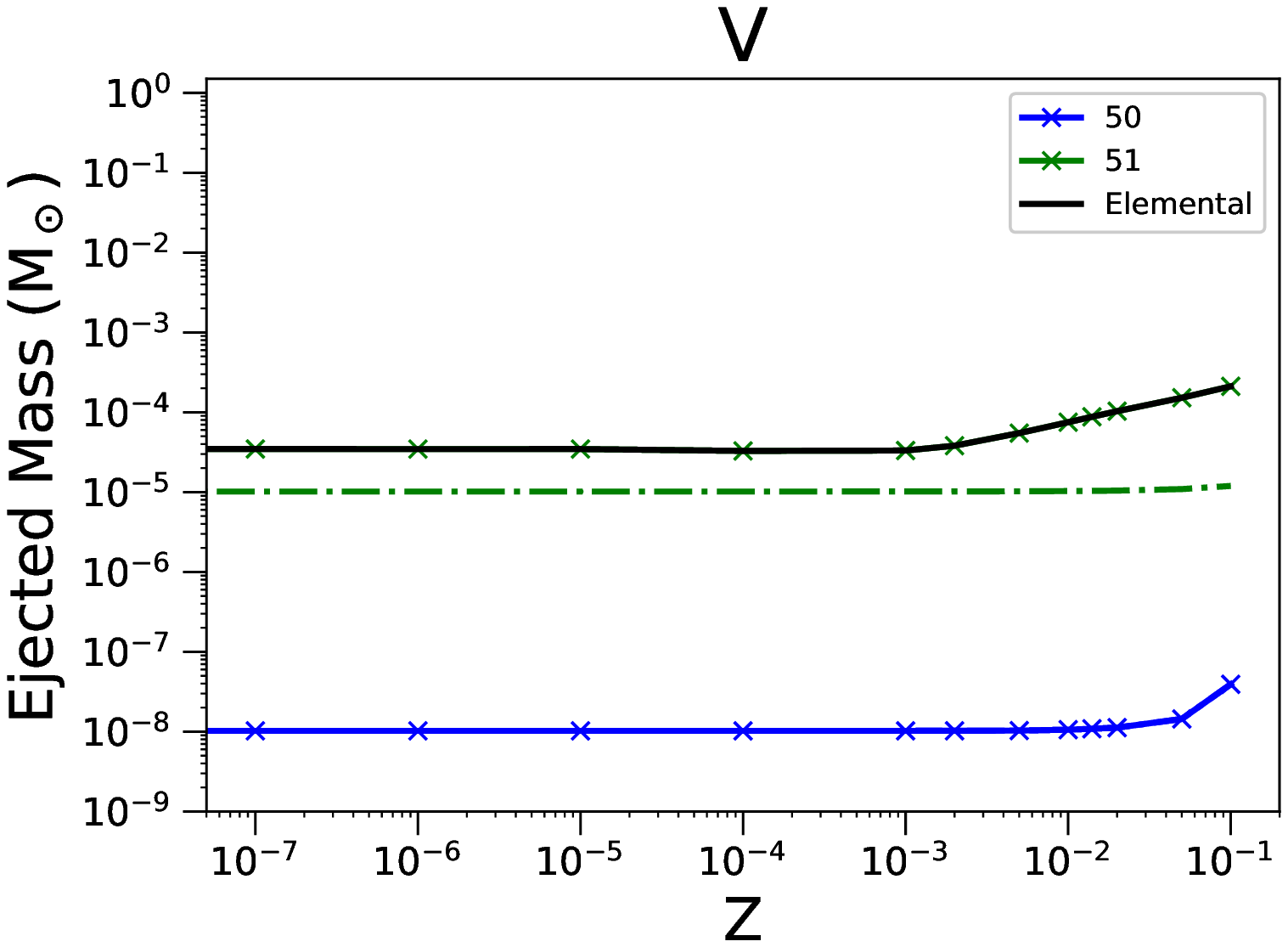}{0.45\textwidth}{T1.4}}
    \gridline{\fig{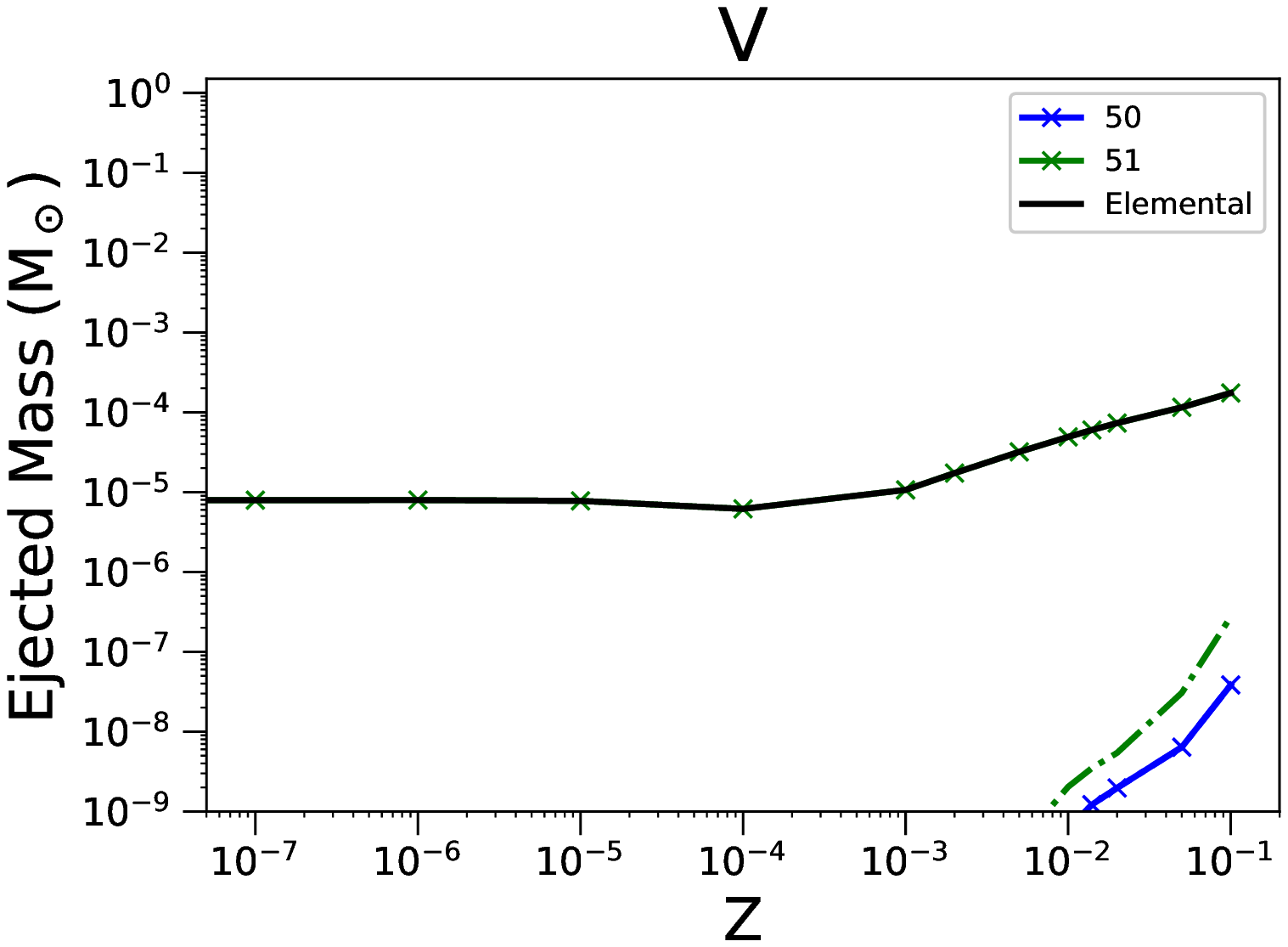}{0.45\textwidth}{S1.0}}
    \gridline{\fig{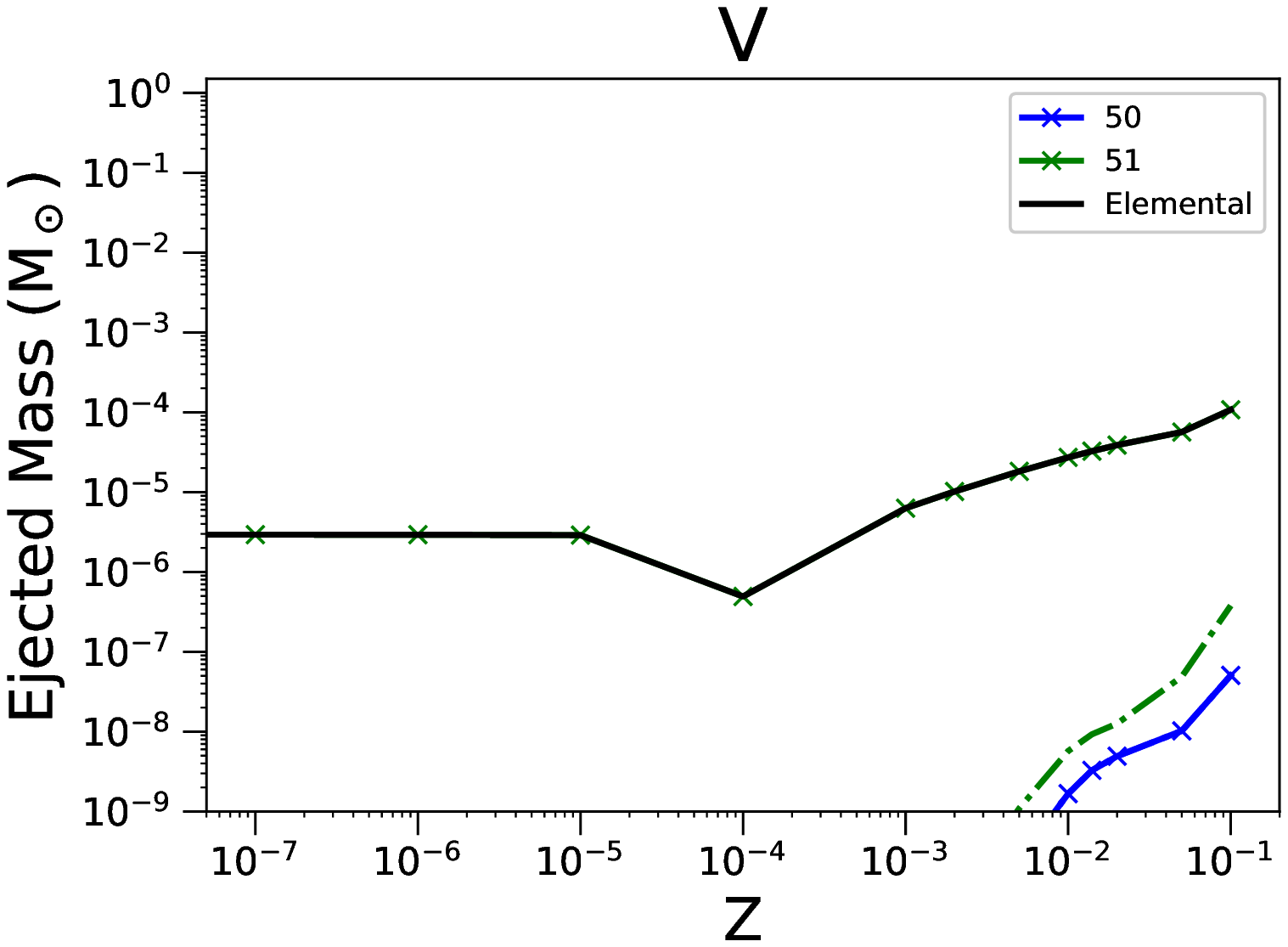}{0.45\textwidth}{M0.8}}
    \caption{As in Figure \ref{fig:Silicon_metalicity_multi} but for ejected mass of V, and its stable isotopes $^{50}$V and $^{51}$V.}
\label{fig:Vanadium_metalicity_multi}
\end{figure*}

We see in Figure \ref{fig:Vanadium_metalicity_multi} that vanadium has a similar trend with metallicity through our three model sets. There is, however, around a factor of 10 difference between the total yields of the T1.4 models compared to the M0.8 yields. We also observe a dip in production for the M0.8 models at Z$_{met}$ = 10$^{-4}$, which is not present in the other models.
Trace amounts of $^{50}$V (0.25\% of the solar budget of V) are produced at all metallicities in the T1.4 models, with a weak secondary component, whereas in the S1.0 and M0.8 models we obtain the same secondary component but without any primary ejecta.

The primary $^{50}$V component (99.75\% of the solar budget) in the T1.4 models is ejected with the hottest particles, with a maximum temperature peak $>$ 8 GK. This is obviously not present in the other sets with smaller SNIa progenitors. The secondary $^{50}$V component arises in the 2.5-4\,GK temperature range, at solar and super-solar metallicity.

Production of $^{51}$V is significantly more complex than $^{50}$V, particularly in the T1.4 models. In model T1.4Z0, the highest peak temperature particles above 9\,GK directly produces $^{51}$V. At lower peak temperatures, from approximately 7.8 to 8.5\,GK, $^{51}$Cr becomes the largest contributor. 
Finally, for temperatures $<$ 7.8 GK, the synthesis of $^{51}$V is mostly dominated by radiogenic $^{51}$Mn. We also see a trace contribution from $^{51}$Fe in this model, but it is small in comparison to the other contributing isotopes. For model T1.4Z0.014 and  T1.4Z0.1, the high temperature contribution to the isotopic abundance of $^{51}$V remains the same. The low temperature contribution from $^{51}$Cr increases, causing an increase in the $^{51}$V yields.
The synthesis of $^{51}$Mn is the main route for production in model S1.0Z0, with a trace contribution from radioactive $^{51}$Cr. This production occurs over a broad range of temperatures, but is much less abundant than in the T1.4 cases.
In models S1.0Z0.014, the production of $^{51}$Cr at lower temperatures begins to contribute significantly to the overall ejected mass of $^{51}$V. This trend increases as we move to model S1.Z0.1, where the contributions from $^{51}$Mn and $^{51}$Cr become comparable. The contribution from $^{51}$Mn occurs above peak temperatures greater than 4.5\,GK and $^{51}$Cr below this.
Production in the M0.8 models follows a similar trend to the S1.0 models. M0.8Z0.1 shows a much larger $^{51}$Cr to $^{51}$Mn production due to the smaller number of particles at higher peak temperatures, and the lower maximum peak temperature in this model.


\subsection{Chromium}

\begin{figure*}
    \gridline{\fig{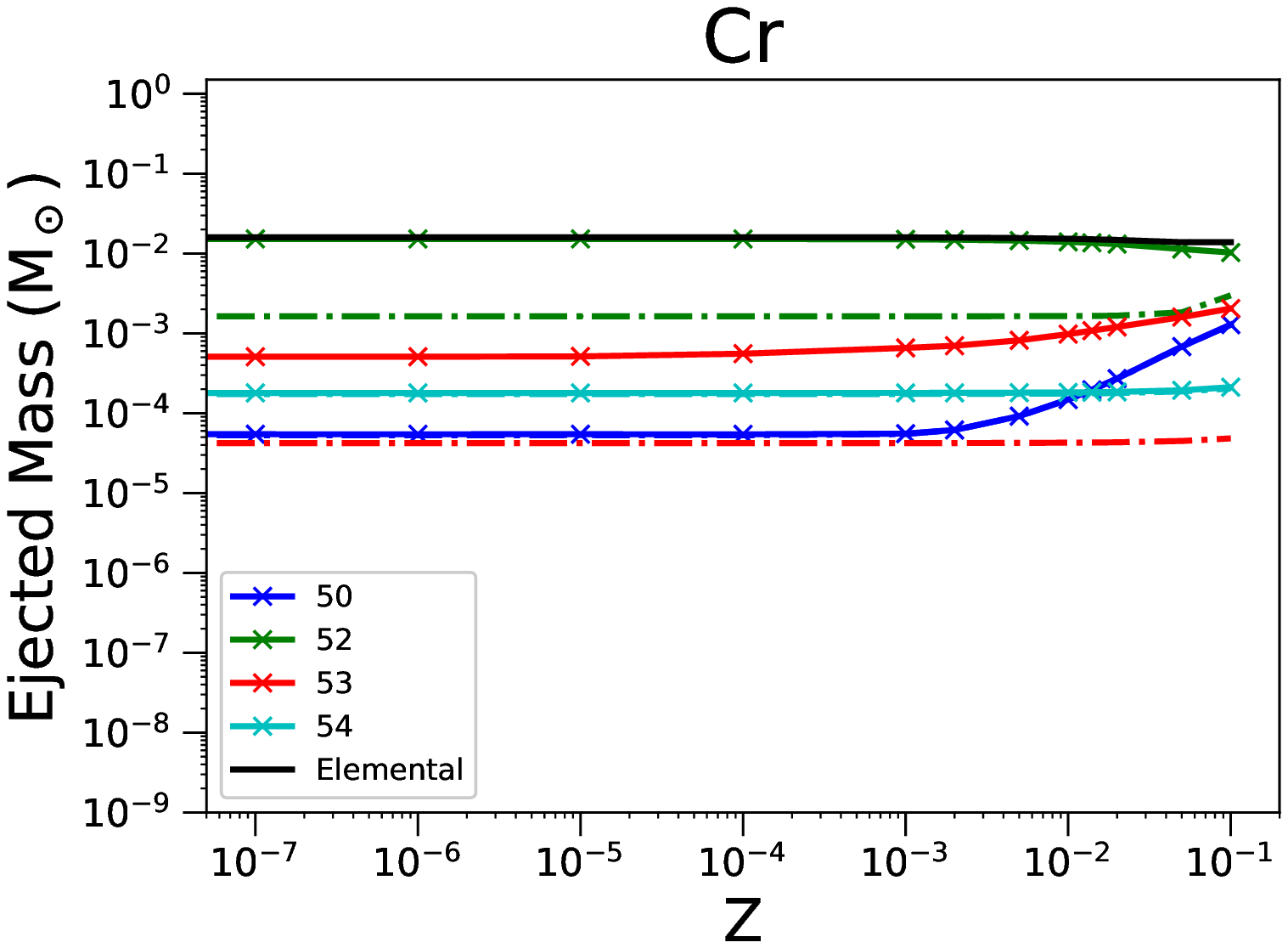}{0.45\textwidth}{T1.4}}
    \gridline{\fig{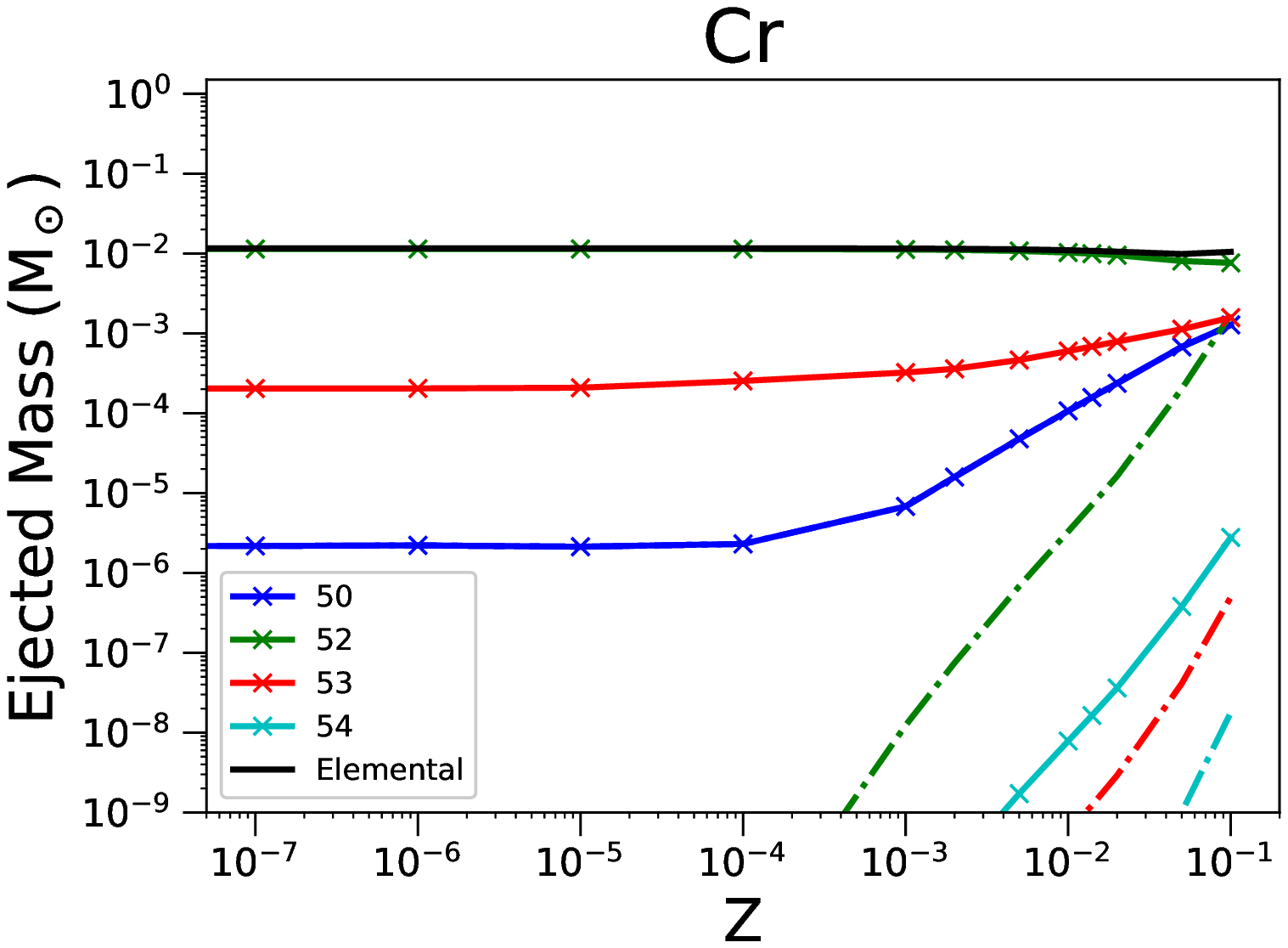}{0.45\textwidth}{S1.0}}
    \gridline{\fig{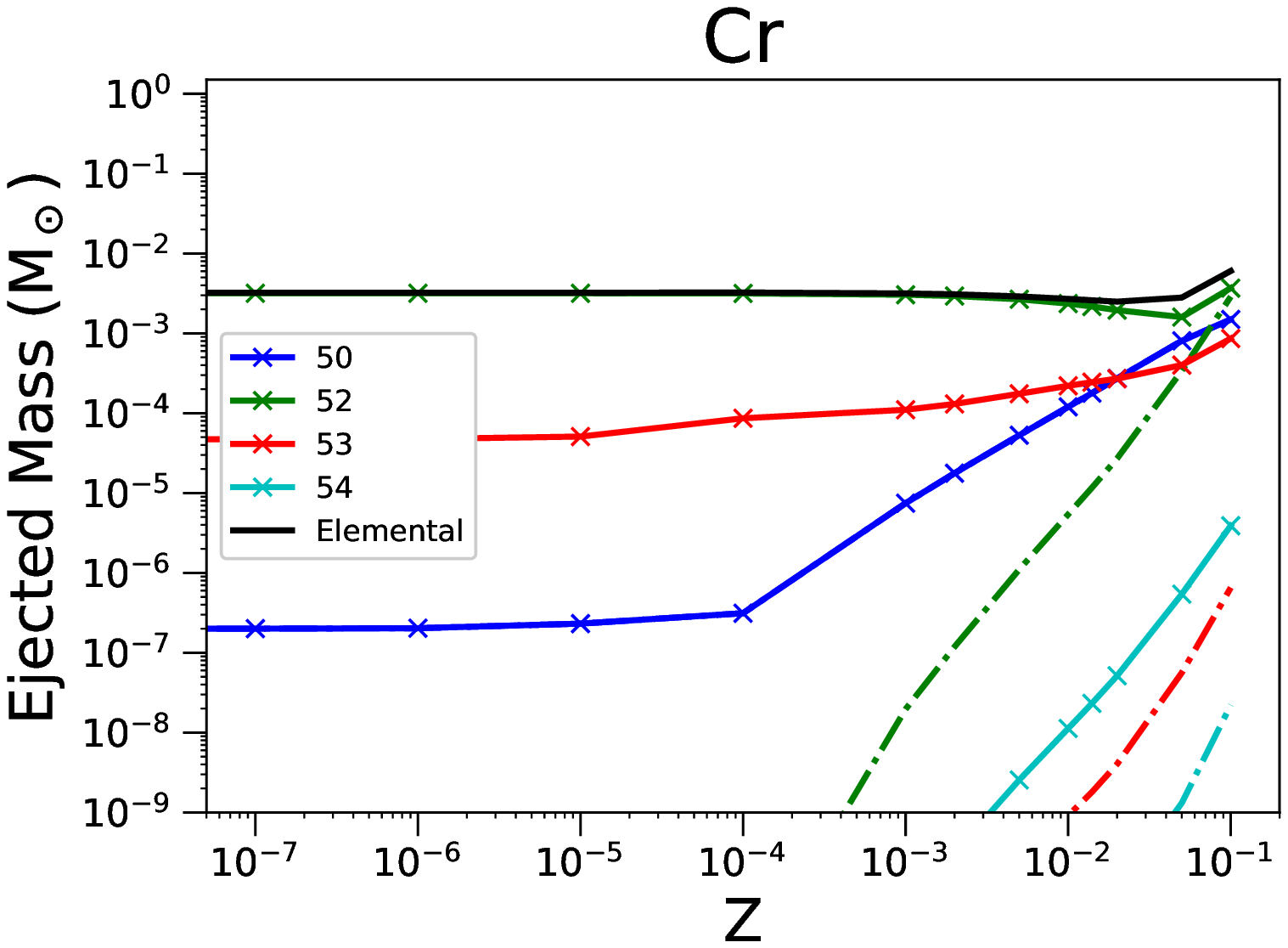}{0.45\textwidth}{M0.8}}
    \caption{As in Figure \ref{fig:Silicon_metalicity_multi} but for ejected mass of Cr, and its stable isotopes $^{50}$Cr, $^{52}$Cr, $^{53}$Cr and $^{54}$Cr.}
\label{fig:Chromium_metalicity_multi}
\end{figure*}

Chromium has 4 stable isotopes: $^{50}$Cr, which has a half-life of 1.3$\times$10$^{18}$ years and we therefore treat as stable. $^{52}$Cr, $^{53}$Cr and $^{54}$Cr. $^{52}$Cr is the most abundant Cr isotope in the Sun (83.789$\%$ of the solar Cr), while $^{54}$Cr has the lowest abundance (2.365$\%$ of the solar Cr) \citep{asplund2009chemical,lodders2003solar}. 

SNIa are major contributors to the solar Cr budget \citep{kobayashi2020origin}. 
In Figure~\ref{fig:Chromium_metalicity_multi}, we show the integrated Cr yields for the element and its isotopes, for the T1.4, S1.0 and M0.8 sets. Cr production is dominated by $^{52}$Cr for all metallicities in the T1.4 and S1.0 sets. At metallicities higher than solar, in the M0.8Z0.1 models $^{50}$Cr and $^{53}$Cr production increases enough as the metallicity of the progenitor increases to affect the Cr abundance, causing an increase of its total ejected mass. Except for this pattern at high metallicities, Cr production appears to be primary in all models following the $^{52}$Cr trend. 
On the other hand, the relative production of different Cr isotopes changes significantly with metallicity and using different model sets.
$^{50}$Cr is the only isotope that is always produced directly. $^{54}$Cr is mostly ejected as itself in T1.4 models, while for other sets it is mostly radiogenic.  $^{52}$Cr and $^{53}$Cr are produced by radiogenic contribution in all the explosion models. 
For all sets the yield of $^{50}$Cr rises with increasing metallicity of the WD progenitor above a threshold value of Z$_{met}$ = 10$^{-4}$. $^{53}$Cr shows a much weaker rise, and $^{54}$Cr is mostly primary for T1.4 models while its (much weaker) production is secondary for other sets.

In the T1.4 set, the $^{50}$Cr yields are dominated by two production peaks, at 7-9 \,GK and 3-5 \,GK temperature, respectively. The first peak is primary, while the low temperature peak is secondary, rising with increasing metallicity. Only the T1.4Z0 model shows a relevant radiogenic contribution by $^{50}$Mn at lower temperatures, but the production is dominated by the first peak.
S1.0 and M0.8 sets only carry the secondary peak, explaining their smaller $^{50}$Cr production at low metallicities compared to T1.4. On the other hand, for all sets the strong secondary component in $^{50}$Cr causes this isotope to form up to about 10\% of all Cr ejecta (M0.8Z0.1 model).
The $^{52}$Cr abundance pattern is significantly more complex than $^{50}$Cr, revealing a larger number of production channels. For the T1.4 set, primary $^{52}$Cr is directly produced in the hottest ejecta, with temperatures above 8 \,GK. Two additional peaks, from the eventual decay of $^{52}$Fe occur at around 7\,GK and 5\,GK. These are insensitive to the initial metallicity of the model, and contribute the majority of $^{52}$Cr. Finally, we obtain a secondary production of $^{52}$Cr at about 4\,GK in the T1.4Z0.014 and T1.4Z0.1 models. Small amounts of $^{52}$Mn are also produced in all models.
The S1.0 and M0.8 models all show the primary $^{52}$Fe production, and the secondary $^{52}$Cr channel at lower temperatures.
The $^{53}$Cr isotope is produced directly above 9\,GK in the innermost ejecta for the T1.4 set. For 8-9 \,GK the production is dominated by a primary peak of radiogenic $^{53}$Mn. Two additional $^{53}$Mn channels appear at lower temperatures, with the colder production peak in oxygen burning conditions becoming particularly relevant at super-solar metallicities.
Two broad primary peaks at 7-8 \,GK and about 5 \,GK are due to the radiogenic contribution from $^{53}$Fe.
$^{53}$Fe continues to be the primary radiogenic source of $^{53}$Cr in the S1.0 and M0.8 sets.
In agreement with T1.4 models, a secondary production of $^{53}$Mn and $^{53}$Cr is obtained at 3-4 GK.
Finally, $^{54}$Cr is produced directly at temperatures above 8 \,GK, while a secondary production peak is present at about 4 \,GK from the radiogenic decay of $^{54}$Mn.


\subsection{Manganese}

\begin{figure*}
    \gridline{\fig{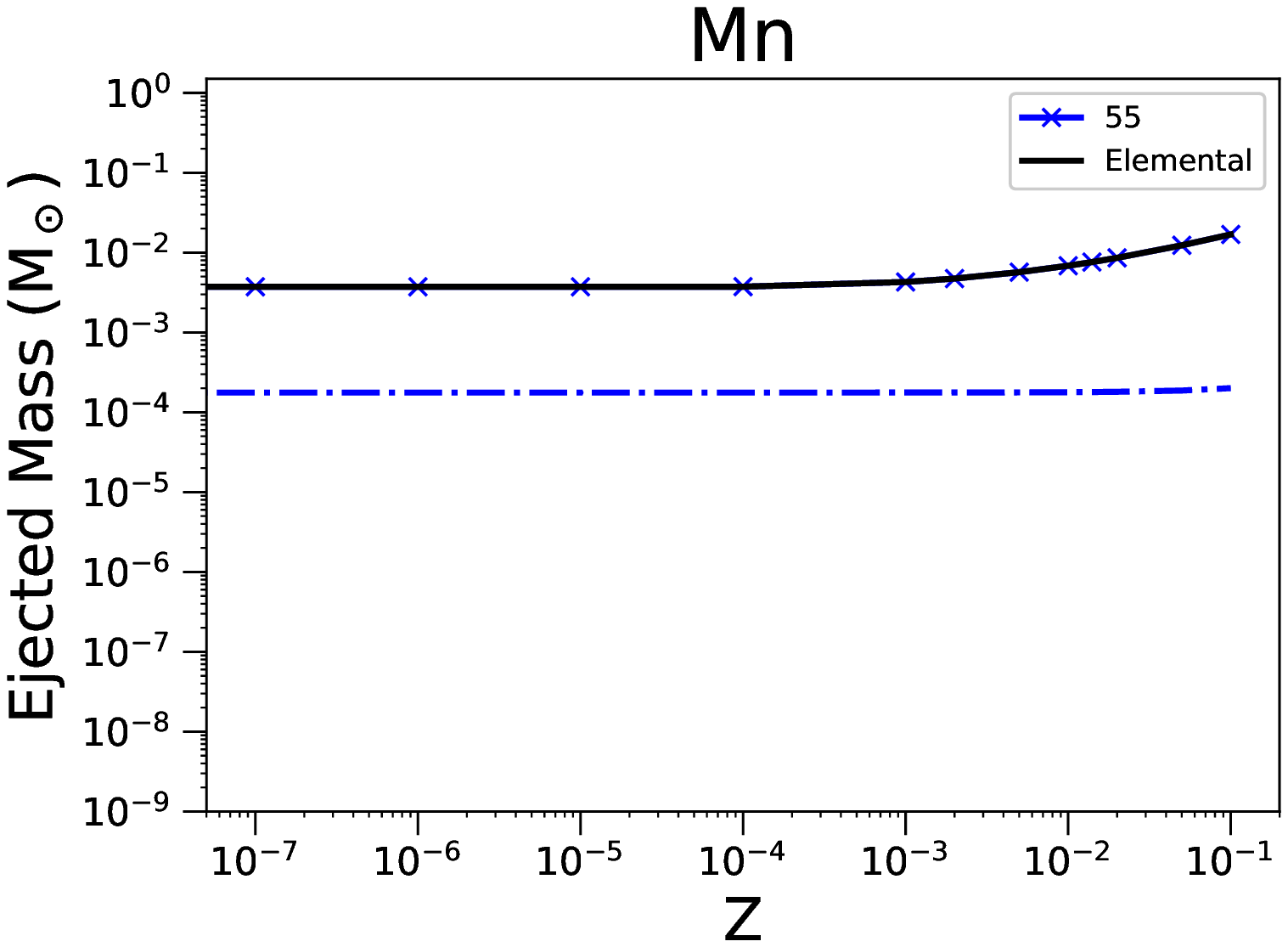}{0.45\textwidth}{T1.4}}
    \gridline{\fig{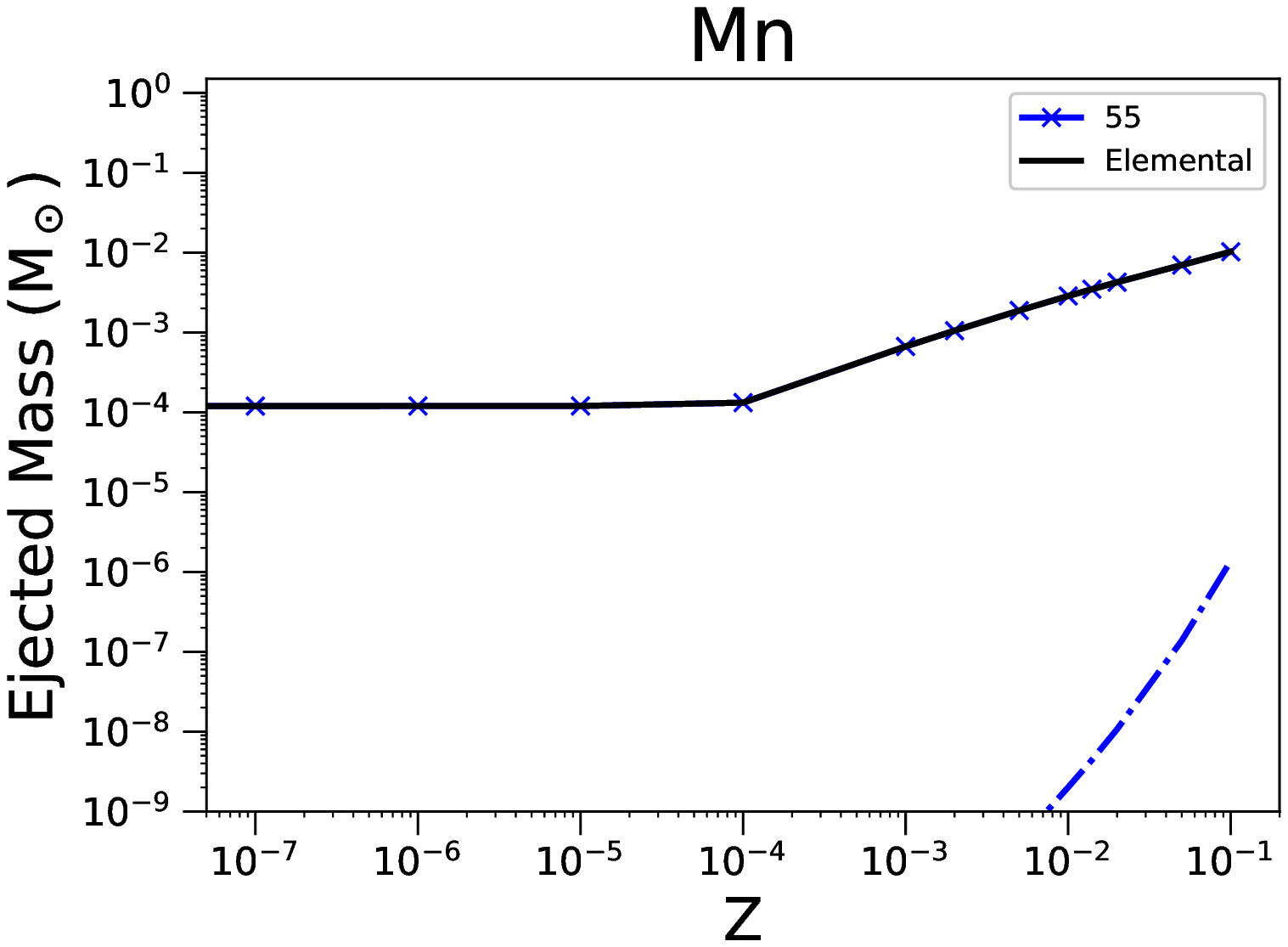}{0.45\textwidth}{S1.0}}
    \gridline{\fig{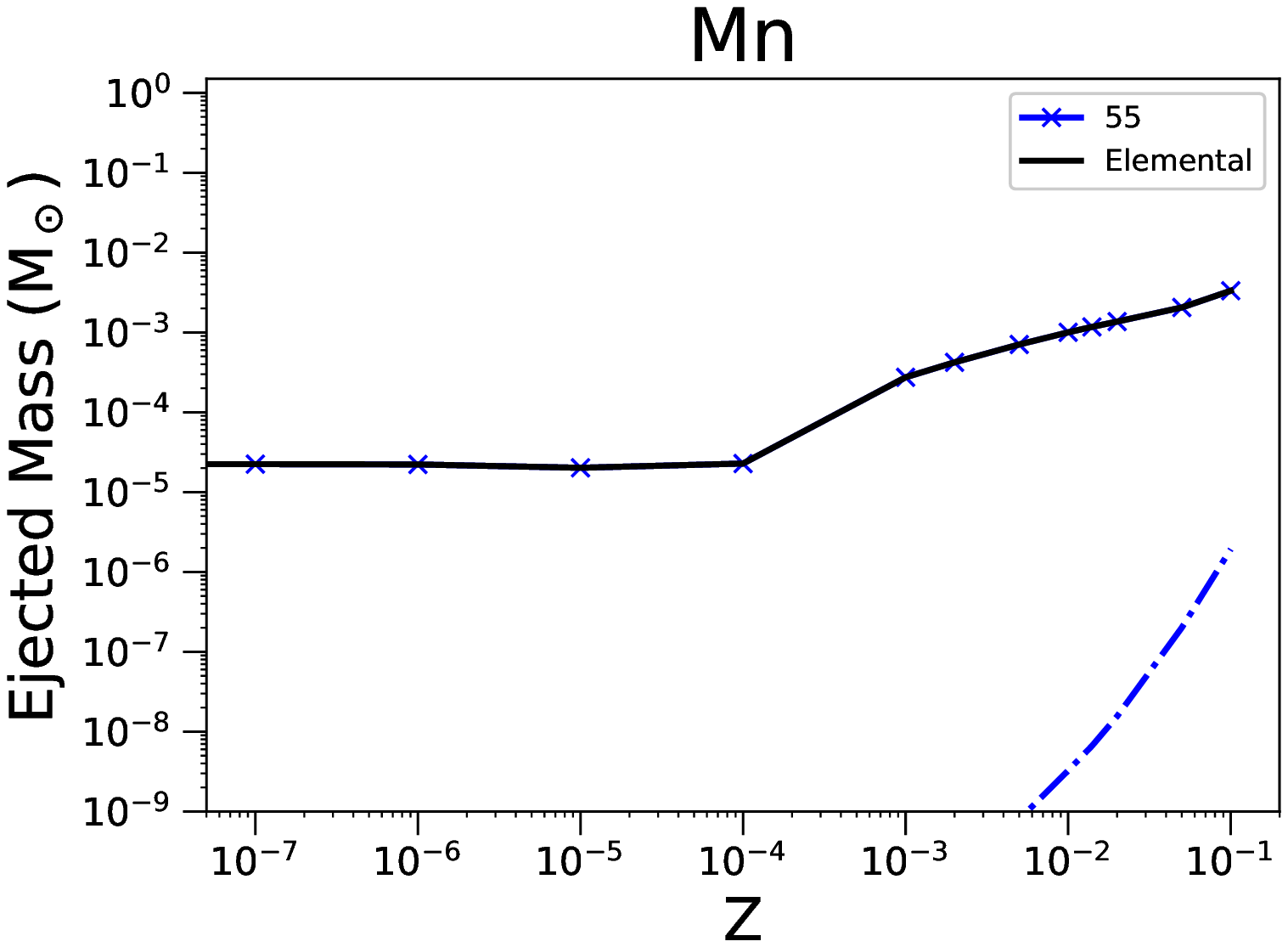}{0.45\textwidth}{M0.8}}
    \caption{As in Figure \ref{fig:Silicon_metalicity_multi} but for ejected mass of $^{55}$Mn.}
\label{fig:Manganese_metalicity_multi}
\end{figure*}

Mn is mono-isotopic, with only $^{55}$Mn being stable. In CCSNe it is mainly formed in explosive silicon burning, with some possible contribution from the neutrino-winds component \citep{woosley2002evolution}. The production of Mn with respect to Fe in SNIa has been the subject of a number of GCE investigations \citep{kobayashi2006galactic, seitenzahl2013solar, kobayashi2020origin, eitner2020observational} as both Mn and Fe in the Milky Way disk and in the Sun are produced primarily by SNIa \citep{matteucci2009effect,de2020manganese}.
Mn has also been proposed as a key tracer for differentiating the relative contribution from SNIa progenitors of different mass \citep[e.g.,][]{seitenzahl2013solar, eitner2020observational}. \citet{kobayashi2020new} identify a positive correlation for the production of Mn with respect to initial metallicity, similar to the results presented here. They, however, see a decrease in the ejected mass of Mn with increasing progenitor mass, at odds with the findings of \citet{seitenzahl2013three} and this work.
Finally, based on GCE simulations, SNIax were discussed as a possible important source to the galactic Mn abundance \citep{kobayashi2020new,cescutti2017manganese}, although this is still a matter of debate \citep{ eitner2022observational}. SNIax are a subclass of SNIa, which are spectroscopically similar to SNIa explosions but with lower ejecta velocities and lower peak magnitudes as described in \citet{foley2013type}. 
Figure \ref{fig:Manganese_metalicity_multi} shows the ejected mass of Mn from our three models with respect to metallicity, indicating a large radiogenic contribution across the range of Z investigated. This result is expected and consistent with previous results in the literature. For instance, \citet{truran1967nucleosynthesis} already identifies Mn as being mainly produced as $^{55}$Co.

In the low metallicity sub-Chandrasekhar mass models, most production of Mn is as $^{55}$Co, with a negligible contribution from $^{55}$Fe. This contribution increases with increasing initial metallicity as the secondary $^{55}$Fe component begins to dominate at lower temperatures (between 3 and 4 GK).
The production of $^{55}$Co is also boosted by this secondary component in both the S1.0 and M0.8 models, increasing the overall production of $^{55}$Mn. In the T1.4 models, we see that the production of $^{55}$Fe is again  boosted at higher metallicities, with a small peak of $^{55}$Mn contributing at lower temperatures in these models although this peak has a negligible impact on the total production of Mn. We see that the trace amount of $^{55}$Ni produced at intermediate temperatures in this model is suppressed with increasing initial $^{22}$Ne abundance while $^{55}$Fe production is boosted throughout the model. $^{55}$Cr production is comparable to the radiogenic $^{55}$Fe contribution, in contrast with the lower mass models where it is the dominant radiogenic contribution. 

\citet{seitenzahl2013solar} also finds that $^{55}$Co is the largest contributor to manganese yields at densities above 2$\times$10$^{8}$ g cm$^{-3}$. In the T1.4 models, these densities cover the full range of peak temperatures (although not all of the particles have a density above this threshold density). In particular, $^{55}$Co is the dominant source of $^{55}$Mn in the T1.4Z0 model, and remains the largest contributor to ejected $^{55}$Mn mass. The relative contribution from $^{55}$Fe in the T1.4 models increases with increasing metallicity and at Z$_{met}$ = 0.1 contributes on the order of 10\% of the ejected $^{55}$Mn mass. Finally, there is also a contribution from direct synthesis of $^{55}$Mn in the highest temperature regions, however this is small - of the order  1\% of total ejected mass of $^{55}$Mn in the T1.4Z0 model.
\citet{eitner2020observational} consider the fraction of sub-Chandrasekhar mass SNIa to be as high as 75\%, due to the large [Mn/Fe] ratios possible in these low mass models.


\subsection{Iron}
\label{ssec:Iron_yields}

\begin{figure*}
    \gridline{\fig{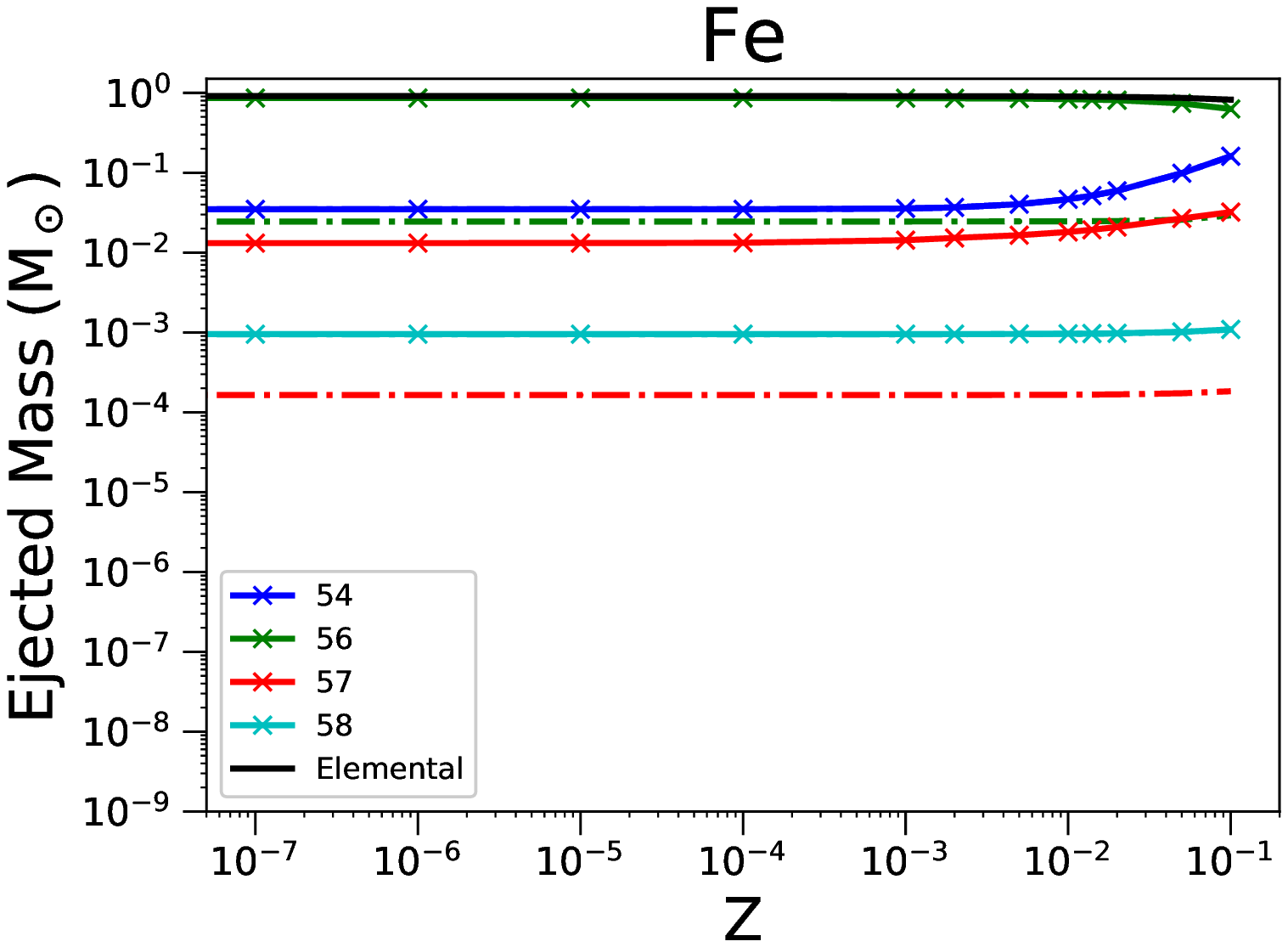}{0.45\textwidth}{T1.4}}
    \gridline{\fig{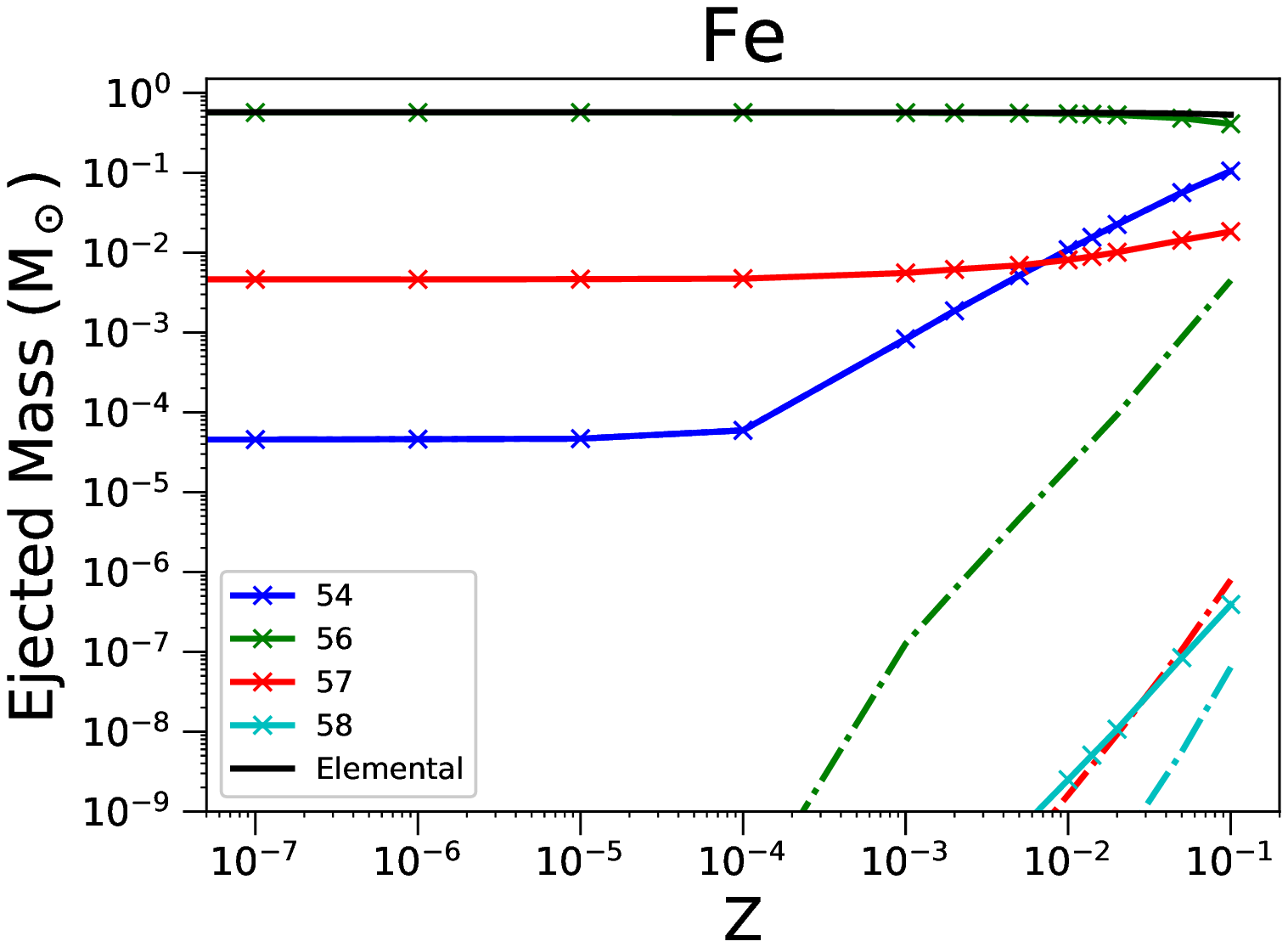}{0.45\textwidth}{S1.0}}
    \gridline{\fig{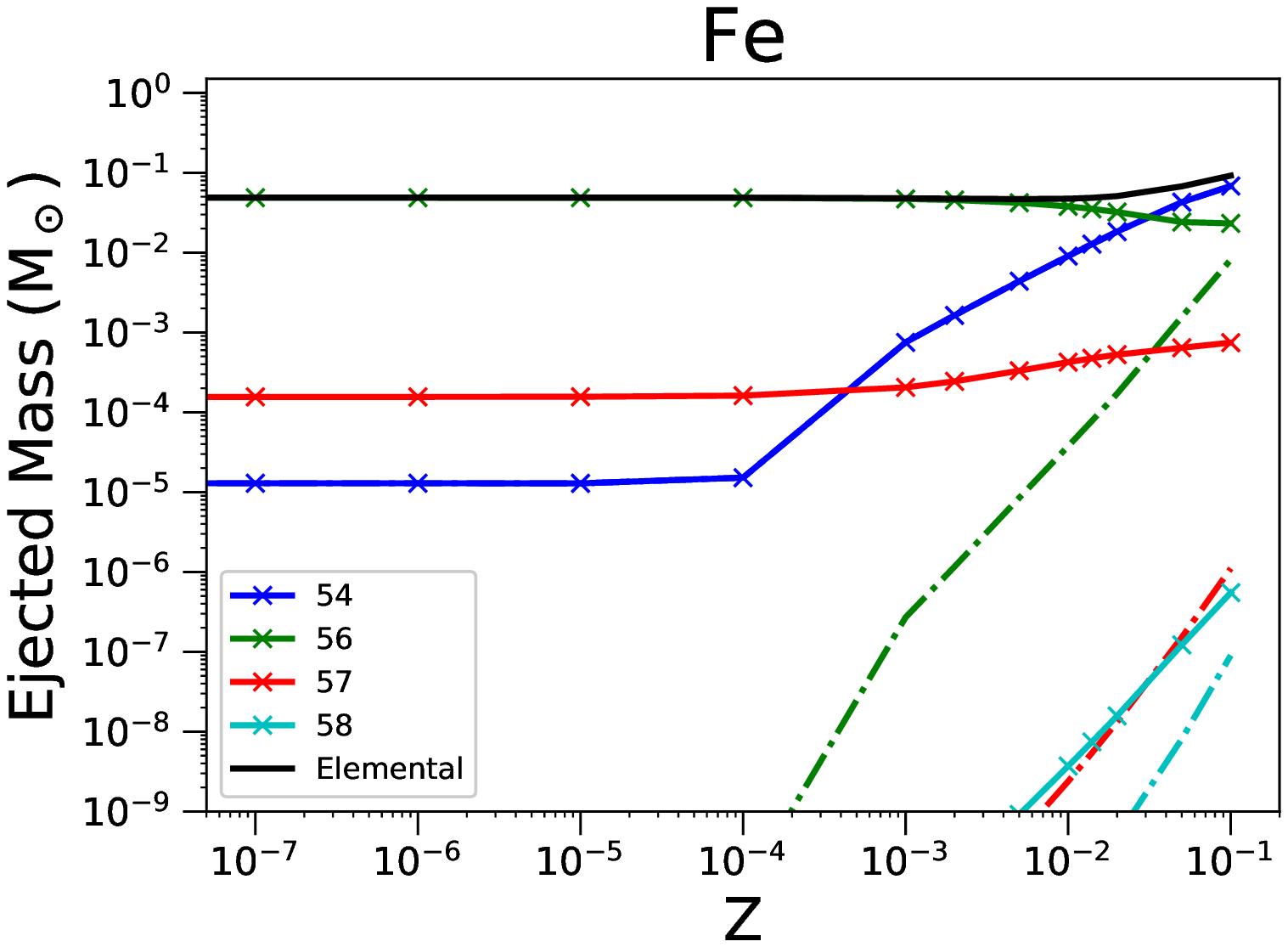}{0.45\textwidth}{M0.8}}
    \caption{As in Figure \ref{fig:Silicon_metalicity_multi} but for ejected mass of Fe, and its stable isotopes $^{54}$Fe, $^{56}$Fe, $^{57}$Fe and $^{58}$Fe.}
\label{fig:Iron_metalicity_multi}
\end{figure*}

Iron has 4 stable isotopes: $^{54}$Fe (5.8\% of the solar budget), $^{56}$Fe (91.75\%), $^{57}$Fe (2.1\%) and $^{58}$Fe (0.3\%). In CCSNe, $^{54,56,57}$Fe are all produced in explosive silicon burning, and $^{58}$Fe is produced efficiently by the s-process during the evolution of the stellar progenitor \citep{woosley2002evolution}. Around 70\% of the solar Fe is produced, however, by SNIa explosions \citep{matteucci1986relative}. 

In Figure \ref{fig:Iron_metalicity_multi}, the ejected mass of the stable Fe isotopes is given, along with their radiogenic contributions, as a function of initial metallicity. There is a small decrease in the abundance of elemental Fe ejected from the T1.4 and S1.0 models due to a reduction in the ejected mass of $^{56}$Fe with increasing initial metallicity. This is partially compensated for by an increase in the abundances of $^{54}$Fe in these two models, which occurs sharply at Z$_{met}$ = 0.0001 in the S1.0 and M0.8 models, and more gradually at super-solar metallicities in the T1.4 model. The increase of over two orders of magnitude in the abundance of $^{54}$Fe becomes a significant contributor to the overall abundance of Fe in T1.4 models at super-solar metallicities. $^{54}$Fe is produced directly in all models. 

$^{56}$Fe is one of the largest components of the ejecta of all the models. While M0.8 has a smaller contribution than the others, most metallicities still have mass fractions of $^{56}$Fe above 10\%. In T1.4 the production is dominated by the $^{56}$Ni yields, with a non-negligible contribution from the direct synthesis of $^{56}$Fe at higher metallicities. This is offset by the reduced $^{56}$Ni production at intermediate peak temperatures from the Z$_{met}$ = 0.014 and Z$_{met}$ = 0.1 metallicity models. Notice also that $^{56}$Fe is produced directly at temperatures above 8\,GK, which component is insensitive to metallicity effects. A more pronounced effect on the overall abundance of $^{56}$Fe with metallicity is obtained in the 0.8M$_{\odot}$ model. However, the iron production actually increases for M0.8 at higher metallicities, due to a significant contribution from $^{54}$Fe.
$^{57}$Fe has a large radiogenic contribution across all models. The $^{57}$Fe amount produced directly by the explosion accounts for ~1\% of the total mass of the $^{57}$Fe yields in the T1.4 models, and it is negligible in the S1.0 and M0.8 sets, where almost all production is through radiogenic contributions. The radioactive isotopes $^{57}$Ni and $^{57}$Co are the main contributors to the $^{57}$Fe yields, particularly in the lower mass models. There is a small contribution to $^{57}$Fe abundance from $^{57}$Mn and $^{57}$Cu in T1.4, but it is negligible. 
Primary production of $^{57}$Fe occurs in models at higher metallicity, in low explosion temperature peaks. This constitutes a small fraction of the total ejected mass of $^{57}$Fe. $^{57}$Co production is boosted at higher metallicities, driving the increasing trend in all models in Figure \ref{fig:Iron_metalicity_multi}.
Finally, not shown in the figures, we have $^{58}$Fe, where only trace abundances are produced in the sub-Chandrasekhar models. Its production in T1.4 models is consistent over a range of metallicities with an ejected mass in the order of 0.001 M$_{\odot}$. At lower temperatures, $^{58}$Fe nucleosynthesis is very similar between our three sets of model as the initial metallicity increases, with radiogenic $^{58}$Co contributing to most of $^{58}$Fe ejecta. At the highest temperature peaks in the T1.4 models (greater than 8.5\,GK) most of $^{58}$Fe is produced directly.

\begin{figure*}
    \gridline{\fig{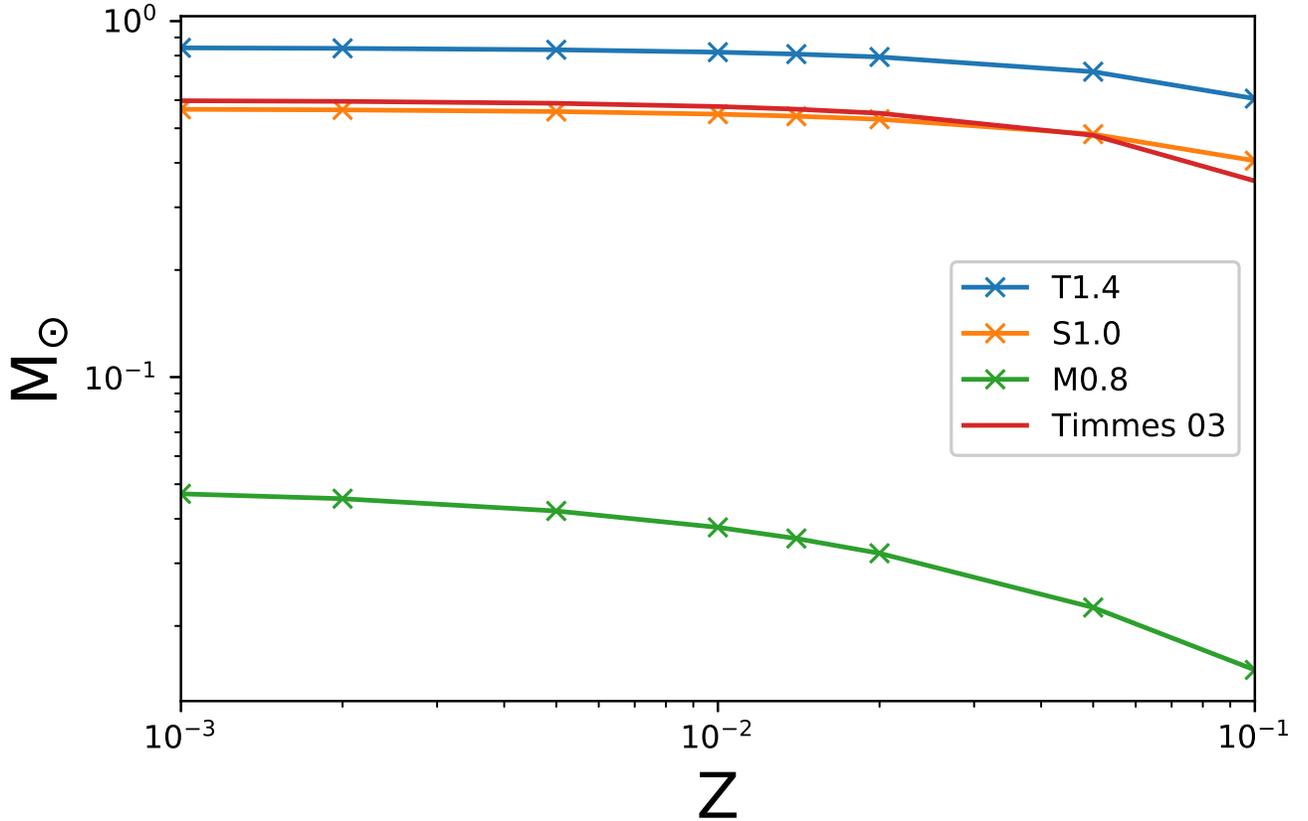}{1\textwidth}{}}
    \caption{Ejected mass of $^{56}$Ni as a function of initial metallicity for our models, compared with \cite{timmes2003variations} calculations.}
\label{fig:ni_56_vs_met}
\end{figure*}

\citet{timmes2003variations} investigated the impact of the initial metallicity of the stellar progenitor on the Fe production in SNIa, and find that by varying the metallicity from 0.1 to 10 times the solar value, the Fe production decreases by 25\%. In this work, models T1.4 and S1.0 show similar Ni trends with metallicity (Figure \ref{fig:ni_56_vs_met}), which is the main component of ejected Fe and is in agreement with the results of \citet{timmes2003variations}. The M0.8 model however does not show the same linear dependence as the higher mass models. We instead see a much sharper decrease in the ejected $^{56}$Ni mass, down to around 30\% of the initial ejected mass at Z$_{met}$ = 0.


\subsection{Cobalt}

\begin{figure*}
    \gridline{\fig{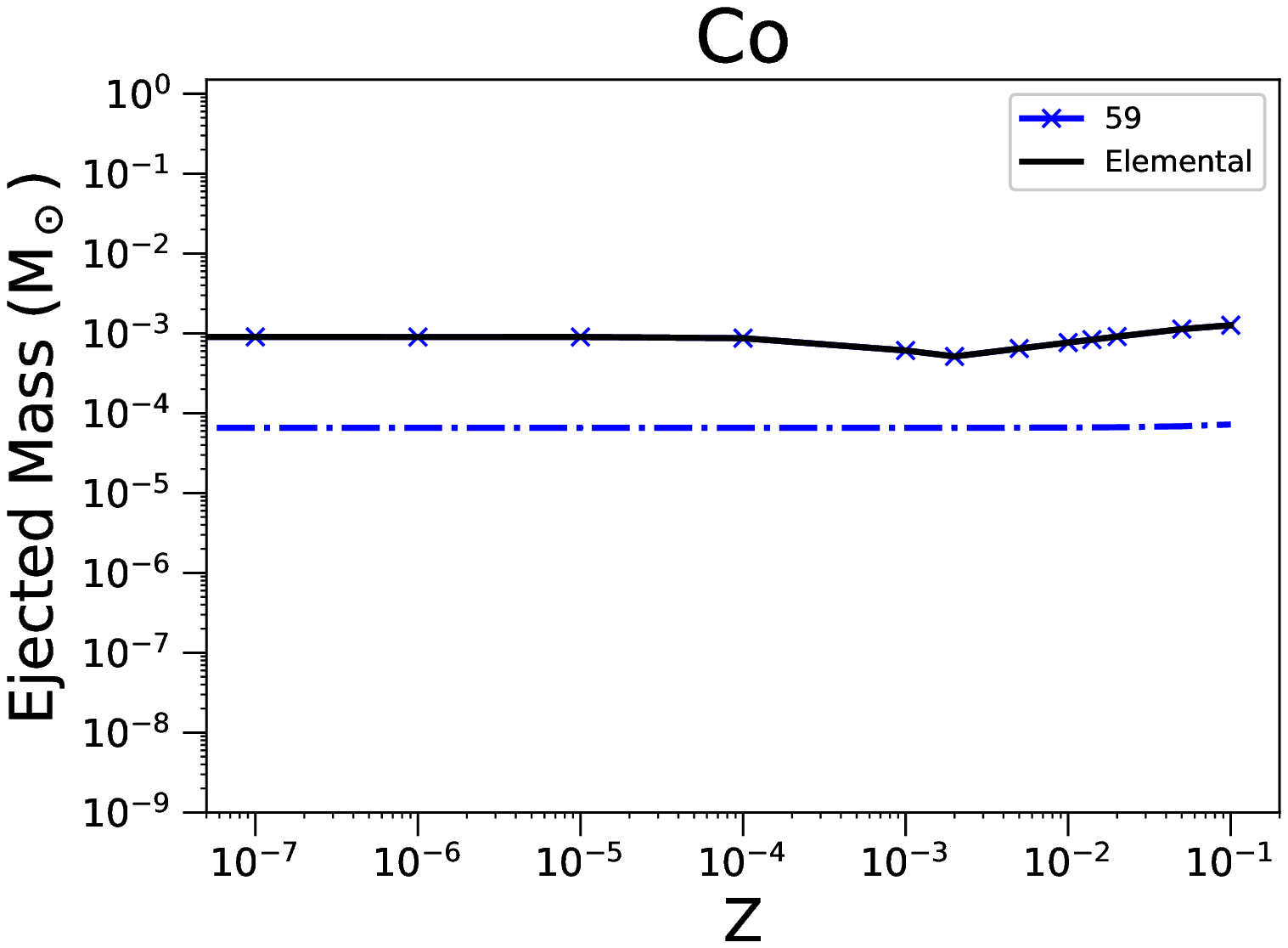}{0.45\textwidth}{T1.4}}
    \gridline{\fig{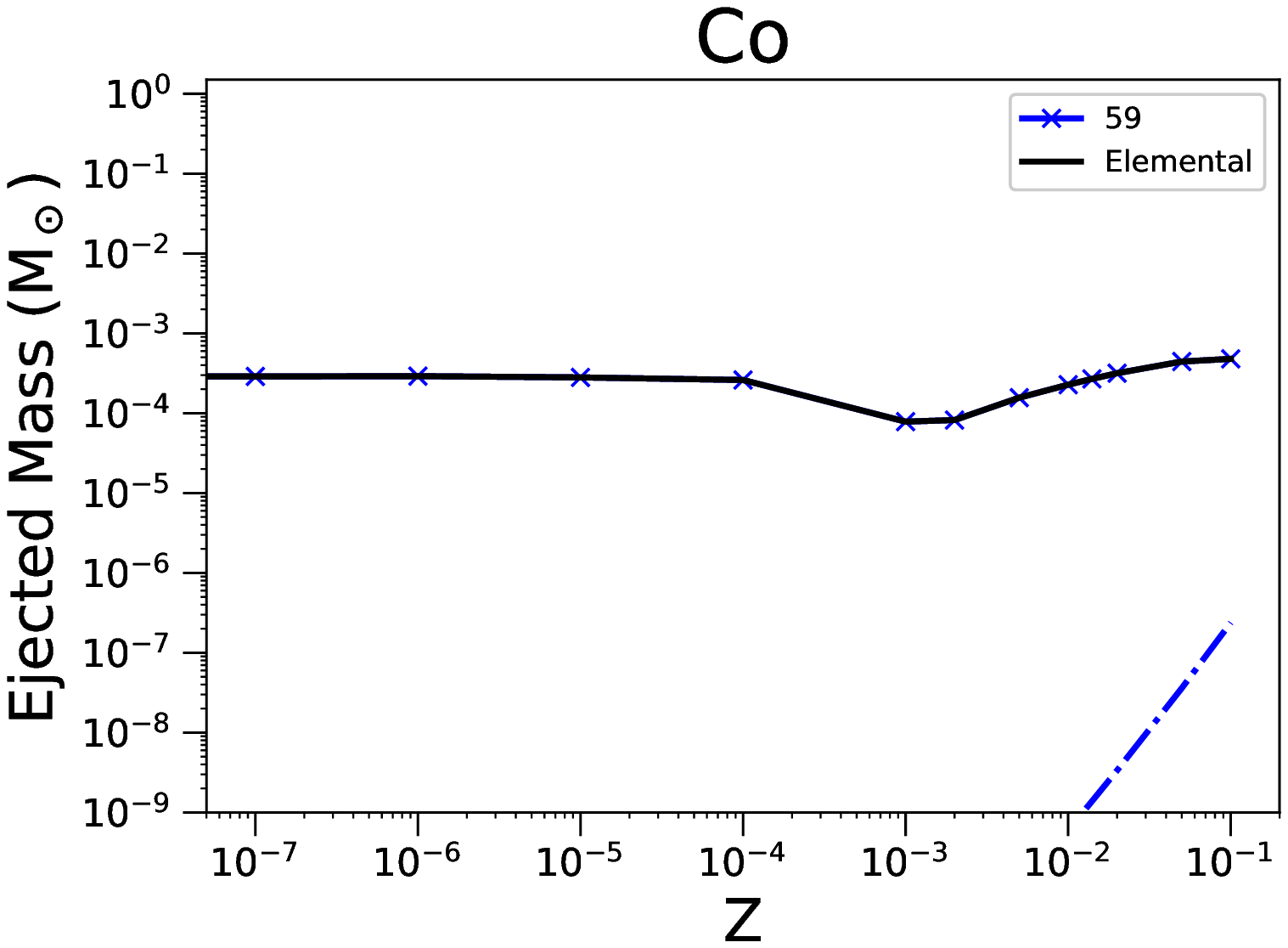}{0.45\textwidth}{S1.0}}
    \gridline{\fig{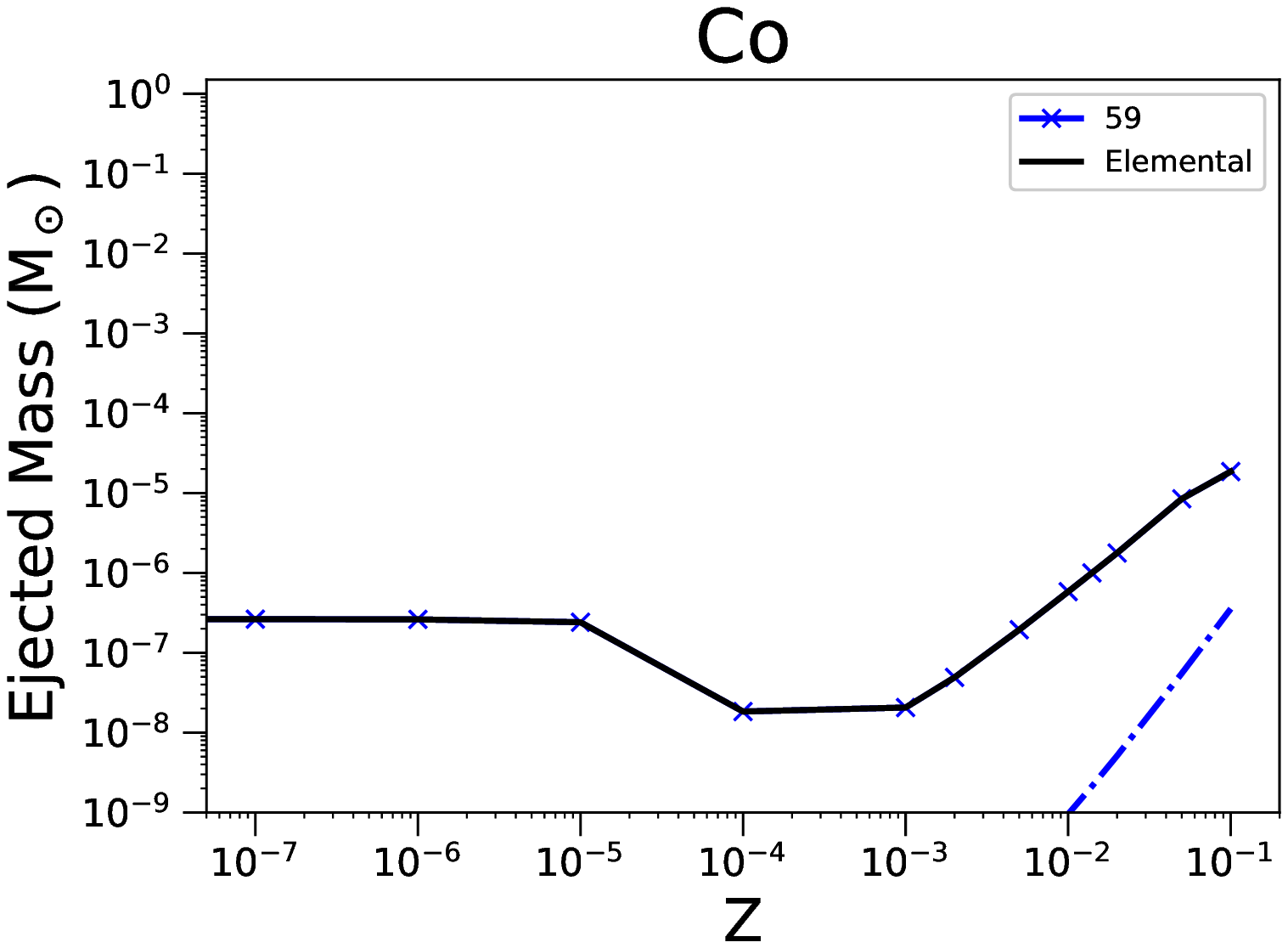}{0.45\textwidth}{M0.8}}
    \caption{As in Figure \ref{fig:Silicon_metalicity_multi} but for ejected mass of $^{59}$Co.}
\label{fig:Cobalt_metalicity_multi}
\end{figure*}

$^{59}$Co is the only stable isotope of cobalt. In CCSNe it can be produced in a variety of stellar conditions - by the s-process, the $\alpha$-rich freeze-out, and by neutrino-winds \citep{woosley2002evolution}. Additionally, \citet{kobayashi2020origin} highlights that Co also has a small contribution from SNIa events. In all of the SNIa models shown in Figure \ref{fig:Cobalt_metalicity_multi}, $^{59}$Co has significant contributions from radiogenic sources. We see that for the case of S1.0 and M0.8, there is a strong secondary effect in both the direct and radiogenic synthesis of $^{59}$Co, starting at Z$_{met}$ = 0.001. In the models T1.4, there is no strong dependency on metallicity. The production of cobalt shows an increasing efficiency with an increase in the mass of the SNIa progenitor. M0.8 models show the strongest metallicity dependence, with an increase in production of almost 3 orders of magnitude between the model at Z$_{met}$ = 0.001 and the maximum at Z$_{met}$ = 0.1.

The radiogenic contribution to $^{59}$Co is dominated by $^{59}$Ni at peak temperatures between 7.5 and 8.5 GK and $^{59}$Cu at intermediate and lower peak temperatures. $^{59}$Co is produced directly only for T $>$ 8.5 GK. The radiogenic production of cobalt is insensitive to metallicity effects at temperatures above 5\,GK. The $^{59}$Ni ejecta is uniform across T1.4 models at temperatures exceeding 7\,GK, and the increase at around T = 4 GK in models T1.40.014 and T1.4Z0.1 is small compared to contributions from deeper layers. In the intermediate peak temperature range, $^{59}$Cu nucleosynthesis is broadly insensitive to varying metallicity. Since this component contributes a majority of the $^{59}$Co production in this model, the Co production will be mostly primary. The small $^{59}$Zn production found in model T1.4Z0 is suppressed at metallicities above Z$_{met}$ = 0. 
The $^{59}$Co yields in the S1.0 models are dominated by two components - a large, narrow peak for T $>$ 5.5 GK, and a broader shallower peak between 3.5 and 5.5 GK. In the S1.0Z0 model, both of these Co peaks are dominated by radiogenic $^{59}$Cu. As initial metallicity rises, the production of $^{59}$Cu is suppressed in the lower temperature peak, and the secondary component of $^{59}$Ni is boosted. In model S1.0Z0.1, production of $^{59}$Ni is boosted to the extent that it is comparable with the total contribution of $^{59}$Cu.
In the M0.8 models, the amount of $^{59}$Co ejected is severely reduced as compared with the previous sets of models. M0.8Z0 has a small peak between 3.2 and 5\,GK, where the main contributor to the ejected cobalt mass is from radioactive $^{59}$Cu. As seen in the S1.0 models, production of $^{59}$Cu is suppressed at higher metallicities, 
where the radioactive isotope $^{59}$Ni provides a crucial contribution to the final Co yields.


\subsection{Nickel}

\begin{figure*}
    \gridline{\fig{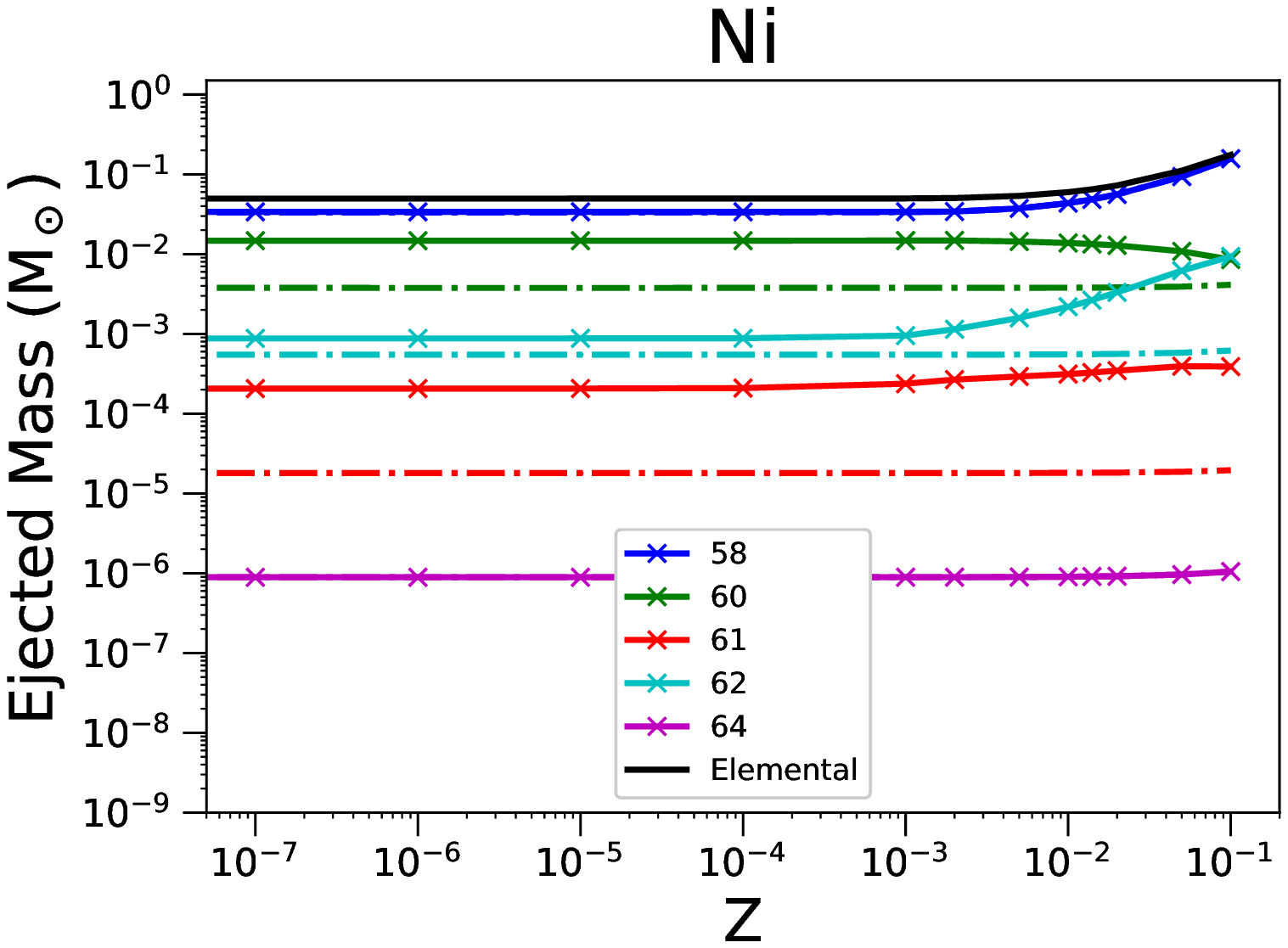}{0.45\textwidth}{T1.4}}
    \gridline{\fig{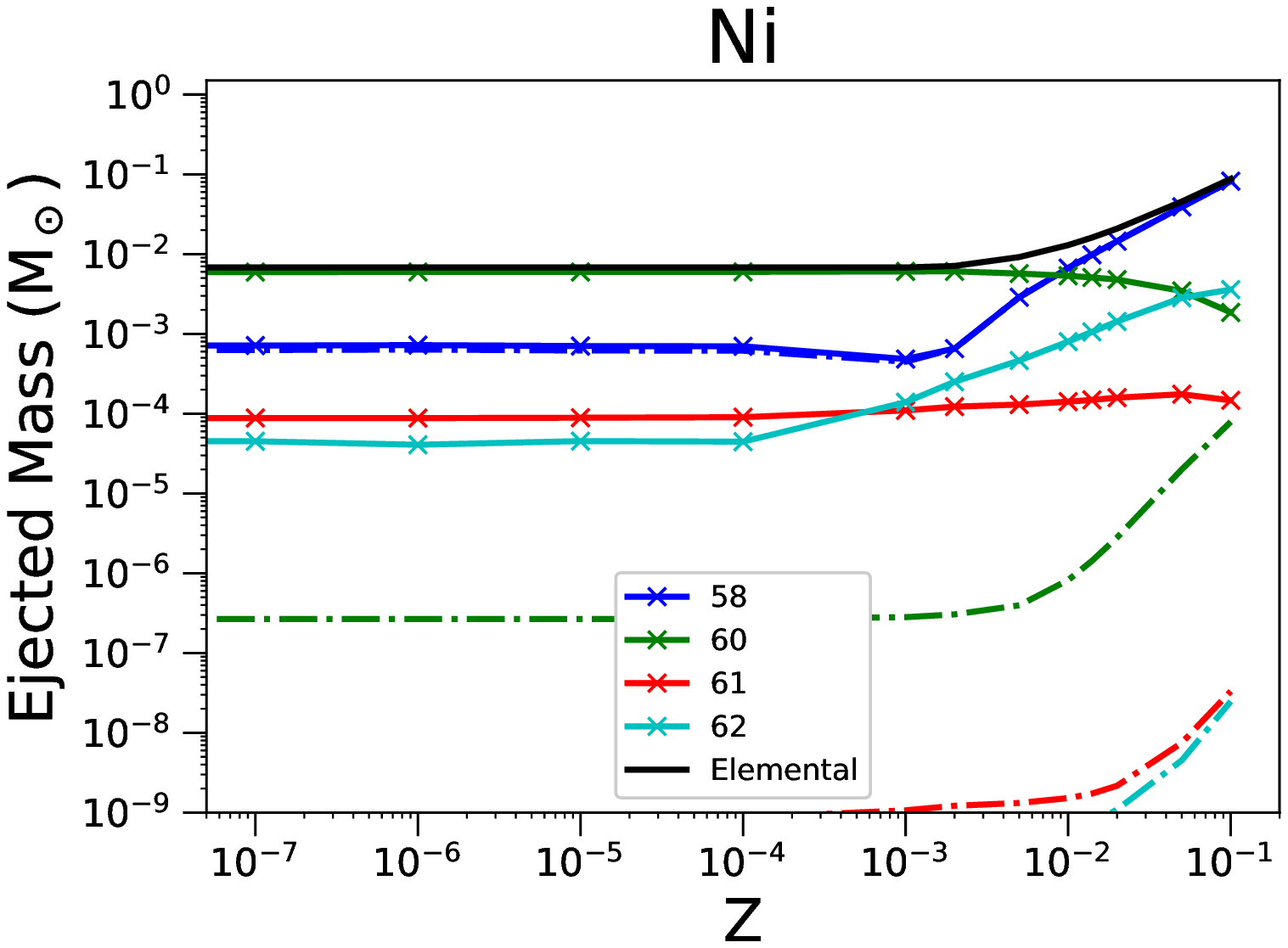}{0.45\textwidth}{S1.0}}
    \gridline{\fig{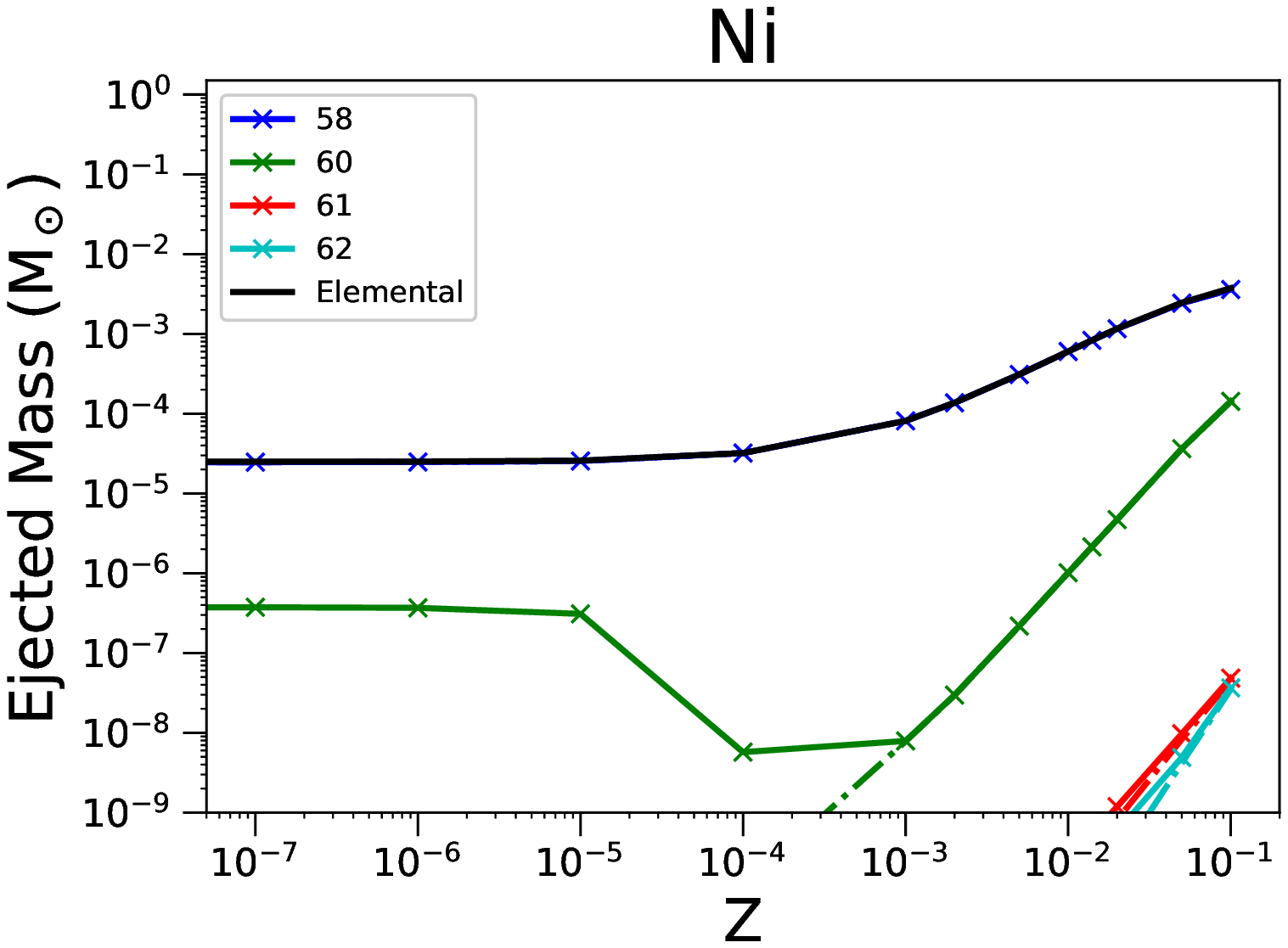}{0.45\textwidth}{M0.8}}
    \caption{As in Figure \ref{fig:Silicon_metalicity_multi} but for ejected mass of Ni, and its stable isotopes $^{58}$Ni, $^{60}$Ni, $^{61}$Ni, $^{62}$Ni and $^{64}$Ni.}
\label{fig:Nickel_metalicity_multi}
\end{figure*}

Nickel is composed of 5 stable isotopes: $^{58}$Ni (68\% of the solar budget), $^{60}$Ni (26\%), $^{61}$Ni (1.1\%), $^{62}$Ni (3.6\%) and $^{64}$Ni (0.9\%). In CCSNe the isotopes $^{58,60,61,62}$Ni are produced during the $\alpha$-rich freeze-out from NSE \citep{woosley2002evolution}, while the s process makes most of the solar $^{62}$Ni and $^{64}$Ni and possibly a significant fraction of $^{60}$Ni and $^{61}$Ni \citep[e.g.,][]{pignatari2016nugrid}. According to GCE simulations, however, most of the solar Ni is made by SNIa \citep[][and references therein]{kobayashi2020origin}.

Figure \ref{fig:Nickel_metalicity_multi} shows the Ni abundance trend with metallicity with the highlighted radiogenic contributions. In the T1.4 models there is an increase in the ejected mass of nickel at around Z$_{\odot}$, moving from about 5x10$^{-2}$ M$_{\odot}$ to over 0.1 M$_{\odot}$ at Z$_{met}$ = 0.1, driven by an increase in the production of $^{58}$Ni. In S1.0, we see a similar increase from a lower initial yields of approximately 6x10$^{-3}$ M$_{\odot}$ to nearly 0.1 M$_{\odot}$. In the M0.8 set, the abundance of $^{58}$Ni increases from a low 2.5$\times$10$^{-5}$ M$_{\odot}$ at Z$_{met}$ = 0 to 3.6$\times$10$^{-3}$ M$_{\odot}$ at Z$_{met}$ = 0.1.

In all models, $^{58}$Ni is mostly produced directly with only a small radiogenic contribution. In the T1.4 models the $^{58}$Ni isotope production in the hottest central ejecta is largely insensitive to the initial composition, although the minor radiogenic contributions from $^{58}$Cu decrease in the 5-6 GK explosion temperature range. The models S1.0 and M0.8 do not include such a high temperature component. As a consequence, their total $^{58}$Ni yields are much smaller, but also more affected by the initial metallicity, increasing the ejected $^{58}$Ni mass by over an order of magnitude over the metallicity range investigated.

The metallicity dependence of $^{60}$Ni yields is similar for models T1.4 and S1.0, where the ejected mass of $^{60}$Ni decreases with increasing metallicity above approximately Z$_{met}$ = 10$^{-2}$ (T1.4 set) and Z$_{met}$ = 2$\times$10$^{-3}$ (S1.0 set) (Figure \ref{fig:Nickel_metalicity_multi}). 
In T1.4Z0 we obtain three areas of production - one above 8 GK, one between 5 and 7 GK and a smaller peak between 4 and 5 GK. In the 8-9 GK region $^{60}$Ni is directly produced, while in the 5-6 GK peak most $^{60}$Ni is synthesized as $^{60}$Zn. This contribution decreases with increasing metallicity, driving the shallow decline of ejected $^{60}$Ni. In the T1.4Z0.014 model, the low temperature peak has been boosted significantly with direct production of $^{60}$Ni. The $^{60}$Ni production also extends further down to lower explosion temperatures around 3.5 GK. 
This trend continues in T1.4Z0.1, where there is a significant boost to the $^{60}$Ni production in the intermediate temperature region. In model S1.0, the highest temperature peak is missing, as no particles experience these conditions. We obtain therefore a much stronger metallicity dependence on the $^{60}$Ni for this model, with the total ejected mass of $^{60}$Ni decreasing as the metallicity increases. This is due to the same effects driving the decrease in the T1.4 models - namely a drop in production of $^{60}$Zn in the intermediate peak temperature region, but as there is not a large contribution from T $>$ 6 GK regions in the S1.0 models, there is proportionally a much larger effect. In the M0.8 models, we obtain even smaller $^{60}$Ni yields. Similar to S1.0Z0, the analogous M0.8 model shows the $^{60}$Ni yields mostly made by radiogenic $^{60}$Zn. At higher metallicity, there are no relevant contributions from radioactive isotopes. 

In the T1.4 model sets the $^{61}$Ni production is dominated by the intermediate and high temperature regions. The ejected mass of $^{61}$Ni is mostly primary, and changes from 2.06$\times$10$^{-4}$ M$_{\odot}$ at Z$_{met}$ = 0 to 3.89$\times$10$^{-4}$ M$_{\odot}$ at Z$_{met}$ = 0.1. In the high temperature region, $^{61}$Ni is directly produced, in the intermediate peak temperature region $^{61}$Zn is the major contributor to ejected $^{61}$Ni mass. 
Finally, in the low temperature region $^{61}$Ni is produced directly.
For the S1.0 models $^{61}$Ni is also primary. Production in the intermediate peak temperature region is insensitive to an increase in the initial $^{22}$Ne mass fraction, with $^{61}$Ni being synthesized as $^{61}$Zn, as in the T.14 models. The low temperature region is directly produced as $^{61}$Ni, and contributes only on the order of 0.1-1\% of the ejected mass of $^{61}$Ni in this model. With the high temperature region missing, the total ejected mass of $^{61}$Ni compared with the T1.4 models is a factor of two lower.
In the M0.8 models, both the intermediate and high temperature peaks are missing, and so we only obtain a small contribution at explosive temperature peaks of about 4 GK.

Not shown in the figures, the nucleosynthesis of $^{62}$Ni is very similar to that for $^{61}$Ni, with a large contribution from the high and intermediate peak temperature regions. In the T1.4 models, we see that the direct production of $^{62}$Ni in the 8-9 GK temperature range is larger than the production of $^{61}$Ni in that same region.
The intermediate peak temperature region from around 5 to 7 GK is very similar to the $^{61}$Ni production region in the T1.4 models, with the mass fraction of $^{62}$Ni being around 0.1\%. $^{62}$Zn is the major contributor to abundances in this region as the direct production drops for temperature peaks lower than 8 GK. The production in this region is sensitive to the initial metallicity, increasing from the initial approximately 0.2\% by mass fraction in T1.4Z0 to around 3\% in model T1.4Z0.1. We also see a small contribution at lower temperatures through direct production of $^{62}$Ni, although this has a negligible effect on the ejected mass.
In the S1.0 series of models, we again find an increase in the production at intermediate peak temperatures with the initial metallicity. As the highest temperatures in this model are at approximately 6 GK, we see only the lower end of the production region at high temperatures, which significantly reduces the ejected mass of $^{62}$Ni. In the models M0.8, there is no production at Z$_{met}$ = 0, and only a marginal production at Z$_{met}$ = Z$_{\odot}$. The 6 GK region is completely absent from this model and so the production through the formation of $^{62}$Zn is not possible. Finally, the $^{64}$Ni production is confined to the high peak temperature region above 8 GK, which is only found in the T1.4 models.


\subsection{Zinc}

\begin{figure*}
    \gridline{\fig{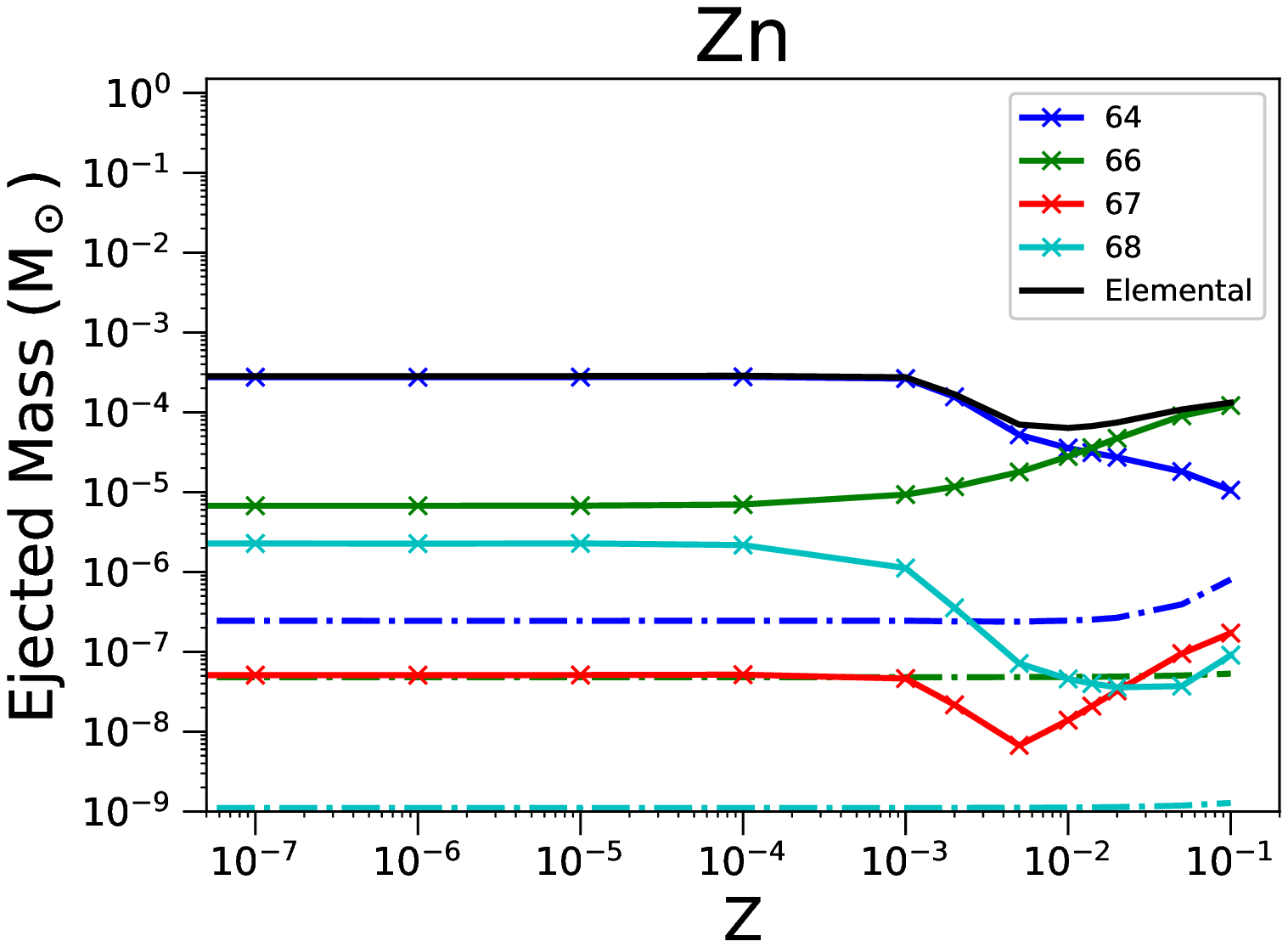}{0.45\textwidth}{T1.4}}
    \gridline{\fig{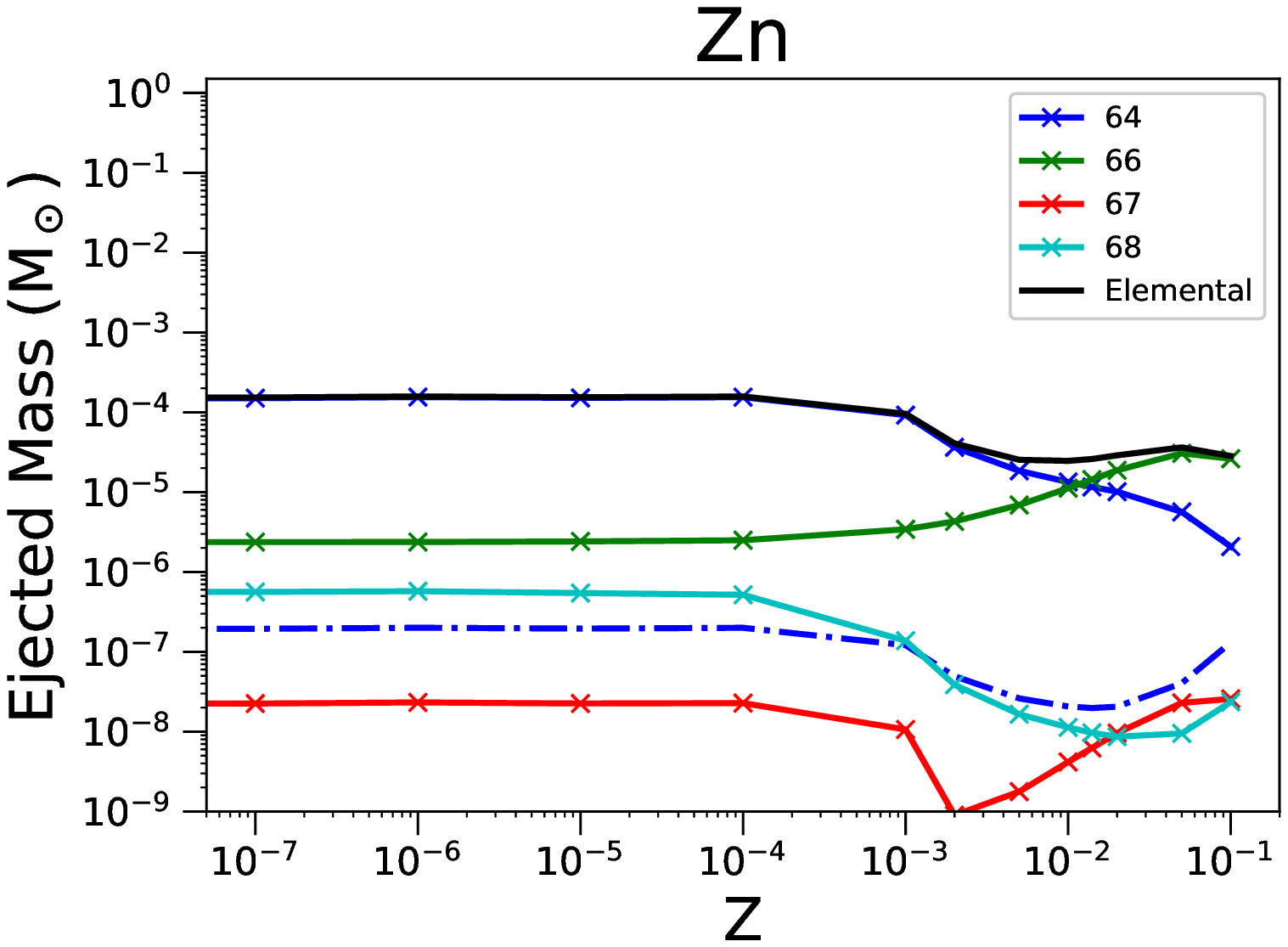}{0.45\textwidth}{S1.0}}
    \caption{As in Figure \ref{fig:Silicon_metalicity_multi} but for ejected mass of Zn, and its stable isotopes $^{64}$Zn, $^{66}$Zn, $^{67}$Zn and $^{68}$Zn. There is no production of zinc in the M0.8 models and therefore the plot is omitted.}
\label{fig:Zinc_metalicity_multi}
\end{figure*}

Zinc has 5 stable isotopes: $^{64}$Zn (49.2\% of the solar budget), $^{66}$Zn (27.7\%), $^{67}$Zn (4\%), $^{68}$Zn (18.5\%) and $^{70}$Zn (0.6\%). None of the models investigated here produce $^{70}$Zn, as such, we will not discuss this isotope further. In CCSNe, $^{64}$Zn is mostly produced by neutrino driven winds, in $\alpha$-rich freeze-out conditions, and a small s-process contribution is also possible \citep{woosley2002evolution}. 
$^{66}$Zn is efficiently produced by the s-process and the $\alpha$-rich freeze-out, and $^{68}$Zn is produced mostly in massive stars through the s-process \citep{woosley2002evolution}. Nevertheless, \citet{kobayashi2020origin} show that Zn also has a small contribution from SNIa.

Zinc production in the models T1.4 and S1.0 is similar (Figure \ref{fig:Zinc_metalicity_multi}). Zn yields decrease at around Z$_{met}$ = 0.001 by approximately a factor of 5, following the trend of $^{64}$Zn. At the same time, $^{66}$Zn production increases becoming the most abundant isotope at higher metallicities.

Not shown in the figures, in all T1.4 models the $^{64}$Zn nucleosynthesis is dominated by the radioactive isotope $^{64}$Ge. The contribution of $^{64}$Ga to the ejected abundance increases at higher metallicities as the intermediate peak temperature range production of $^{64}$Ge shrinks.
In the S1.0 models the $^{64}$Zn production is also primarily through synthesis of $^{64}$Ge. The decrease in the overall production of $^{64}$Zn follows the reduction in $^{64}$Ge and $^{64}$Ga produced at peak temperatures around 5.5-6 GK. We also observe the rise in direct production of $^{64}$Zn with increasing metallicity, although the absolute abundance is only on the order of 5\% in the S1.0Z0.1 model.
The production of $^{66}$Zn is largely centered in the intermediate peak temperature range starting at approximately 5.2\,GK and continuing to 8\,GK. $^{66}$Zn is mostly produced as $^{66}$Ge in the T1.4Z0 model, with trace amounts of other isotopes. Production is nearly identical in the T1.4Z0.014 model, with only a slight increase in the production of $^{66}$Ga (but still only a negligible contribution). 
Production in all three S1.0 models is similar, with a strong radiogenic contribution to $^{66}$Ge. Trace amounts of $^{66}$Ga do not affect the overall yields.

$^{67}$Zn has a similar trend as for $^{66}$Zn. In model T1.4Z0 there is a comparable contribution to the ejected mass of $^{67}$Zn from both $^{67}$Ge and $^{67}$As. As the initial metallicity increases, both of these contributions become small compared to the increased production of $^{67}$Zn, culminating in T1.4Z0.1, where only trace amounts of $^{67}$Ge and $^{67}$Cu contribute to the abundance of $^{67}$Zn.

The production of $^{68}$Zn occurs mainly at intermediate peak temperatures from 5.2 to 7\,GK. In the model T1.4Z0, there is a small contribution to $^{68}$Zn from $^{68}$As, with the bulk of material being synthesized as $^{68}$Se. $^{68}$Se production is reduced as the metallicity increases, leading to an overall decrease in the ejected mass of $^{68}$Zn and and increase in the $^{68}$Ge to $^{68}$Se ratio. In model T1.4Z0.1, $^{68}$Se is no longer the most abundant isotope as $^{68}$Ge has become more favorable to produce. In the S1.0 models the nucleosynthesis is similar, with $^{68}$Se being the most abundant isotope in the S1.0Z0 models. There is a decrease in the production of $^{68}$Se as the initial metallicity increases, and then production becomes dominated by $^{68}$Ge, with $^{68}$Se produced only in trace amounts. 

The decayed yields from the T1.4, S1.0 and M0.8 models are available online at at \dataset[Zenodo: Yields]{https://doi.org/10.5281/zenodo.8060323}. 

\subsection{Metallicity Trends Summary}

The isotopes with the largest change in production between Z$_{met}$ = 0 and Z$_{met}$ = 0.1 for models T1.4, S1.0 and M0.8 are shown in tables \ref{table:dean_met_factors}, \ref{table:shen_met_factors} and \ref{table:brox_met_factors} respectively. We first selected for isotopes with more than a 10$^{3}$ increase from Z$_{met}$ = 0 to Z$_{met}$ = 0.1 from elements with a contribution from SNIa events as identified in \cite{kobayashi2020origin}, and then compared the change in abundances from Z$_{met}$ = Z$_{\odot}$ to Z$_{met}$ = 0.1. We see for the T1.4 case, the isotopes with the largest change in production are from the intermediate mass region. From the production factors given in Figure \ref{fig:dean_production_factors} we can see that $^{34}$S, $^{36}$S, $^{40}$K and $^{42}$Ca begin to make significant contributions to the solar budget of these isotopes with an initial metallicity close to the solar value. In a similar manner, the S1.0 yields show significant contributions from $^{34}$S, $^{38}$Ar, $^{40}$K, $^{42}$Ca, $^{50}$V and $^{54}$Cr. While other abundances also increase significantly across the metallicity space they are in trace amounts. $^{40}$Ar may have a significant contribution to solar abundances at super-solar metallicities for S1.0. Metallicity dependent yields of SNIa, as well as a well constrained distribution of the true initial metallicity of SNIa progenitors and their relative contribution to the solar budget is therefore vital in determining the origins of $^{34}$S, $^{36}$S, $^{40}$K and $^{42}$Ca in the Chandrasekhar and $^{34}$S, $^{38}$Ar, $^{40}$K, $^{42}$Ca, $^{50}$V, $^{54}$Cr and, to a lesser extent  $^{40}$Ar, in the sub-Chandrasekhar cases.

\begin{table}
\centering
\begin{tabular}{c c c c} 
 \hline
 Element & Mass & X$_{0.1}$/X$_{0}$ & X$_{0.1}$/X$_{_\odot}$ \\ 
 \hline
 S  & 34 & 1.473e+03 & 7.167e+00 \\
 S  & 36 & 4.213e+06 & 3.333e+02 \\
 Ar & 40 & 1.104e+05 & 3.016e+02 \\
 Ca & 42 & 2.199e+03 & 6.060e+00 \\
 \hline
\end{tabular}
\caption{Relative production in the T1.4 models for elements with a significant contribution from SNIa explosions. We present only those isotopes with $X_{0.1}/X_0>10^{-3}$ to highlight those isotopes with a strong metallicity dependence. Column 1 shows the change in production between the zero metallicity and the case with maximum metallicity, column 2 the same of column 1 but between solar and the maximum metallicity. 
}
\label{table:dean_met_factors}
\end{table}

\begin{table}
\centering
\begin{tabular}{c c c c} 
 \hline
 Element & Mass & X$_{0.1}$/X$_{0}$ & X$_{0.1}$/X$_{_\odot}$ \\ 
 \hline
 S  & 34 & 3.725e+03 & 7.600e+00 \\  
 S  & 36 & 7.749e+08 & 3.118e+02 \\  
 Ar & 38 & 2.893e+03 & 7.000e+00 \\  
 Ar & 40 & 3.582e+08 & 2.653e+02 \\  
 Ca & 42 & 1.283e+04 & 5.408e+00 \\  
 Ca & 46 & 9.842e+11 & 5.858e+02 \\  
 Ca & 48 & 7.955e+17 & 1.489e+05 \\  
 Ti & 50 & 1.594e+11 & 3.477e+01 \\  
 V  & 50 & 6.174e+08 & 3.148e+01 \\  
 Cr & 54 & 5.804e+09 & 1.695e+02 \\  
 Fe & 54 & 3.281e+03 & 6.818e+00 \\  
 Fe & 58 & 2.131e+07 & 7.692e+01 \\  
 Ni & 64 & 1.777e+08 & 1.668e+02 \\  
 \hline
\end{tabular}
\caption{Relative production in the S1.0 models for elements with a significant contribution from SNIa explosions. We present only those isotopes with $X_{0.1}/X_0>10^{-3}$ to highlight those isotopes with a strong metallicity dependence. Column 1 shows the change in production between the zero metallicity and maximum metallicity case, column 2 between solar and maximum}
\label{table:shen_met_factors}
\end{table}

\begin{table}
\centering
\begin{tabular}{c c c c} 
 \hline
 Element & Mass & X$_{0.1}$/X$_{0}$ & X$_{0.1}$/X$_{_\odot}$ \\ 
 \hline
 S  & 34 & 3.379630e+03 & 6.347826e+00 \\  
 S  & 36 & 9.438503e+08 & 4.114219e+02 \\  
 Ar & 38 & 1.733668e+04 & 6.900000e+00 \\  
 Ar & 40 & 5.169082e+08 & 3.635334e+02 \\  
 Ca & 42 & 2.109705e+04 & 3.759398e+00 \\  
 Ca & 46 & 1.066946e+12 & 5.930233e+02 \\  
 Ca & 48 & 4.714286e+15 & 4.400000e+05 \\  
 Ti & 46 & 3.937107e+04 & 4.852713e+00 \\  
 Ti & 50 & 1.183013e+11 & 1.914894e+01 \\  
 V  & 50 & 4.912621e+08 & 1.547401e+01 \\  
 Cr & 50 & 7.035176e+03 & 1.400000e+01 \\  
 Cr & 54 & 1.111111e+10 & 1.681034e+02 \\  
 Fe & 54 & 5.271318e+03 & 5.354331e+00 \\  
 Fe & 58 & 2.083333e+07 & 7.402423e+01 \\  
 Ni & 62 & 6.336806e+03 & 1.196721e+02 \\  
 Ni & 64 & 1.744898e+11 & 8.829604e+01 \\  
 Zn & 64 & 1.140351e+05 & 3.762663e+01 \\  
 Zn & 66 & 3.486486e+05 & 1.425414e+02 \\  
 Zn & 67 & 1.435374e+03 & 7.429577e+01 \\  
 \hline
\end{tabular}
\caption{Relative production in the M0.8 models for elements with a significant contribution from SNIa explosions. We present only those isotopes with $X_{0.1}/X_0>10^{-3}$ to highlight those isotopes with a strong metallicity dependence. Column 1 shows the change in production between the zero metallicity and maximum metallicity case, column 2 between solar and maximum}
\label{table:brox_met_factors}
\end{table}

\section{Literature Yields Comparisons}\label{sec:lit_yield_comparison}

In this section the results of the nucleosynthesis calculations are compared with other yields in the literature, namely \citet{nomoto2018single} and \citet{seitenzahl2013three} with the T1.4 models, and \citet{sim2010detonations} and \citet{gronow2020sne} with the S1.0 models. Results are first compared with their parent model to highlight differences due to the nuclear reaction network.

\subsection{Comparison With Parent Models}

\begin{figure*}
    \gridline{\fig{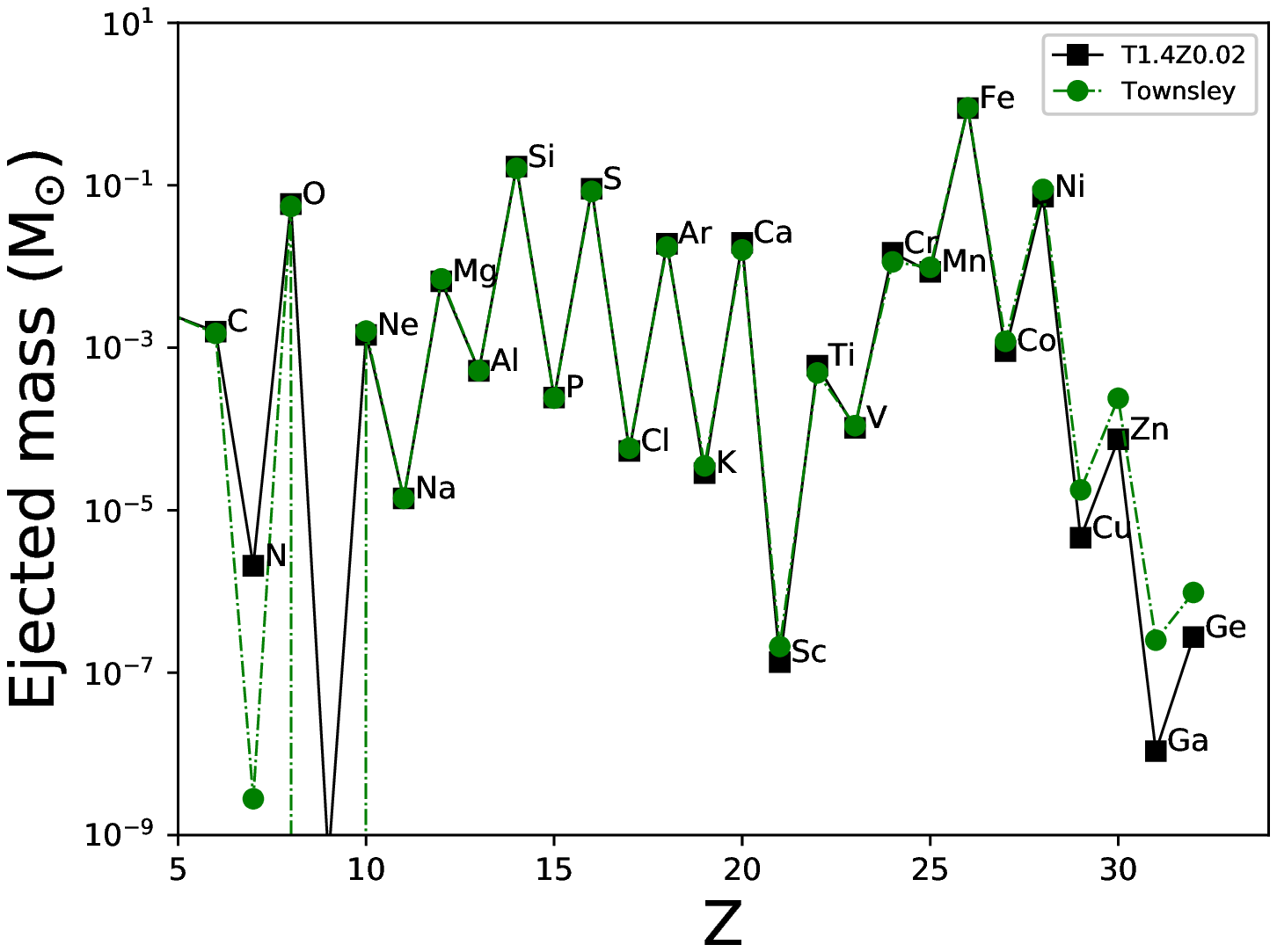}{0.45\textwidth}{T1.4Z0.02}}
    \gridline{\fig{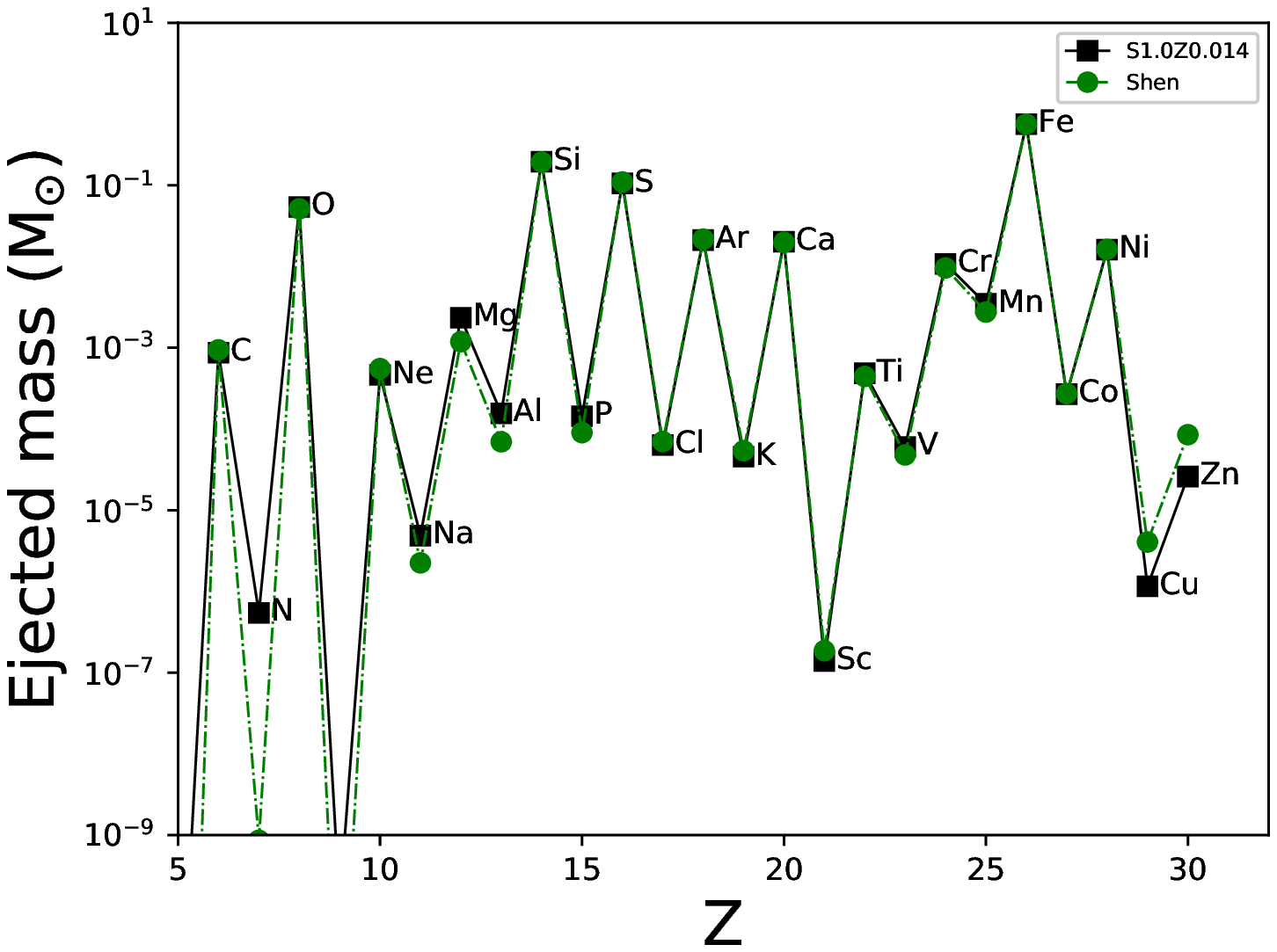}{0.45\textwidth}{S1.0Z0.014}}
    \gridline{\fig{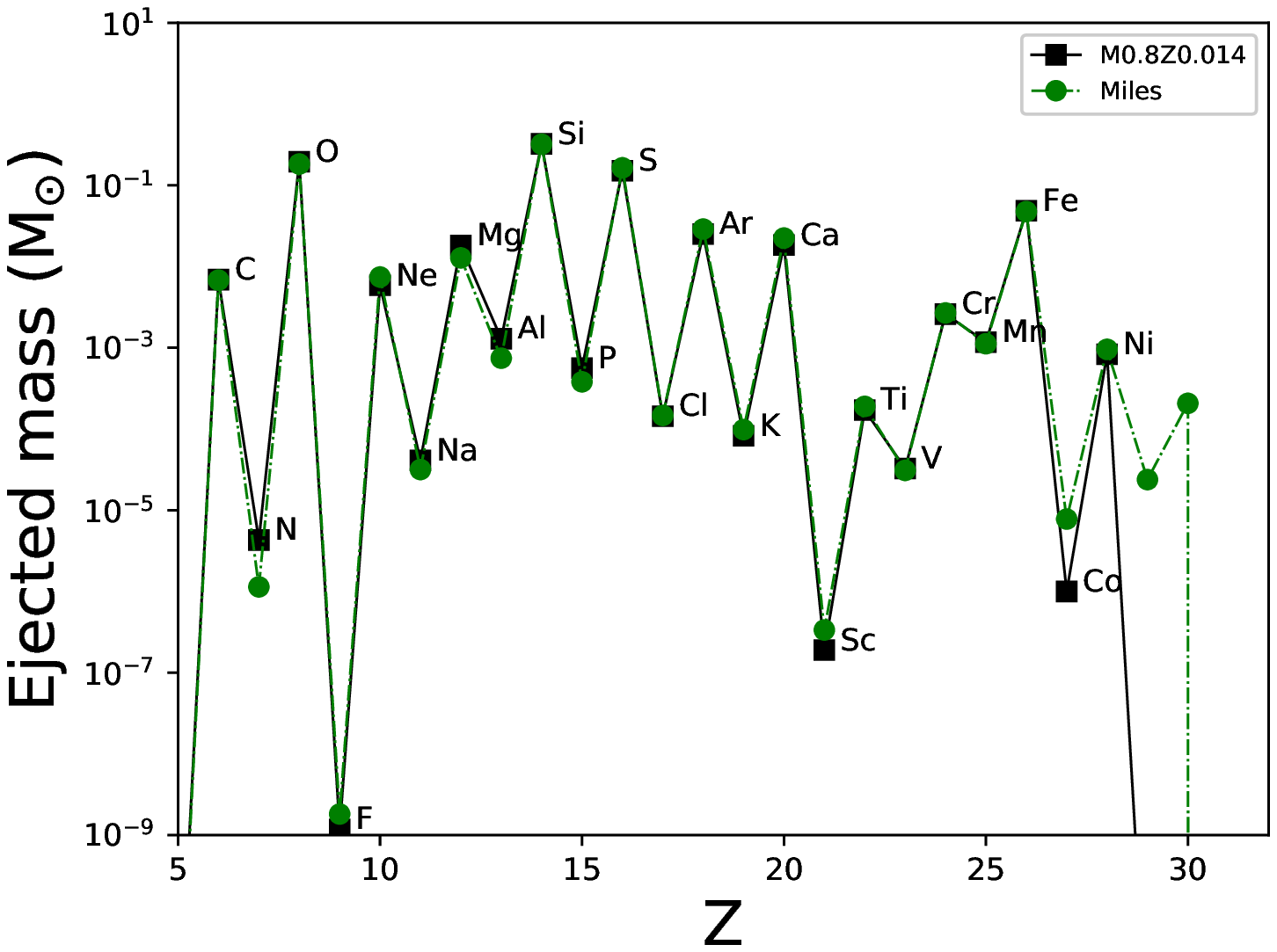}{0.45\textwidth}{M0.8Z0.014}}
    \caption{Comparison with the T1.4 yields T1.4Z0.014 (upper panel), S1.0Z0.014 (middle panel) and M0.8Z0.014 (lower panel). In each case, the metallicity closest to the metallicity used in the parent model is selected from our grid of initial $^{22}$Ne mass fractions}
\label{fig:parent_model_comparison}
\end{figure*}

Shown in \ref{fig:parent_model_comparison} is the comparison between our post-processed models and their parent models from the literature \citep{townsley2016tracer,shen2018sub,miles2019quantifying}. For the T1.4 case, we have chosen an $^{22}$Ne mass fraction of 0.02 which most closely matches the previously published model. we see that there are some differences, particularly in the heavier iron group region where the ejected mass of Cu is around a factor of five less in our post-processed yields, a factor of two for ejected Zn and Ge, and more than an order of magnitude for Ga. Other than these differences, the production in the post-processed model follows that in the parent model very closely. The smaller ejected masses in the upper iron group are caused by less complete burning, as we see a small increase in the ejected mass of Si, S, Ar and Ca. Ejected masses for the S1.0Z0.014 model are closer to the parent model than the T1.4Z0.02 case throughout the iron group, with only a significant underproduction of Cu and Zn as compared with the published model, and then by a factor of two to three. In addition, Mg is also slightly overproduced, and Al and Na slightly under-produced. Broadly however, the abundances are very close between the two treatments. The same is true for M0.8Z0.014 up to Ni, with only Co under-produced as compared to the parent model, here by around a factor of five. Cu and Zn, however, are not produced in our post-processed results as they are in the parent model, likely due to small increases in the ejected mass of Mg, Al and P indicating less complete burning than is present in \citet{miles2019quantifying}. We conclude that the results of our post-processing, for the metallicities closest to the parent models, are in each case close to the expected values, with reasonable differences which we attribute to different choices in reaction networks, our treatment of NSE and our choice of a flat $^{22}$Ne abundance.

\subsection{Comparison with the T1.4 yields}

\begin{figure*}
     \gridline{\fig{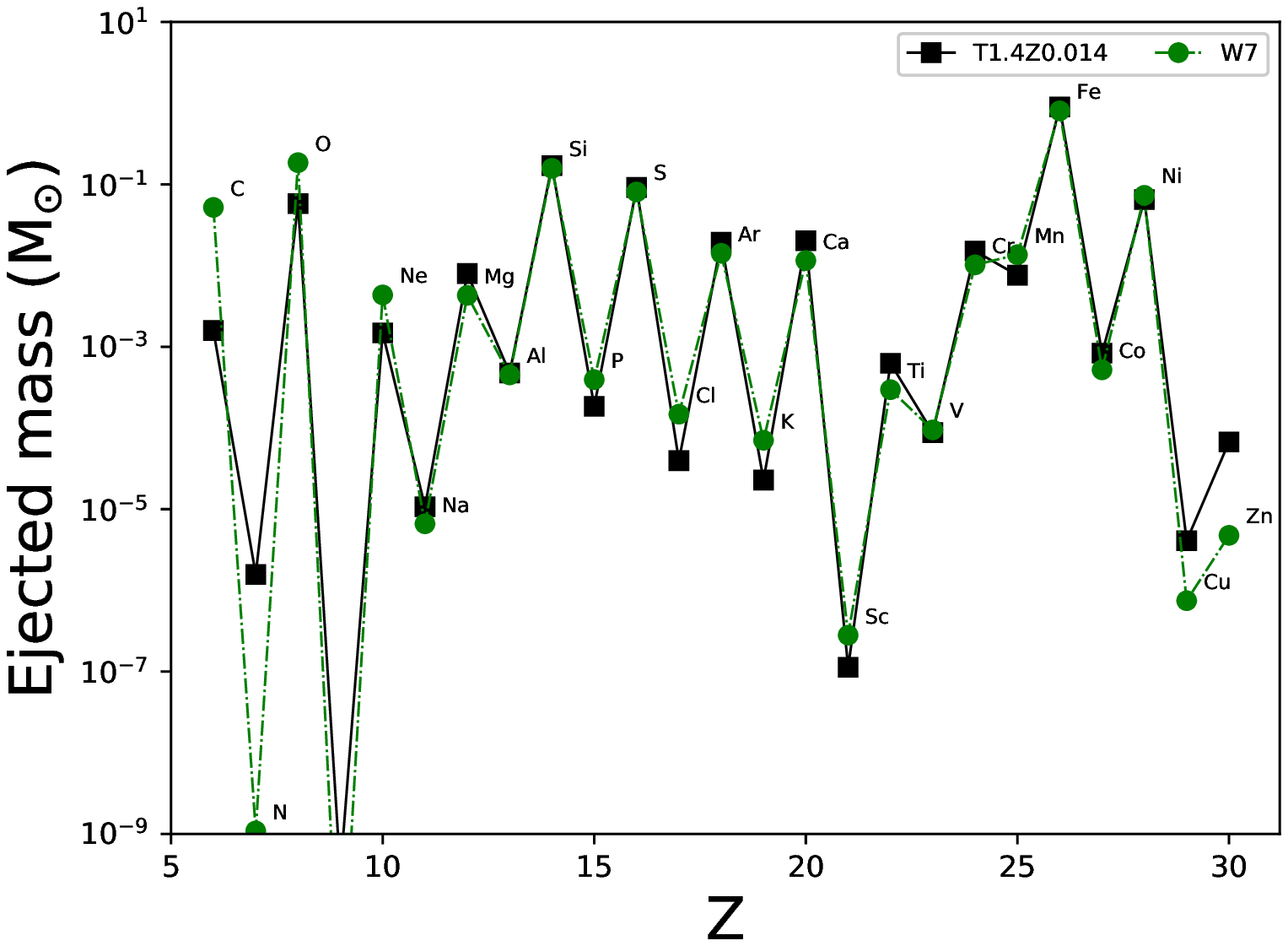}{0.72\textwidth}{}}
     \gridline{\fig{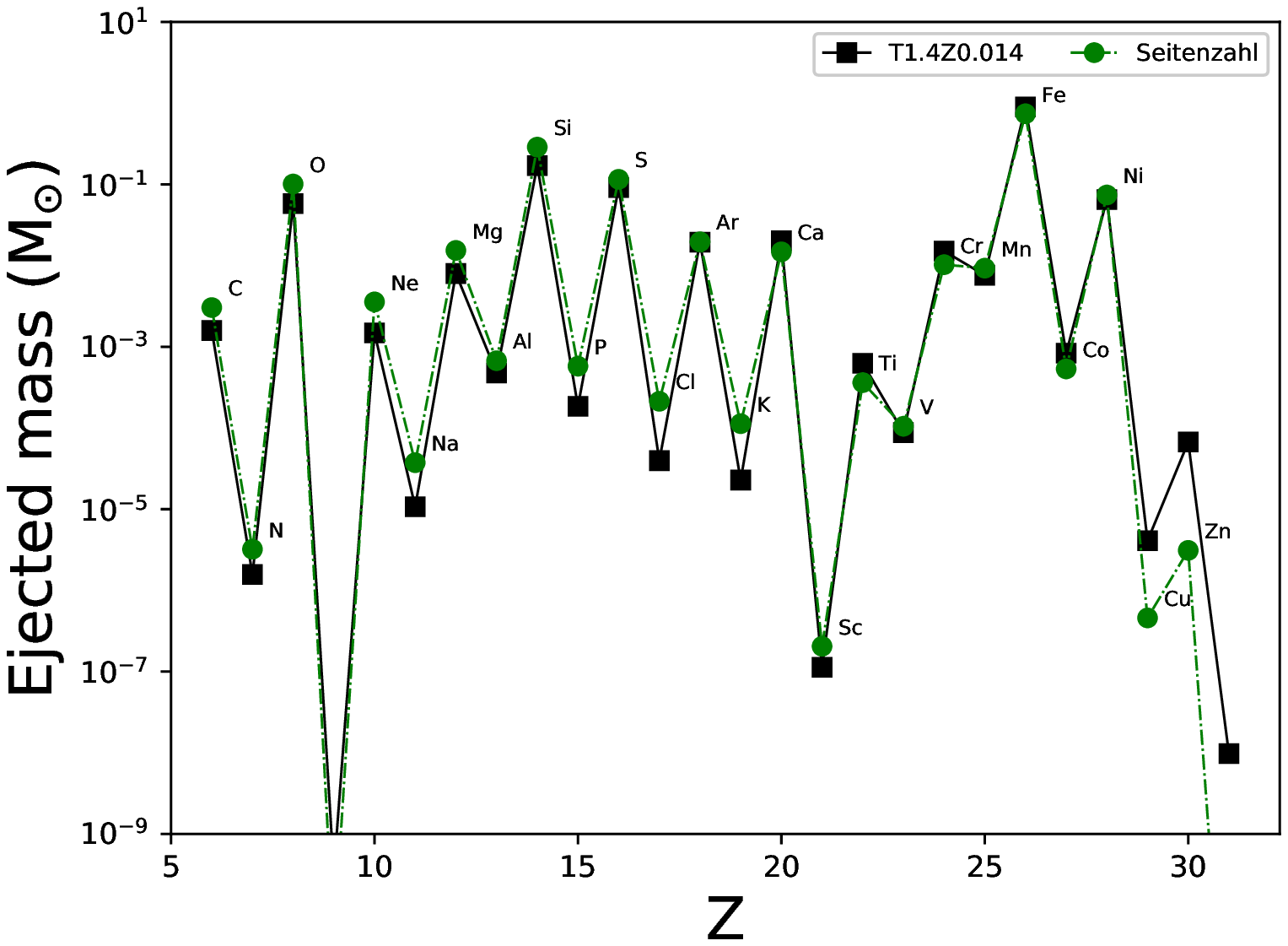}{0.72\textwidth}{}}
\caption{In the upper panel shows W7 decayed elemental abundances from \citet{nomoto2018single} (green circles) compared with T1.4Z0.014 (black line, square markers). In the lower panel, the yields from \cite{seitenzahl2013three} are compared instead. 
}
\label{fig:W7_ele_comparison_j}
\end{figure*}

Figure \ref{fig:W7_ele_comparison_j} shows the comparison of our T.4Z0.014 model with the W7 yields of \citet{nomoto2018single} (upper panel) and with the N100 model of \citet{seitenzahl2013three} (lower panel). The W7 yields have been updated from the original yields of \citep{nomoto1982accreting} and later \citet{nomoto1997nucleosynthesis} papers with improved nuclear physics. All models are shown at solar metallicity, which is set in all cases by the initial $^{22}$Ne content. The agreement between the models is good, with some differences in the ejected mass of Cl, K and S where the W7 and Seitenzahl models are enriched by around a factor of 2 compared with the T1.4Z0.014 model. We see that the T1.4Z0.014 model produces more Cu by a factor of around 5, and more Zn by a factor of around 10 than the W7 model. Our agreement with the Seitenzahl yields is generally better than with the W7 model, for example the Cr/Mn ratio for W7 is $<$ 1 but $>$ 1 for the T1.4Z0.014 and Seitenzahl models.

\citet{nomoto2018single}, use the same 1D structure of the original W7 model, the \citet{townsley2016tracer} model is 2D and the \cite{seitenzahl2013three} models are 3D. For the most abundant isotopes, e.g. $^{28}$Si, $^{32}$S, $^{56}$Fe the agreement is very good. There is, however, a systematic under-production of the intermediate mass, odd Z elements, as compared with the W7 model. We also see that W7 over-produces Mn as compared with our post-processed results by a factor of 2, as well as a slightly reduced abundance of Fe leading to a larger [Mn/Fe]. \citet{nomoto2018single} attribute the relative insensitivity of the Chandrasekhar mass SNIa yields to initial metallicity, due to the synthesis of neutron rich isotopes in the hot NSE region. The sensitivity of the outer layers of the WD to the initial metallicity is important, however, as we see a significant metallicity dependence in the intermediate mass elemental yields.

Note that neither W7 nor \citet{seitenzahl2013three} produces the trace abundances of Ga and Ge reported in \citet{townsley2016tracer}. The network of \citet{seitenzahl2013three} contains 384 isotopes, and so is at least as complete as the \citet{townsley2016tracer} network. This discrepancy is therefore a consequence of the hydrodynamics modeling or the nuclear physics inputs. Zn has a significant ejected mass and is produced in non-negligible amounts; however, Ga is over 2 orders of magnitude less abundant than Zn, and is unlikely to have a large effect on the abundances of the iron group through leakage of material above the networks in W7 and \citet{seitenzahl2013three}.

The [Cr/Mn] ratio for our post-processed models is closer to the value of \citet{seitenzahl2013three} than to that of the W7 results. Without further study, it is not possible to conclude whether this is due to the hydrodynamical modeling or the differences in reaction networks. More complete burning throughout the T1.4Z0.014 model is indicated by the significantly reduced abundances of C and O.

\subsection{Comparison with the S1.0 yields}

\begin{figure*}
     \gridline{\fig{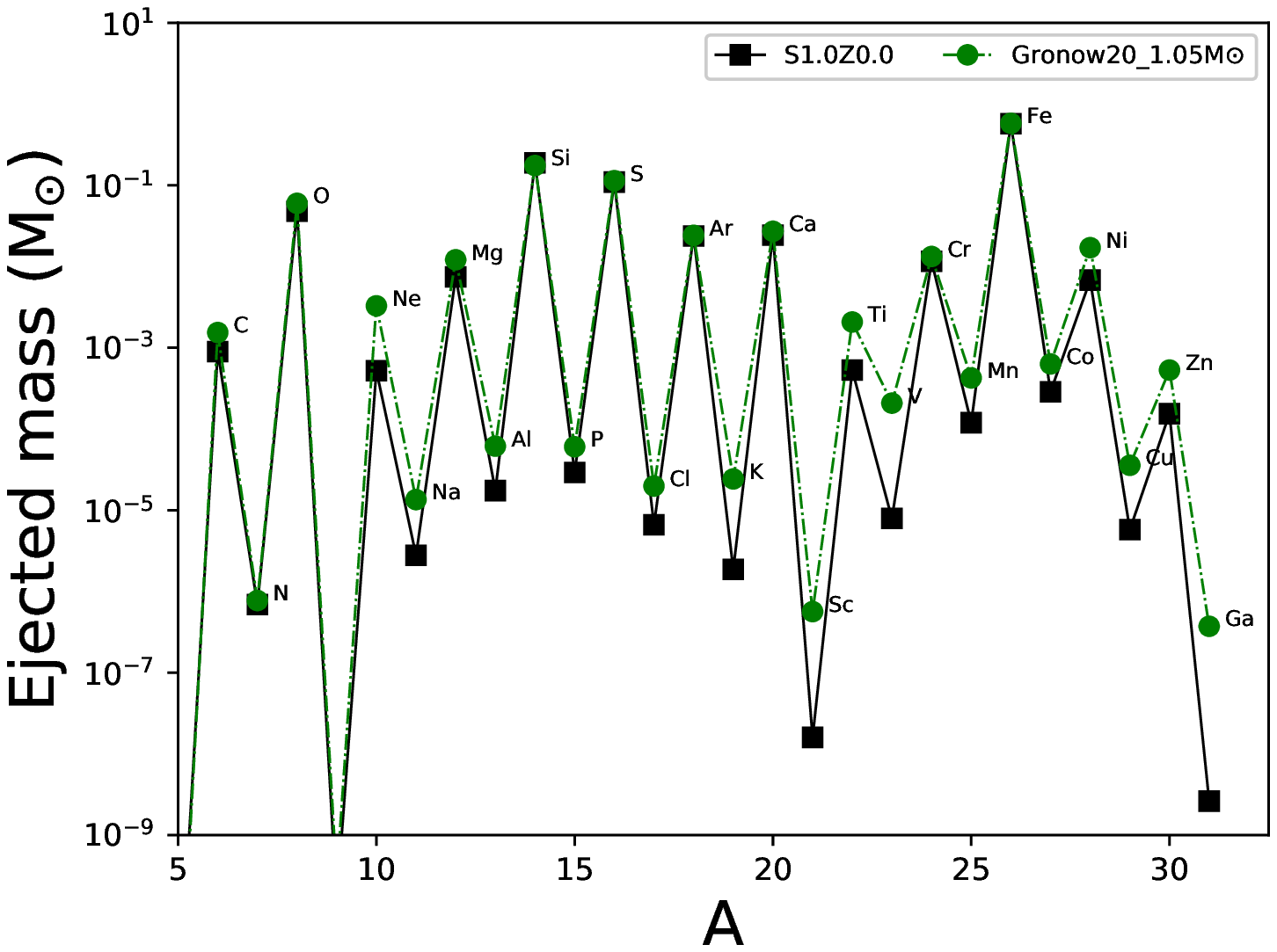}{0.74\textwidth}{}}
     \gridline{\fig{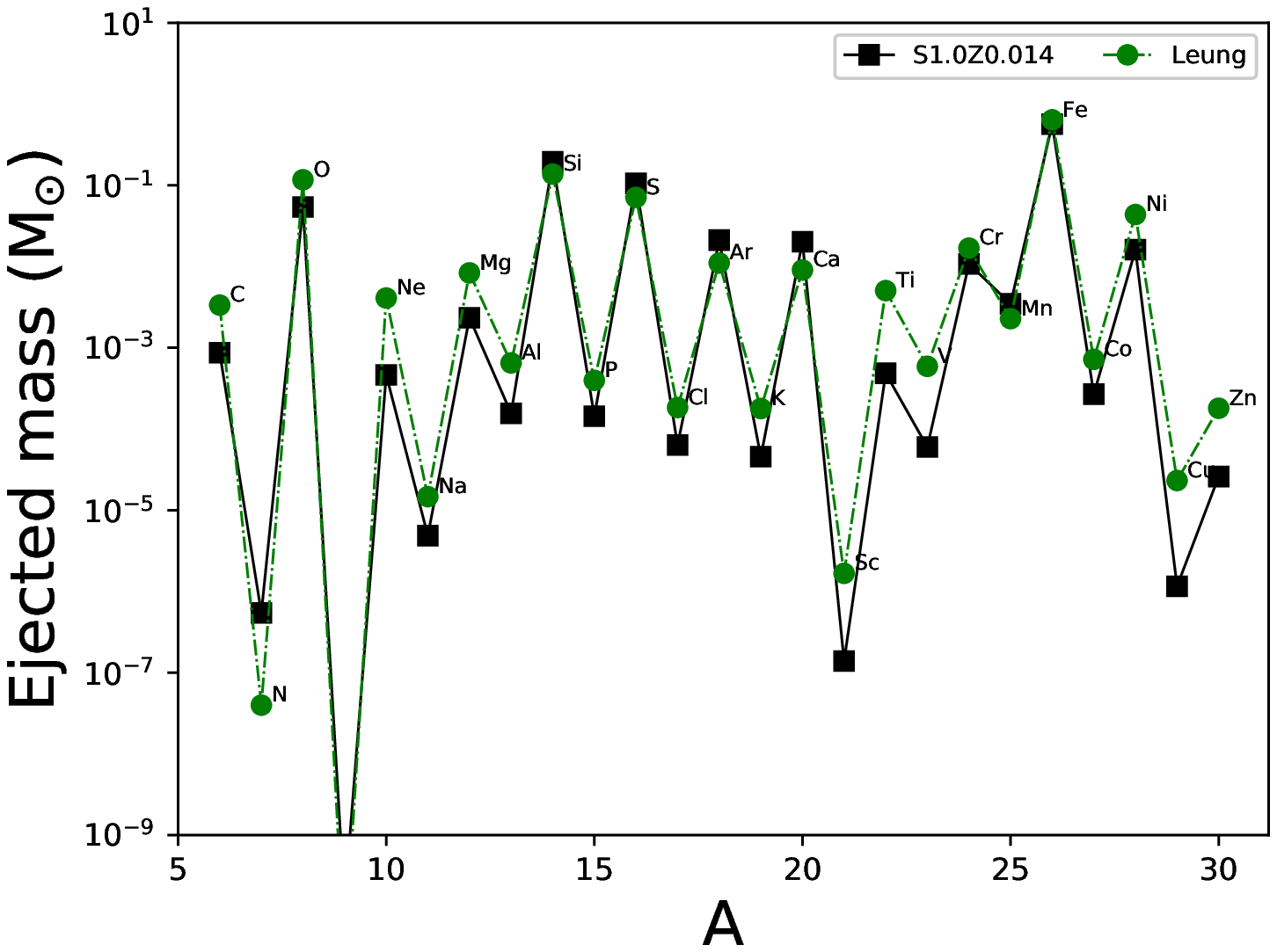}{0.74\textwidth}{}}
\caption{Yields from \citet[][green circles]{gronow2020sne} compared with our S1.0Z0 model (black squares, upper panel) and \citet{leung2020explosive} compared with model S1.0Z0.014 (lower panel).
}
\label{fig:gronow_2020_S_0}
\end{figure*}

Figure \ref{fig:gronow_2020_S_0} (upper panel) shows a comparison between our post-processed results for the S1.0Z0 model and the \citet{gronow2020sne} SNIa model. There are some key differences in the model parameters, particularly in the treatment of the accreted helium layer. In the \citet{gronow2020sne} work, the helium layer is fully modeled 
however there is a large degree of artificial numerical mixing between the helium layer and the underlying C/O; causing an additional 30-40\% of C/O to be added into the helium layer before the start of the explosion, which is likely to increase the impact of the helium shell nucleosynthesis significantly.

Further comparison for the 1.0 M$_{\odot}$ against model \citet{leung2020explosive} 1.05M$_{\odot}$ SNIa are shown in Figure \ref{fig:gronow_2020_S_0} lower panel.
C, N, O, F, Si, S, Ar, Cr, Fe and Co are all within a factor of 2 between the two models. The iron group distribution matches more closely, although the production of Mn in the \citet{shen2018sub} post-processed model is increased by a factor of 15 \citep{gronow2020sne}. The detonation mechanism therefore significantly influences the distribution and ejected yields of the SNIa explosion, both in the iron group region and the intermediate mass elements. 
Significant differences in the abundances of Ti and V are observed, with this post-processed work being a factor of 5 under-produced compared to that work. A factor of 2 overproduction of the much more abundant Ca accounts for many of the differences. We again see higher production of even Z intermediate mass elements, and a reduction in the ejected mass of odd Z intermediate mass elements as compared with \citet{leung2020explosive}.

In \citet{shen2018sub} the helium layer is treated as being minimally thin, and contributing nothing to the ejected mass of the SNIa. The different treatment of the outer layers of the WD may explain the discrepancies in production of the lower Z elements (e.g. Al, P, Cl) as these odd-Z elements are produced primarily in the cooler outer regions of the explosion in the S1.0Z0 model (whilst the even-Z elements have a larger contribution from the intermediate temperature regions). A more massive cooler component may therefore lead to an increase in production, as seen in the Gronow yields however as the composition of the He shell is different to the CO core of the S1.0Z0 model, further testing of this result is required. We see a significant increase in the production of elemental Ne, which is again produced in the cool outer regions of the star. Other discrepancies may be due simply to the fact that the models are of different masses of progenitor, with a difference of 0.05M$\odot$. For the less abundant species, this will have a large impact. $^{51}$V is the largest component of the ejected elemental vanadium. and has a broad contribution from ejecta with maximum peak temperatures between 3 and 6 G. Further investigation is needed to determine the origin of this large difference.

\subsection{Comparison with the M0.8 yields}

\begin{figure*}
     \gridline{\fig{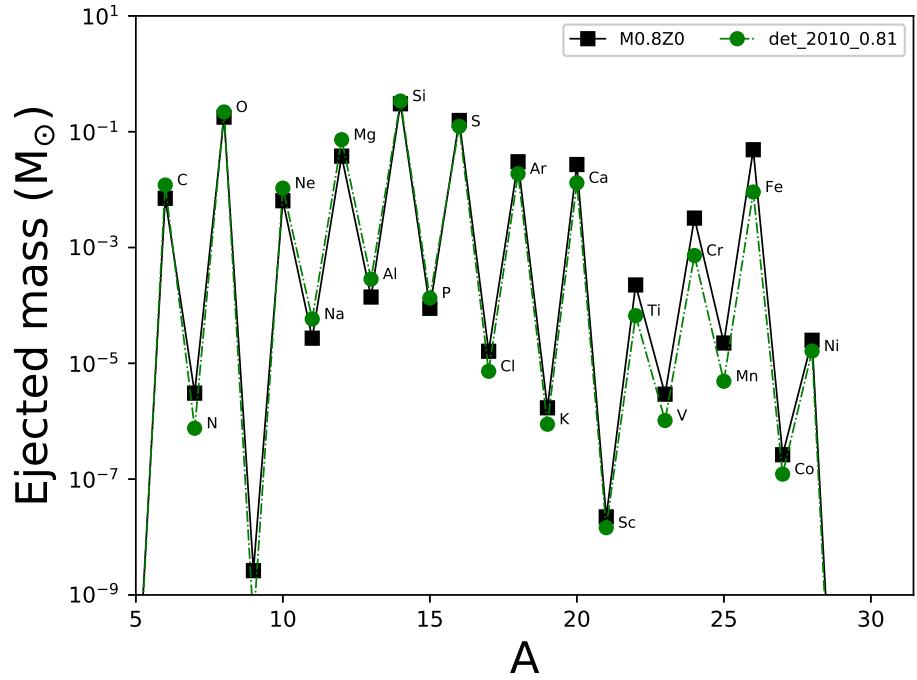}{0.74\textwidth}{}}
     \gridline{\fig{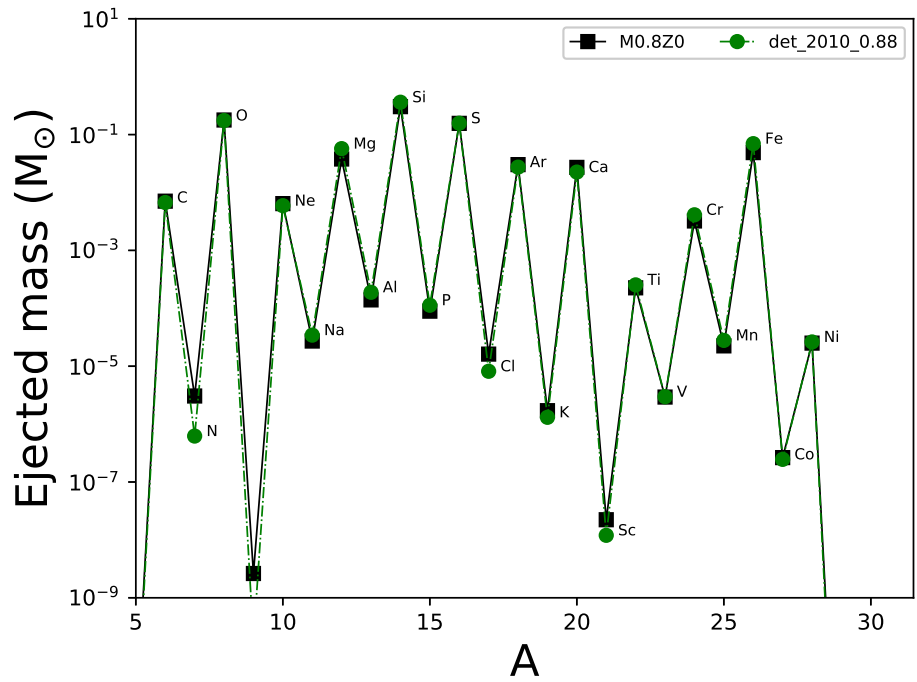}{0.74\textwidth}{}}
\caption{0.81 M$_{\odot}$ model from \citet[][green circles]{sim2010detonations} compared with our M0.8Z0 model (black squares, upper panel) and the 0.88 M$_{\odot}$ model S1.0 from the same publication (lower panel). 
}
\label{fig:sim_2010_comparison}
\end{figure*}

Comparison with the 0.81 and 0.88 M$_{\odot}$ models of \citet{sim2010detonations} in Figure \ref{fig:sim_2010_comparison} show better agreement with the higher mass model, where only N and Sc differ by a factor of 2 or more. There are larger differences as compared with the 0.81 M$_{\odot}$ model, where the low Z elements Al, Cl and K are overproduced in the M0.8Z0 model by around a factor of two which may be due to differences in the nuclear network between the runs. It is unclear why the more massive model more closely matches our results however as the yields from all models are consistent we are satisfied that the parameters of our models are reasonable.

Our 0.8 M$_{\odot}$ is predicted to be undetectable observationally due to the small mass of $^{56}$Ni produced. \citet{sim2010detonations}  show that their 0.88 M$_{\odot}$ mass progenitor lies at the limit of observable production of $^{56}$Ni - ejecting a total mass of 0.07M$_{\odot}$ of $^{56}$Ni. Their 0.81 M$_{\odot}$ model ejects a total of 0.01M$_{\odot}$ of $^{56}$Ni, is not discussed further in their paper due to the absence of a possible observational counterpart. Our M0.8Z0 produces 0.048 M$_{\odot}$ of $^{56}$Ni, close to the observable limit of the 0.88 M$_{\odot}$ model however the observational counterpart of this explosion may not resemble a typical SNIa explosion.

\section{Conclusions} \label{sec:conclusions}

In this work we have presented yields from 39 models of type Ia supernovae, at 3 masses (1.4, 1.0 and 0.8 M$_{\odot}$) and 13 metallicities. The metallicities were varied by changing the $^{22}$Ne mass fraction of each tracer particle to a fixed value before post-processing. In this way, we have investigated the metallicity dependence on the production of all stable isotopes in the SNIa explosion.

Metallicity dependencies of the ejected elemental yields are negligible in all cases below a mass fraction of $^{22}$Ne of 10$^{-4}$, corresponding to a metallicity 1/100 that of solar. Significant metallicity effects of a factor of 2 or greater start at a $^{22}$Ne mass fraction of around 10$^{-3}$. Elemental abundances change, in most cases, by a factor of a few however for some elements, such as V, with close to mono-isotopic contributions the change in the ejected elemental abundances can be a factor of 5 or 10.

Isotopic abundances are generally more sensitive to the changing $^{22}$Ne content, and occur at lower initial $^{22}$Ne mass fractions; however this is a general statement and each isotope should be considered separately. The impact of a changing initial metallicity is mostly observed in the cooler outer regions of the WD below around 4 GK. There is a limited effect on isotopic distribution in the Chandrasekhar mass models as electron captures in the high density central regions cause the electron fraction to converge to a value determined by the central density, washing out the effect of initial composition. Our results agree well with the current literature available for a range of SNIa masses, and provide much needed metallicity dependence to the available suite of yields. 

This work presents a comprehensive suite of models for galactic chemical evolution applications, with a consistent nuclear reaction network between the various masses of model. In this way, we have disentangled the nuclear physics uncertainties from those associated with the explosion models. Further work should focus on fully modeling the accreted helium shell of the double detonation models (S1.0, M0.8), with full realistic abundance distributions. While the effect of seed nuclei in the majority of the CO core will be negligible, due to the same considerations as for the electron fraction, there can be a significant impact on the nucleosynthesis in the cooler layers of the WD, resulting in p-nuclei production. This should be fully investigated in order to accurately describe the contributions of SNIa to GCE models.

Our models provide an important first step in producing detailed, metallicity dependent models of SNIa explosions which will be used to identify tracers for the sub- to Chandrasekhar mass progenitor ratio, and identify the burning conditions involved in the production of the full range of SNIa ejecta. It is noted that, due to the small production of $^{56}$Fe in the M0.8 model - with a total ejected mass ranging from 4.87$\times$10$^{-2}$ to 2.32$\times$10$^{-2}$ in mass fraction between Z$_{met}$ = 0 to Z$_{met}$ = 0.1 - that the explosion of the M0.8 model would not resemble a standard SNIa explosion, as the production of $^{56}$Ni during the explosion is too small.

\section{Acknowledgments}

The authors thank the {\tt viper} High Performance Computing Facility team at the University of Hull for the allocation of computing time. We acknowledge support from the National Science Foundation (Grant: PHY-1430152), the Science and Technology Facilities Council (through the
University of Hull’s Consolidated Grant ST/R000840/1), the European Union’s Horizon 2020 research and innovation program (ChETEC-INFRA -- Project no. 101008324) and ChETEC COST Action CA16117 through grant E-COST-GRANT-CA16117-a8de8b83. MP thanks the "Lendulet-2014" Program of the Hungarian Academy of Sciences (Hungary) and the ERC Consolidator Grant (Hungary) funding scheme (Project RADIOSTAR, G.A. n. 724560). This work was supported by the U.S. Department of Energy through the Los Alamos National Laboratory. Los Alamos National Laboratory is operated by Triad National Security, LLC, for the National Nuclear Security Administration of U.S. Department of Energy (Contract No. 89233218CNA000001).


\bibliography{bib_SNIa_yields}{}
\bibliographystyle{aasjournal}

\end{document}